\documentclass[reprint, superscriptaddress]{revtex4-1}
\usepackage{amsmath}
\usepackage{amssymb}
\usepackage{mathtools}
\usepackage{color}
\usepackage{url, hyperref}
\usepackage{bbold}
\usepackage{color}
\usepackage{graphicx}

\begin{document}
\title{Coherent photon manipulation in interacting atomic ensembles}

\begin{abstract}
Coupling photons to Rydberg excitations in a cold atomic gas yields unprecedentedly large optical nonlinearities at the level of individual light quanta, where the formation of nearby dark-state polaritons is blocked by the strong interactions between Rydberg atoms. This blockade mechanism, however, realizes an inherently dissipative nonlinearity, which limits the performance of practical applications. In this work, we propose a new approach to strong photon interactions via a largely coherent mechanism at drastically suppressed photon losses. Rather than a polariton blockade, it is based on an interaction induced conversion between distinct types of dark-state polaritons with different propagation characteristics. We outline a specific implementation of this approach and show that it permits to turn a single photon into an effective mirror with a robust and continuously tuneable reflection phase. We describe potential applications, including a detailed discussion of achievable operational fidelities. 
\end{abstract}

\author{Callum R. Murray}
\author{Thomas Pohl}
\affiliation{Max Planck Institute for the Physics of Complex Systems, N\"othnitzer Stra\ss e 38, 01187 Dresden, Germany}
\affiliation{Department of Physics and Astronomy, Aarhus University, Ny Munkegade 120, DK 8000 Aarhus C, Denmark}

\maketitle

\section{Introduction}

The notion that photons are devoid of mutual interactions in vacuum is well-rooted in our elementary understanding of light. Nevertheless, the ability to engineer such interactions synthetically would hold profound implications for both fundamental and applied Science \cite{OBrien2007, OBrien2009,Kok2007,Kimble2008}, and has since ushered in a new era of research into nonlinear optics at the ultimate quantum level \cite{Chang2014}. Intense efforts have been directed towards enhancing light-matter coupling through tight mode confinement \cite{Reiserer2015, Reiserer2013, Reiserer2014, Shahmoon2014, Lodahl2015, Javadi2015, Sollner2015, Hwang2009, Maser2016, Shomroni2014, Thompson2013, Tiecke2014, Goban2014} in order to achieve local nonlinearities by interfacing photons with a single quantum emitter. A complementary strategy, which is rapidly gaining momentum, exploits the collective coupling of light to particle ensembles with finite-range interactions \cite{Hofferberth2016, Murray2016, Douglas2015, Shi2015, Shahmoon2016, Douglas2016} to establish large and nonlocal nonlinearities.

Interfacing light with strongly interacting atomic Rydberg ensembles \cite{Singer2004,Tong2004,Liebisch2005,Heidemann2007,Isenhower2010,Wilk2010,Viteau2011,Ebert2014,Schempp2014,Malossi2014,Urvoy2015,Ebert2015,Faoro2016,Saffman2010} under conditions of electromagnetically induced transparency (EIT) \cite{Fleischhauer2005} has emerged as a particularly promising way to implement this new type of mechanism \cite{Friedler2005, Pritchard2010,Schempp2010, Ates2011, Sevincli2011,Petrosyan2011,Gorshkov2011,Gorshkov2013,Petrosyan2016,Das2016,Hofferberth2016, Murray2016}. EIT in these systems is based on the formation of Rydberg dark-state polaritons, which correspond to coherent superposition states of light and matter that are immune to dissipation \cite{Fleischhauer2000, Fleischhauer2002}. Whilst this polariton formation supports the lossless and form-stable propagation of single photons, the strong mutual interaction between two such polaritons can easily perturb and break the underlying EIT condition, thereby rendering light propagation highly nonlinear \cite{Dudin2012, Peyronel2012, Firstenberg2013}. Indeed, there have now been a number of experiments that demonstrated controllable photon-photon interactions of unprecedented strength in such systems \cite{Pritchard2010, Parigi2012, Dudin2012, Peyronel2012, Firstenberg2013, Maxwell2013, Maxwell2014, Baur2014, Tiarks2014, Gorniaczyk2014, Gorniaczyk2016, Tiarks2016, Tresp2016a}.

Key to this nonlinearity is the destruction of EIT conditions that originates from an effective polariton blockade, whereby multiple proximate photons are prevented from simultaneously forming dark-state polaritons. As an immediate consequence, the emergent photon interactions inevitably carry an intrinsic dissipative component. Nevertheless, the nonlinear quantum optical response achieved in this way can be utilised to facilitate a broad range of applications, from imaging \cite{Gunter2012,Gunter2013} all-optical switches and transistors \cite{Baur2014, Tiarks2014, Gorniaczyk2014, Gorniaczyk2016, Murray2016a} to quantum gates \cite{Gorshkov2011, Tiarks2016} as well as single photon sources \cite{Dudin2012, Gorshkov2013} and subtractors \cite{Tresp2016a}. Yet, it turns out that high fidelity operations require conditions (e.g., high atomic densities) where the performance of such applications is ultimately eclipsed by additional decoherence effects \cite{Murray2016a, Zeuthen2016, Baur2014, Gaj2014}.

\begin{figure}[t]
\begin{center}
\includegraphics[width=0.9\columnwidth]{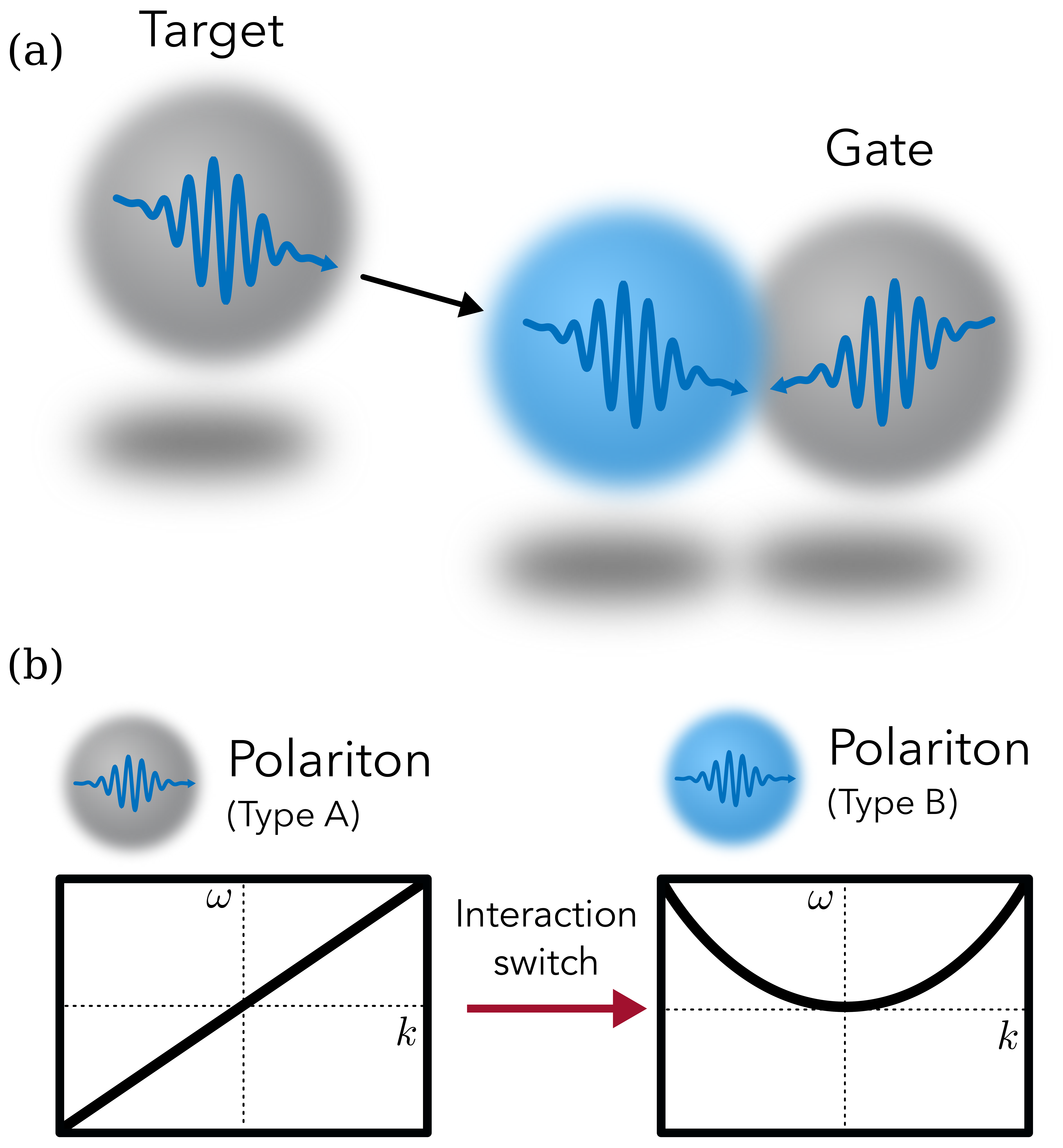}
\end{center}
\caption{\label{fig: Polariton switch} Illustration of the basic principle of nonlinear polariton switching. (a) A photon (target) propagates initially as one type of dark-state polariton (type A, grey sphere), but is subsequently converted to another kind of dark-state polariton (type B, blue sphere) upon interacting with a second (gate) polariton. (b) This induces a change in the dispersion relation that governs the propagation of light and thereby mediates an effective photon interaction at greatly suppressed losses.}
\end{figure}

As a solution to this outstanding issue, we describe here a novel approach to quantum optical nonlinearities in a Rydberg-EIT medium \emph{without the polariton-blockade}. It exploits the atomic interactions to modify EIT conditions, rather than destroying them entirely. Generally, the devised strategy can thus be understood as a dark-state polariton switch, as opposed to the existing schemes based on the polariton-blockade, Fig.\ref{fig: Polariton switch}(a). Consequently, this new mechanism globally preserves EIT conditions such that nonlinear dissipation is intrinsically suppressed, thereby alleviating the decoherence related hindrances discussed in \cite{Murray2016a, Zeuthen2016}.

We outline a specific implementation that can be realized with minimal extensions to current experiments \cite{Peyronel2012, Tiarks2014, Gorniaczyk2014} and is shown to yield a conditional coupling between two distinct photonic modes. In particular, we show how this can be used to establish a \textit{reflective} nonlinearity, in which a single photon stored in a Rydberg spin wave excitation acts as an effective mirror, capable of reflecting photons with an arbitrary and continuously tunable reflection-phase. The described realization of interaction-induced polariton switching can thus function as a single photon router, facilitating a broad range of applications from quantum transistors to photonic gate operations. Finally, we discuss the performance of such applications based on current technology and in relation to previous blockade-based approaches.

\section{Nonlinear polariton switching}
Rydberg dark-state polaritons acquire the properties of their constituents, inheriting kinetics from their photonic admixture, and interactions from their atomic Rydberg state component. Typically, these interactions are of a van der Waals type, causing a level shift $V(z)=C_6/z^6$ of the Rydberg state attached to one polariton when it interacts with another at a distance $z$. The van der Waals coefficient $C_6\sim n^{11}$ increases rapidly with the principal quantum number $n$ of the chosen Rydberg state \cite{Saffman2010}, ultimately providing interaction strengths that vastly exceed any other energy scale in the system. It can thus be used to achieve a polariton blockade by breaking EIT conditions in current approaches to nonlinear optics based on Rydberg-EIT \cite{Friedler2005, Pritchard2010, Ates2011, Sevincli2011,Petrosyan2011,Gorshkov2011,Gorshkov2013,Hofferberth2016, Murray2016}. 

On the contrary, we will consider here a situation where this level shift is rather used to establish a switching mechanism between different types of dark-state polaritons. Consequently, the corresponding nonlinear optical response should thus be associated with minimal refraction and absorption, but instead modify the dispersion relation which characterises the photon propagation. As a specific example we will consider a situation in which the onset of interactions serves to cancel the linear dispersion of light and establish a locally quadratic dispersion, Fig.\ref{fig: Polariton switch}(b). This corresponds to a nonlinear switching between so-called slow-light \cite{Harris1992, Fleischhauer2000, Fleischhauer2002} and stationary-light \cite{Andre2002, Andre2005, Fleischhauer2008, Zimmer2008, Hafezi2012, Iakoupov2016} polaritons, both of which have been separately demonstrated in \cite{Hau1999, Kash1999} and \cite{Bajcsy2003, Lin2009, Everett2016} respectively. 

To leading order in their bandwidth, slow-light polaritons are predominantly governed by a linear dispersion relation \cite{Fleischhauer2000,Fleischhauer2002}, such that their one-dimensional propagation dynamics obeys the following simple equation 
\begin{equation}
\label{eq: slow light eom}
\partial_t \hat{\Psi}(z,t) = v \partial_z \hat{\Psi}(z,t),
\end{equation}
which describes the form-stable linear propagation of the polariton anihilation operator $\hat{\Psi}(z,t)$ with a group velocity $v$ (typically much less than the vacuum speed of light $c$ \cite{Hau1999}). 

Stationary-light polaritons, on the other hand, feature a quadratic dispersion relation and are realised in situations where EIT is achieved for coupled pairs counter-propagating optical modes \cite{Andre2002, Bajcsy2003, Lin2009, Andre2005, Fleischhauer2008, Zimmer2008, Hafezi2012, Everett2016, Iakoupov2016}. The quadratic dispersion in this case effectively endows polaritons with a kinetic energy and mass \cite{Fleischhauer2008, Hafezi2012}, akin to massive particles. To leading order in the polariton bandwidth, the corresponding operator $\hat{\Phi}(z,t)$ describing the annihilation of a stationary-light polariton, thus, obeys the one-dimensional evolution equation \cite{Fleischhauer2008, Hafezi2012}
\begin{equation}
\label{eq: stationary light eom}
\partial_t \hat{\Phi}(z,t) = -\frac{1}{2m} \partial_z^2 \hat{\Phi}(z,t),
\end{equation}
where $m$ is the effective mass acquired by $\hat{\Phi}(z,t)$. 

\begin{figure*}[]
\begin{center}
\includegraphics[width=0.9\textwidth]{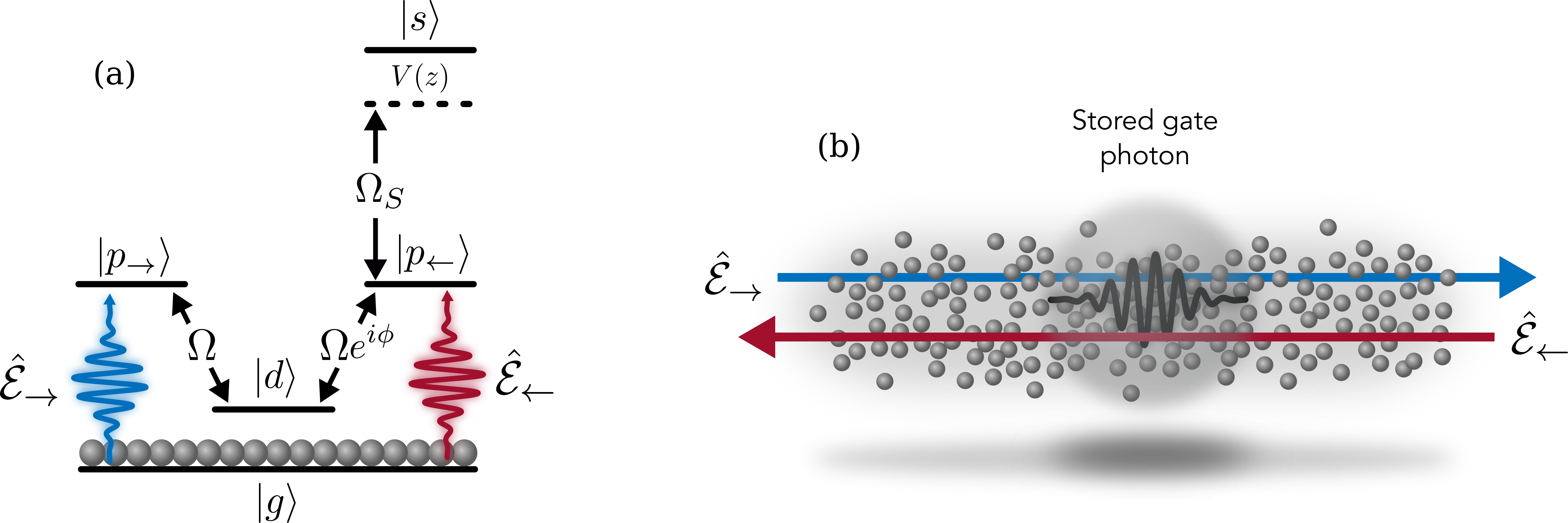}
\end{center}
\caption{\label{fig: Schematic}(a) Schematics of the considered coupling scheme. Classical fields with indicated Rabi frequencies $\Omega$ and $\Omega_S$ establish EIT conditions for the counter propagating photonic modes $\hat{\mathcal{E}}_{\rightarrow}$ and $\hat{\mathcal{E}}_{\leftarrow}$. The Rydberg state $|s\rangle$ is subject to a spatially dependent level shift $V(z)$ upon interacting with the stored gate excitation as illustrated in panel (b). This shift modifies the underlying EIT conditions for $\hat{\mathcal{E}}_{\rightarrow}$ and $\hat{\mathcal{E}}_{\leftarrow}$, rather than perturbing them.}
\end{figure*}

In order to engineer a Rydberg mediated switching between the different types of polaritons described by eq.(\ref{eq: slow light eom}) and eq.(\ref{eq: stationary light eom}), we propose the level structure shown in Fig.\ref{fig: Schematic}(a). Here, EIT is achieved for two counter-propagating light fields described by the field operators $\hat{\mathcal{E}}_{\rightarrow}^{\dagger}$ and $\hat{\mathcal{E}}_{\leftarrow}^{\dagger}$, which create a probe photon in the right and left moving mode respectively. As we will see below, the precise nature of the resulting dark state is however controlled by the interaction induced level shift of the Rydberg state $|s\rangle$ to which the fields are coupled. Specifically,  our proposed coupling scheme is shown to support slow-light EIT conditions for $\hat{\mathcal{E}}_{\rightarrow}$ and $\hat{\mathcal{E}}_{\leftarrow}$ separately in the limit of weak interactions, while it facilitates stationary-light EIT for fully blocked Rydberg excitations.

\section{Interaction with a stored spin wave}

The proposed polariton switching mechanism is best analysed by considering the conceptually simplest type of photon-photon interaction, whereby a (gate) photon is first stored in the atomic ensemble \cite{Fleischhauer2002, Novikova2007, Gorshkov2007, Gorshkov2007a} as a collective Rydberg spin wave excitation. Subsequently, a second (target) photon is sent through the medium and made to interact with the stored gate excitation [see Fig.\ref{fig: Schematic}(b)]. This approach provides a well controlled way to engineer two-photon interactions \cite{Gorshkov2011}, and has been demonstrated in a number of recent experiments \cite{Baur2014, Tiarks2014, Gorniaczyk2014, Gorniaczyk2016, Tiarks2016}. 

The relevant atomic excitations are described by the continuous field operators \cite{Fleischhauer2002} $\hat{P}^{\dagger}_{\rightleftarrows}(z,t)$, $\hat{D}^{\dagger}(z,t)$ and $\hat{S}^{\dagger}(z,t)$ which create an excitation in $|p_{\rightleftarrows}\rangle$, $|d\rangle$ and $|s\rangle$,  respectively, at position $z$. Moreover, we introduce the operator $\hat{S}_g^{\dagger}(z,t)$ that creates a stored gate excitation in an auxiliary Rydberg state $|s_g\rangle$ that is not laser coupled during the probe stage \cite{Gorshkov2011,Baur2014, Tiarks2014, Gorniaczyk2014, Gorniaczyk2016, Tiarks2016}. Along with $\hat{\mathcal{E}}_{\rightarrow}$ and $\hat{\mathcal{E}}_{\leftarrow}$, all of these field operators satisfy  Bosonic commutation relations \cite{Fleischhauer2002}. 

In a rotating frame, the one-dimensional dynamics of this system are governed by the following (non-Hermitian) Hamiltonian
\begin{equation}
\label{eq: Hamiltonian}
\begin{split}
\hat{H} &= -ic \int_{-\infty}^{\infty} dz \left( \hat{\mathcal{E}}^{\dagger}_{\rightarrow}(z) \partial_z \hat{\mathcal{E}}_{\rightarrow}(z) - \hat{\mathcal{E}}^{\dagger}_{\leftarrow}(z) \partial_z \hat{\mathcal{E}}_{\leftarrow}(z) \right) \\
& ~~~ + G  \int_0^L dz\left( \hat{\mathcal{E}}^{\dagger}_{\rightarrow}(z) \hat{P}_{\rightarrow}(z) + \hat{\mathcal{E}}^{\dagger}_{\leftarrow}(z) \hat{P}_{\leftarrow}(z) + \text{h.c.} \right) \\
& ~~~ - i \gamma \int_0^L dz\left(\hat{P}^{\dagger}_{\rightarrow}(z) \hat{P}_{\rightarrow}(z) + \hat{P}^{\dagger}_{\leftarrow}(z) \hat{P}_{\leftarrow}(z) \right) \\
& ~~~ + \Omega \int_0^L dz \left( \hat{D}^{\dagger}(z) \hat{P}_{\rightarrow}(z)  + e^{i\phi} \hat{D}^{\dagger}(z) \hat{P}_{\leftarrow}(z) + \text{h.c.} \right) \\
& ~~~ + \Omega_S \int_0^L dz \left( \hat{S}^{\dagger}(z)\hat{P}_{\leftarrow}(z) + \text{h.c.} \right)  \\
& ~~~ + \int_0^L \int_0^L dz dx V(z - x) \hat{S}_g^{\dagger}(x) \hat{S}^{\dagger}(z) \hat{S}(z) \hat{S}_g(x).
\end{split}
\end{equation}
where $L$ is the length of the medium. For simplicity, we assume that $|p_{\rightarrow}\rangle$ and $|p_{\leftarrow}\rangle$  decay with the scattering rate $2\gamma$ and that the probe photon modes couple to their respective transitions with $G\equiv g\sqrt{\rho_{\rm a}}$, where $\rho_{\rm a}$ is the homogenous atomic density and $g$ is the single photon coupling strength. The state $|d\rangle$ is coupled to $|p_{\rightarrow}\rangle$ and $|p_{\leftarrow}\rangle$ by classical control fields with identical Rabi frequencies $\Omega_\rightarrow=\Omega_\leftarrow=\Omega$, while we allow for a relative relative phase difference $\phi$ between them. Finally, the state $|p_{\leftarrow}\rangle$ is coupled by another classical field to the Rydberg state $|s\rangle$ with a Rabi frequency $\Omega_S$. The last term in eq.(\ref{eq: Hamiltonian}) accounts for the spatially dependant level shift of the Rydberg state $|s\rangle$ due to its van der Waals interaction with the stored gate excitation. The typical range over which this shift affects the probe photon propagation can be characterized by the blockade radius $z_b$ according to $V(z_b)=\Omega_S^2/\gamma$ \cite{Gorshkov2011}.

\section{Polariton analysis}

\begin{figure*}[t]
\begin{center}
\includegraphics[width=0.9\textwidth]{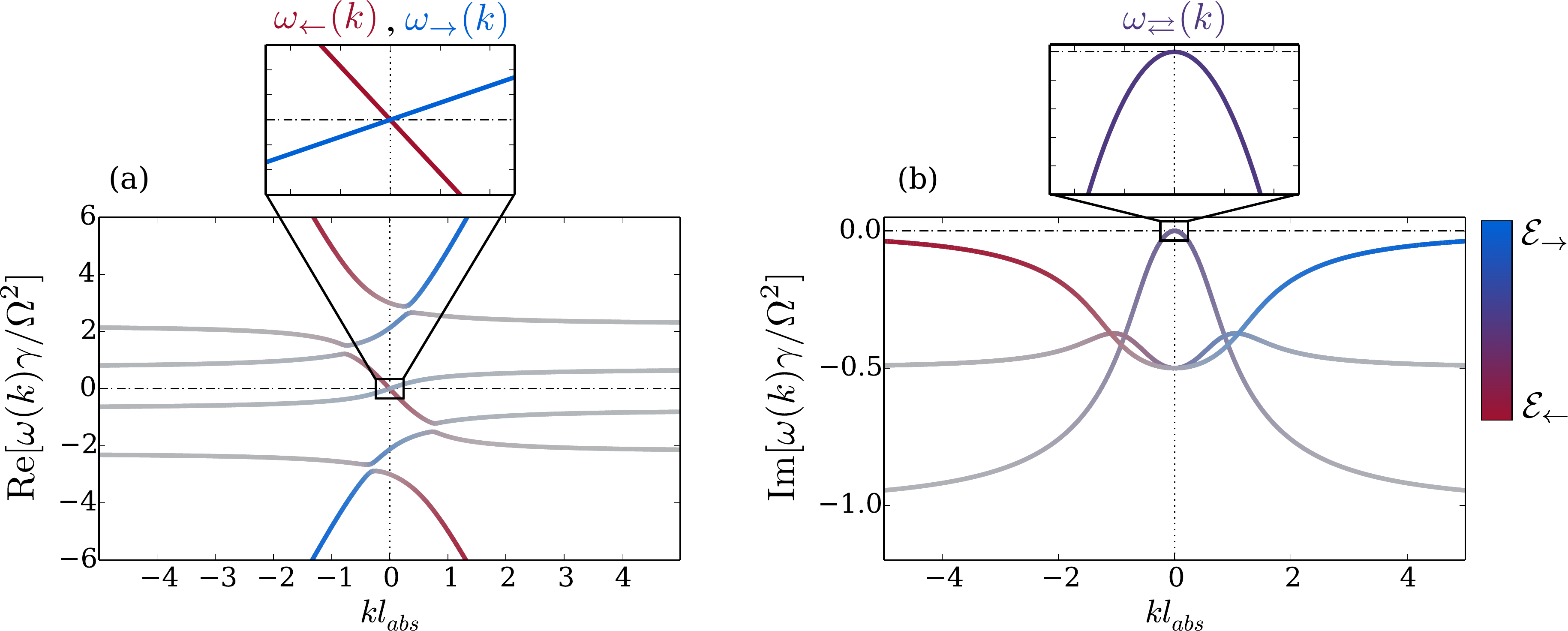}
\end{center}
\caption{\label{fig: dispersion relations} (a) Real part of the polariton spectrum in the absence of interactions, indicating the emergence of two slow-light dark state polaritons. These are separately governed by the linear dispersion relations $\omega_{\rightarrow}(k)$ and $\omega_{\leftarrow}(k)$ as given by eqs.(\ref{eq: Dispersion relation - right}) and (\ref{eq: Dispersion relation - left}), respectively. Here, $\Omega_S/G=1$, $\Omega/G=0.5$ and $\gamma/\Omega=1$. (b) Imaginary part of the polariton spectrum for strong interactions, i.e., under conditions of a complete Rydberg blockade. In this case one finds a single stationary-light dark state polariton described by a quadratic dispersion relation $\omega_{\leftrightarrow}(k)$, eq.(\ref{eq: Dispersion relation - stationary}). Here, $G/\Omega=1$ and $\gamma/\Omega=1$. For each bright state branch, there are two solutions with identical $\text{Im}[\omega(k)]$. The blue-red color coding indicates the relative fraction of $\hat{\mathcal{E}}_{\rightarrow}$ and $\hat{\mathcal{E}}_{\leftarrow}$ comprising the underlying state of each polariton branch, while the greyscale indicates the overall atomic fraction.}
\end{figure*}

Having established the basic idea and the specifics of the considered setup, let us now discuss the characteristics of the dark state polaritons involved in the underling switching protocol. The relevant dark-state polaritons can be identified as the zero-energy eigenstate solutions of the Hamiltonian eq.(\ref{eq: Hamiltonian}) in the two limiting cases $V(z)\to0$ and $V(z)\to\infty$, i.e., for vanishing interactions and in the limit of a complete Rydberg-state blockade.

Focusing first of all on the non-interacting situation, diagonalising the system Hamiltonian in the absence of photon dispersion yields two dark state polaritons of the form
\begin{align}
\label{eq: right polariton}
\hat{\Psi}_{\rightarrow} & = \frac{1}{\mathcal{N}_{\rightarrow}} \left[ \Omega \Omega_S \hat{\mathcal{E}}_{\rightarrow} - G \left( \Omega_S\hat{D}  - \Omega e^{i\phi} \hat{S} \right) \right], \\
\label{eq: left polariton}
\hat{\Psi}_{\leftarrow} & = \frac{1}{\mathcal{N}_{\leftarrow}} \left[ \Omega_S \hat{\mathcal{E}}_{\leftarrow} - G \hat{S} \right],
\end{align}
where $\mathcal{N}_{\rightarrow}=\sqrt{\Omega^2 \Omega_S^2 + G^2\left(\Omega^2 + \Omega_S^2 \right)}$ and $\mathcal{N}_{\leftarrow}=\sqrt{\Omega^2 + \Omega_S^2}$ are the normalisation factors required to obtain standard Bosonic commutation relations for $\hat{\Psi}_{\rightarrow}$ and $\hat{\Psi}_{\leftarrow}$. These polaritons can be accredited to two separate slow-light EIT schemes supported simultaneously by the level structure in  Fig.\ref{fig: Schematic}(a): the 5-level system formed by $|g\rangle$, $|p_{\rightarrow}\rangle$, $|d\rangle$, $|p_{\leftarrow}\rangle$ and $|s\rangle$ (establishing EIT for $\hat{\mathcal{E}}_{\rightarrow}$), and the 3-level system formed by $|g\rangle$, $|p_{\leftarrow}\rangle$ and $|s\rangle$ (establishing EIT for $\hat{\mathcal{E}}_{\leftarrow}$). We emphasise that the dark-state nature of $\hat{\Psi}_{\rightarrow}$ and $\hat{\Psi}_{\leftarrow}$ ensures that there is no coupling between the two underlying photonic modes. Hence, a right moving input photon will undergo low-loss and form-stable propagation through the medium, and so will a left moving photon.

The dispersion relations, $\omega_{\rightarrow}(k)$ and $\omega_{\leftarrow}(k)$, governing this propagation dynamics are readily obtained from a momentum space formulation of eq.(\ref{eq: Hamiltonian}). To leading order in the photon momentum $k$ (and the ratio $\Omega/\Omega_S$) one finds
\begin{align}
\label{eq: Dispersion relation - right}
\omega_{\rightarrow}(k) & \approx  c \left[ \frac{\Omega^2}{G^2 + \Omega^2} \right] k + \mathcal{O}[k^2]\:, \\
\label{eq: Dispersion relation - left}
\omega_{\leftarrow}(k) & \approx - c \left[ \frac{\Omega_S^2}{G^2 + \Omega_S^2} \right] k + \mathcal{O}[k^2]\: .
\end{align}
As expected, one finds linear dispersion relations describing a form-stable propagation of the slow-light polaritons with group velocities $v_{\rightarrow} = \frac{d \omega_{\rightarrow}}{dk}$ and $v_{\leftarrow} = \frac{d \omega_{\rightarrow}}{dk}$ respectively. The two polaritons propagate in opposite directions with $v_\rightarrow=-\frac{\Omega^2 }{ \Omega_S^2} v_\leftarrow$ under the typical condition $G\gg \Omega,\Omega_S$. This is further illustrated in Fig.\ref{fig: dispersion relations}(a), where we plot the complete polariton spectrum admitted in this non-interacting situation, indeed revealing the emergence of two dark-state polariton branches at $k=0$ corresponding to $\hat{\Psi}_{\rightarrow}$ and $\hat{\Psi}_{\leftarrow}$.

Now we consider the polariton spectrum admitted well within a blockade radius away from the stored spin wave, i.e. under strong blockade conditions corresponding to $V(z)\rightarrow \infty$. In this case, the shifted Rydberg state exposes a modified effective level structure corresponding to a so-called dual-V coupling scheme \cite{Fleischhauer2008, Iakoupov2016}, which can support stationary-light phenomena. For this system one finds the emergence of only a single dark-state polariton $\hat{\Phi}$ of the type in eq.(\ref{eq: stationary light eom}). Diagonalising the underlying system Hamiltonian (\ref{eq: Hamiltonian}), again in the absence of photon kinetics, one finds $\hat{\Phi}$ to be of the following form,
\begin{equation}
\label{eq: stationary light polariton 1}
\hat{\Phi} = \frac{1}{\mathcal{N}} \left[ \Omega \left(\hat{\mathcal{E}}_{\rightarrow} + e^{i\phi} \hat{\mathcal{E}}_{\leftarrow} \right) - G \hat{D} \right],
\end{equation}
where $\mathcal{N}=\sqrt{\Omega^2 + G^2}$ is the normalisation factor. In contrast to the non-interacting limit, a coherent coupling is now established between the two optical modes $\hat{\mathcal{E}}_{\rightarrow}$ and $\hat{\mathcal{E}}_{\leftarrow}$. This is reflected in the photonic composition of $\hat{\Phi}$, which is composed of the symmetric superposition state of the optical fields, $\hat{\mathcal{E}}_+=\frac{1}{\sqrt{2}}(\hat{\mathcal{E}}_{\rightarrow} + e^{i\phi} \hat{\mathcal{E}}_{\leftarrow})$.

The corresponding dispersion relation, $\omega_{\leftrightarrow}(k)$, that governs the propagation of $\hat{\Phi}$ can be determined in a similar fashion to before, and reads 
\begin{equation}
\label{eq: Dispersion relation - stationary}
\omega_{\leftrightarrow}(k) \approx -i2 l_{\rm abs} \frac{c\Omega^2}{G^2 + 2 \Omega^2} k^2 + \mathcal{O}[k^3] \\
\end{equation}
to lowest order in the photon momentum $k$, where $l_{\text{abs}}=c\gamma/G^2$ is the resonant two-level absorption length. Indeed, the obtained dispersion is quadratic in $k$, such that $\hat{\Phi}$ behaves as a stationary-light polariton. Fig.\ref{fig: dispersion relations}(b) shows the complete polariton spectrum for $V(z)\to\infty$, and illustrates the above discussion of the dark-state polariton.

\section{Photon propagation}
In order to develop an intuitive physical picture of the target photon dynamics, we will first model the stored gate excitation as a spatially localized Rydberg impurity, and generalise this analysis to the consideration of a collective spin wave state in Sec.\ref{sec:coherence}. First, we transform into the Schr\"odinger picture. Introducing $|\psi(t)\rangle$ as the general time dependent wave function of the system, we define the two-body amplitudes $\mathcal{E}_{\rightarrow}(z, x, t)=\langle 0 | \hat{\mathcal{E}}_{\rightarrow}^{\dagger}(z, t) \hat{S}_g^{\dagger}(x, t) |\psi\rangle$ and $\mathcal{E}_{\leftarrow}(z, x, t)=\langle 0 | \hat{\mathcal{E}}_{\leftarrow}^{\dagger}(z, t) \hat{S}_g^{\dagger}(x, t) |\psi\rangle$ corresponding to a stored gate excitation at position $x$ and a target photon at position $z$ in right and left moving mode, respectively.
Denoting the temporal Fourier transform of $\mathcal{E}_{\rightarrow}(z, x, t)$ and $\mathcal{E}_{\leftarrow}(z, x, t)$ by $\tilde{\mathcal{E}}_{\rightarrow}(z, x, \omega)$ and $\tilde{\mathcal{E}}_{\leftarrow}(z, x, \omega)$, the photon dynamics can be formulated in terms of a matrix equation of the form 
\begin{equation}\label{eq: MatEq}
i\partial_z \mathbf{E}(z, x,\omega) = \mathbf{M}(z - x, \omega) \mathbf{E}(z, x,\omega)\:,
\end{equation}
where $\mathbf{E}(z, x,\omega)=\{ \tilde{\mathcal{E}}_{\rightarrow}(z, x, \omega), \tilde{\mathcal{E}}_{\leftarrow}(z, x, \omega) \}$ and the propagation matrix is given by
\begin{equation}
\label{eq: Propagation matrix}
 \mathbf{M}(z,\omega) = \begin{bmatrix}
  \chi_{\rightarrow}(z, \omega) & \chi(z, \omega) e^{i\phi} \\
  -\chi (z, \omega) e^{-i\phi} & \chi_{\leftarrow} (z, \omega)
 \end{bmatrix}\:.
\end{equation}
The susceptibilities $\chi_{\rightarrow}(z, \omega)$ and $\chi_{\leftarrow}(z, \omega)$ characterize the propagation of $\tilde{\mathcal{E}}_{\rightarrow}$ and $\tilde{\mathcal{E}}_{\leftarrow}$ respectively, whilst $\chi(z, \omega)$ describes the coupling between the two modes. A derivation of eq.(\ref{eq: MatEq}) is provided in  Appendix \ref{appendix: Photon propagation equations}, along with the explicit expressions for the susceptibilities. 

In the continuous wave (CW) limit ($\omega\rightarrow0$) the propagation matrix in eq.(\ref{eq: Propagation matrix}) can be parameterised in terms of a single susceptibility $\chi_0(z) \equiv \chi_{\rightarrow}(z, 0) = - \chi_{\leftarrow}(z, 0) = -\chi(z, 0) $), given by
\begin{equation}
\label{eq: CW chi}
\chi_0(z) = \frac{d_b}{(z/z_b)^6 + 2i}\:.
\end{equation}
Here, we have defined $2d_b = 2 z_b / l_{\text{abs}}$ as the medium's optical depth per blockade radius. $\chi_0(z)$ basically characterises an effective potential through which the stored spin wave can affect the target photon propagation dynamics. In particular, one finds that $\chi_0(z)\to 0$ outside the blockade radius of the stored excitation, $|z|>z_b$, consistent with the slow-light EIT conditions supported in this region and the associated decoupling of the photonic modes. However, $\chi_0(z)$ approaches $-id_b/2$ within the blockade volume, reflecting the fact that a coupling between these modes is established, which gives rise to stationary-light EIT conditions.

Considering a target photon incident on the medium from the left at $z=0$, its transmission and reflection can then be characterised by the following relations
\begin{align}
\label{eq: Transmission relation}
\tilde{\mathcal{E}}_{\rightarrow}(L, x, \omega) & = T_n(\omega, x)\tilde{\mathcal{E}}_{\rightarrow}(0, x, \omega)\:, \\
\label{eq: Reflection relation}
\tilde{\mathcal{E}}_{\leftarrow}(0, x, \omega) & = R_n(\omega, x)\tilde{\mathcal{E}}_{\rightarrow}(0, x, \omega)\:,
\end{align}
where $T_n(\omega, x)$ and $R_n(\omega, x)$ are the transmission and reflection coefficients of the medium containing $n \in [0, 1]$ stored gate excitations at position $x$. 

Let us first consider the situation in which the gate excitation is absent. In this case, the target photon will initially generate the slow-light polariton described by $\hat{\Psi}_{\rightarrow}$ at the entrance of the medium. As described above, the formed polariton will then traverse the medium with a vanishing coupling to the counter propagating mode and experience full transmission under perfect EIT conditions. The actual mechanism underlying this decoupling of $\hat{\mathcal{E}}_{\rightarrow}$ and $\hat{\mathcal{E}}_{\leftarrow}$ can be traced back to quantum interference effects involving the dressed states of the laser-driven Rydberg transition. Specifically, the resonant coupling of $|p_{\leftarrow}\rangle$ and $|s\rangle$ via the classical field $\Omega_S$ [see Fig.\ref{fig: Schematic}(a)] establishes a pair of light shifted states, $|f_{\pm}\rangle=\left[|p_{\leftarrow}\rangle \pm |s\rangle \right]/\sqrt{2}$, which are shifted in energy by $\pm \Omega_S$ respectively. It is the destructive interference between competing excitation pathways involving these states that ultimately decouples the two modes of the target photon.

We note that this decoupling is exact on EIT resonance ($\omega=0$) for any finite value of $\Omega_S$. However, this is not true for a finite bandwidth of the target photon. In this case, a non-vanishing coupling is established between the off-resonant frequency components of $\tilde{\mathcal{E}}_{\rightarrow}(z,x,\omega)$ and $\tilde{\mathcal{E}}_{\leftarrow}(z,x,\omega)$. Such bandwidth limitations exist for any realistic EIT setting, but can be minimised through a proper choice of parameters. To establish these conditions, we first expand $|T_0(\omega)|$ to lowest order in $\omega$ as $|T_0(\omega)| \approx 1 - (\omega/\Delta\omega_0)^2$. $\Delta\omega_0$ then corresponds to the characteristic width of the transmission resonance, defining the range of frequencies over which the target photon is transparent, and is given explicitly as,
\begin{equation}
\label{eq: Transparency window}
\Delta\omega_0 = \Gamma \left[ \left( 1 + 2\frac{\Omega^2}{\Omega_S^2} + 2\frac{\Omega^4}{\Omega_S^4} \right)   + \frac{d}{2} \frac{\Omega^4}{\Omega_S^4} \right]^{-\frac{1}{2}},
\end{equation}
where $2d=2d_b L/z_{\rm b}$ is the total optical depth of the medium, and $\Gamma = \Omega^2/\gamma\sqrt{d}$.

\begin{figure}[!h]
\begin{center}
\includegraphics[width=0.8\columnwidth]{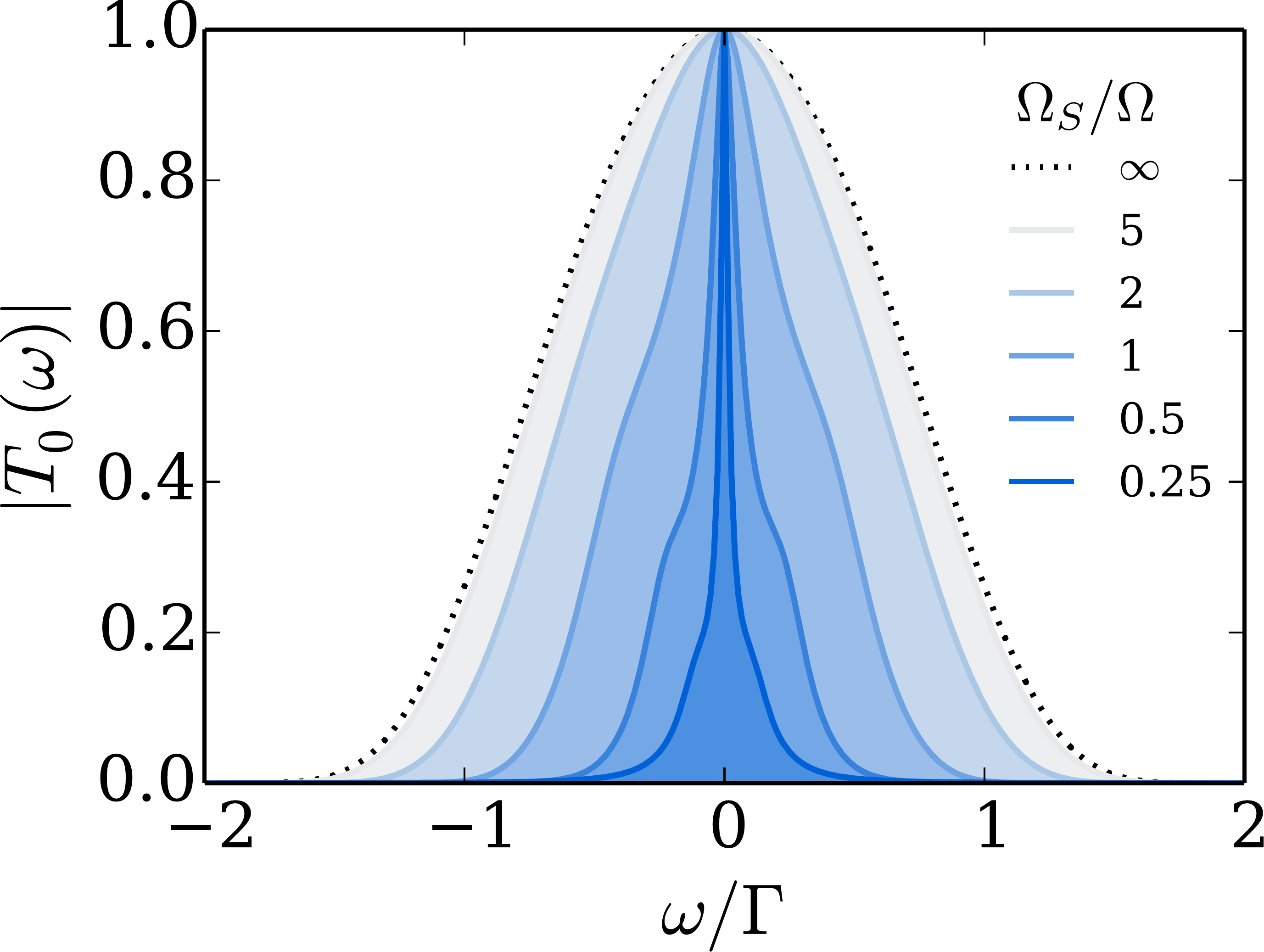}
\end{center}
\caption{\label{fig: Transmission} Transmission coefficient in the absence of a stored gate excitation for various indicated values of $\Omega_S/\Omega$. The total optical depth is $2d=50$, while $\gamma/\Omega=0.5$ and $G/\Omega=0.1$.}
\end{figure}

\begin{figure*}[t]
\begin{center}
\includegraphics[width=0.95\textwidth]{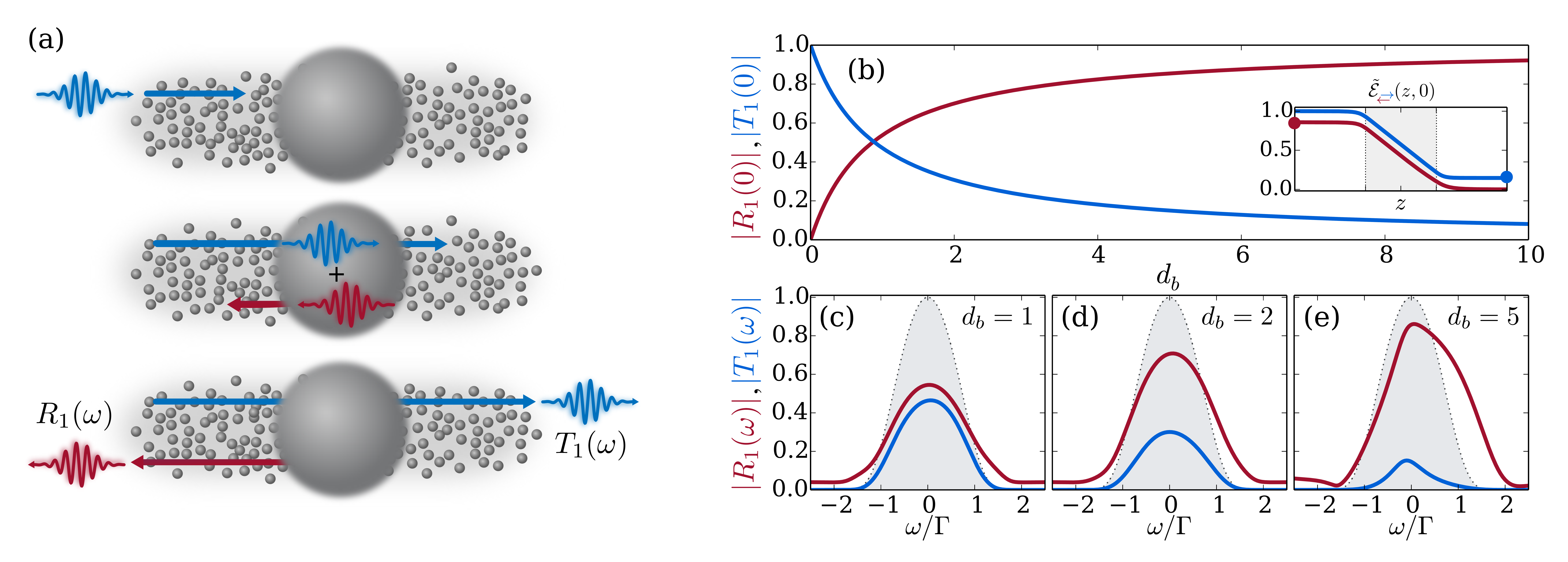}
\end{center}
\caption{\label{fig: Spinwave_present} (a) Illustration of target photon propagation in the presence of a stored gate excitation, where $R_1(\omega)$ and $T_1(\omega)$ are the reflection and transmission coefficients respectively. (b) The resonant values of $R_1(\omega=0)$ and $T_1(\omega=0)$ are shown as a function of $d_b$ in red and blue, respectively. The inset shows the spatial dependence of the photonic amplitudes $\hat{\mathcal{E}}_{\rightarrow}(z)$ and $\hat{\mathcal{E}}_{\leftarrow}(z)$ within the medium for $d_b=5$, where the grey shaded region indicates the extent of the blockade volume established by the stored gate excitation. (c-e) The transmission and reflection spectra $T_1(\omega)$ and $R_1(\omega)$ are shown in blue and red, respectively, for various indicated values of $d_b$, where the stored excitation is located at the centre of the medium. The total optical depth is $2d=50$, and the remaining parameters are $\gamma/\Omega = 0.5$, $\Omega/G = 0.1$, $\Omega_S/G = 0.5$. The transmission coefficient $T_0(\omega)$ in the absence of interactions is indicated by the grey shaded region for reference.}
\end{figure*}

According to eq.(\ref{eq: Transparency window}), the transparency width tends to a maximum value of $\Delta\omega_0 \to \Gamma$ when the conditions $2 \Omega^2/\Omega_S^2\ll1$ and $d ^2 \Omega^4/2\Omega_S^4\ll1$ are satisfied. The condition $d ^2 \Omega^4/2\Omega_S^4\ll1$ accounts for a weak oscillatory behaviour in $|T_0(\omega)|$. However, the effects of this only become observable at significant values of $d$, and can be safely neglected for our purposes. To understand the second condition, $2 \Omega^2/\Omega_S^2\ll1$, first notice that in the limit $\Omega^2/\Omega_S^2\to\infty$ the Rydberg state admixture of the $\hat{\Psi}_{\rightarrow}$-polariton vanishes, and the effective coupling scheme establishing EIT for $\hat{\mathcal{E}}_{\rightarrow}$ reduces to a $\Lambda$-system formed from $|g\rangle \leftrightarrow |p_{\rightarrow}\rangle \leftrightarrow |d\rangle$ [see Fig.\ref{fig: Schematic}(a)]. In this ideal limit, the optical response of the medium is then set solely by this effective EIT-scheme with an effective transparency width of $\Omega^2/\gamma\sqrt{d}$ \cite{Fleischhauer2002}, which is consistent with the limiting value of $\Delta\omega_0$. Physically though, $\Omega_S$ is required to be large since the effective decoupling of the photonic modes ultimately demands a large level splitting of the dressed states $|f_{\pm}\rangle$, and this is given by $\pm\Omega_S$.

To verify this picture, we plot the solution for $T_0(\omega)$ in Fig.\ref{fig: Transmission} for various ratios of $\Omega_S/\Omega$, and indeed find that the transmission spectrum converges to that of the $|g\rangle \leftrightarrow |p_{\rightarrow}\rangle \leftrightarrow |d\rangle$ $\Lambda$-system as $\Omega_S$ increases. Importantly, Fig.\ref{fig: Transmission} demonstrates that near optimal transmission is already reached for remarkably small ratios of $\Omega_S/\Omega$, that are well within current experimental capabilities. 

Let us now consider the propagation dynamics in the presence of a stored gate excitation. As before, upon entering the medium, the target photon propagates according to the linear dispersion relation $\omega_{\rightarrow}(\omega)$ in the form of a slow-light polariton, $\hat{\Psi}_{\rightarrow}$. Upon entering the blockade volume established by the stored excitation, however, the target photon is subject to stationary-light EIT conditions, as described in the preceding sections. In this case, a coherent coupling between the $\hat{\mathcal{E}}_{\rightarrow}$ and $\hat{\mathcal{E}}_{\leftarrow}$ modes is established and the symmetric photonic state $\hat{\mathcal{E}}_+$ is generated. Due to the lack of a linear dispersion, the photon can now only traverse the blockade region through a slow dispersion following $\omega_{\leftrightarrow}(k)=-ik^2/2m$. Upon crossing the blockade region the photon can reestablish the slow-light polariton, $\hat{\Psi}_{\rightarrow}$, and be transmitted through the medium. However, there is typically a larger amplitude for the photon to diffuse into the counter-propagating direction and exit the medium as a $\hat{\Psi}_{\leftarrow}$ polariton, corresponding to reflection by the stored excitation. This reflection bias is essentially due to a boundary effect; the photon is more likely to diffuse the much shorter distance in the counter-propagating direction upon entering the interaction volume than diffuse the full $2z_b$ length in the forward direction. The schematics of this overall process are depicted in Fig.\ref{fig: Spinwave_present}(a). 

To determine the relative importance of reflection and transmission, we can solve the propagation dynamics of the target photon exactly in the CW limit, for which the propagation matrix $\mathbf{M}$ in eq.(\ref{eq: Propagation matrix}) is parameterised solely by $\chi_0(z)$, eq.(\ref{eq: CW chi}). With the boundary conditions $\tilde{\mathcal{E}}_{\rightarrow}(z=0,x,0)=\tilde{\mathcal{E}}_0$ and $\tilde{\mathcal{E}}_{\leftarrow}(z=L,x,0)=0$, the solutions for the photonic fields can be readily obtained as
\begin{align}
\label{eq: Er(z)}
\tilde{\mathcal{E}}_{\rightarrow}(z,x,0) & = \left[ 1 - \frac{\nu(z,x)}{1 + \nu(L,x)}  \right] \tilde{\mathcal{E}}_0 \\
\label{eq: El(z)}
\tilde{\mathcal{E}}_{\leftarrow}(z,x,0) & = e^{-i\phi} \left[\frac{\nu(L,x) - \nu(z,x)}{1 + \nu(L,x)} \right] \tilde{\mathcal{E}}_0
\end{align}
where we have introduced $\nu(z,x) = i\int_{0}^{z}\chi_0(z^{\prime} - x)dz^{\prime}$. Assuming that the gate excitation is stored farther than $z_{\rm b}$ away from the medium boundaries, the problem effectively becomes independent off the spin wave position $x$. One can then write $\nu(L,x) \approx d_b \nu$, with $\nu =  i \int_{-\infty}^{\infty} dz/(z^6 + 2i) = (\pi\sqrt[3]{1 + i}/3) \approx 1.1 + 0.3i$, and obtain (from eq.(\ref{eq: Transmission relation}-\ref{eq: Reflection relation})) the following simple expressions for the transmission and reflection coefficients,
\begin{align}
\label{eq: T1}
T_1(\omega=0) & = \frac{1}{1 + \nu d_b}\;, \\
\label{eq: R1}
R_1(\omega=0) & = \frac{\nu d_b}{1 + \nu d_b} e^{-i\phi}\;.
\end{align}
We plot $T_1(0)$ and $R_1(0)$ in Fig.\ref{fig: Spinwave_present}(b), along with the spatial solutions of $\tilde{\mathcal{E}}_{\rightarrow}(z, 0)$ and $\tilde{\mathcal{E}}_{\leftarrow}(z, 0)$ within the medium. 

The reflection amplitude increases with $d_b$, and eventually dominates the transmission beyond $d_b\approx1$. Physically, this can understood from the effective mass $m\propto 1/l_{\rm abs}$ \cite{Fleischhauer2008} of the stationary-light polariton and the length, $2z_b$, of the blockade region. Since both increase with $d_b=z_b/l_{\rm abs}$, reflection dominates at large $d_b$ where traversing the entire blockade region is strongly suppressed such that photons predominantly exit in the opposite direction right at the incident boundary. 

Additionally, eq.(\ref{eq: R1}) shows that one can imprint an arbitrary phase onto the reflected photon. This phase, $\phi$, is the relative phase difference between the classical control fields that establish stationary-light conditions within the interaction volume [see Fig.\ref{fig: Schematic}] and, therefore, can be well controlled and tuned in a continuous manner. For large $d_b$, the phase is a pure result of the reflection physics and does not depend on any system parameter other than the relative phase between the two classical control fields. Our setup can thus function as a photonic quantum router with an arbitrary and continuously tunable reflection phase. This \emph{robust} conditional phase presents an important distinguishing feature compared to previous Rydberg-EIT based protocols \cite{Gorshkov2011, Tiarks2016}.

Fig.\ref{fig: Spinwave_present}(c) shows numerically obtained spectra $T_1(\omega)$ and $R_1(\omega)$ for various values of $d_b$. These are compared to the  transmission spectrum $T_0(\omega)$ in the absence of the stored spin wave. As $d_b$ increases one finds that the reflection spectrum develops an asymmetry. This asymmetry emerges since the Rydberg level shift, $V(z)$, induces a non-symmetric optical response, where the positive $(\omega > 0)$ and negative $(\omega < 0)$ frequency components are affected differently. However, the effects of this can be minimised by choosing a sufficiently narrow bandwidth of the target photon, as discussed in Sec.\ref{sec: Experimental considerations}.

\section{Coherence properties}\label{sec:coherence}

\begin{figure*}[t]
\begin{center}
\includegraphics[width=0.95\textwidth]{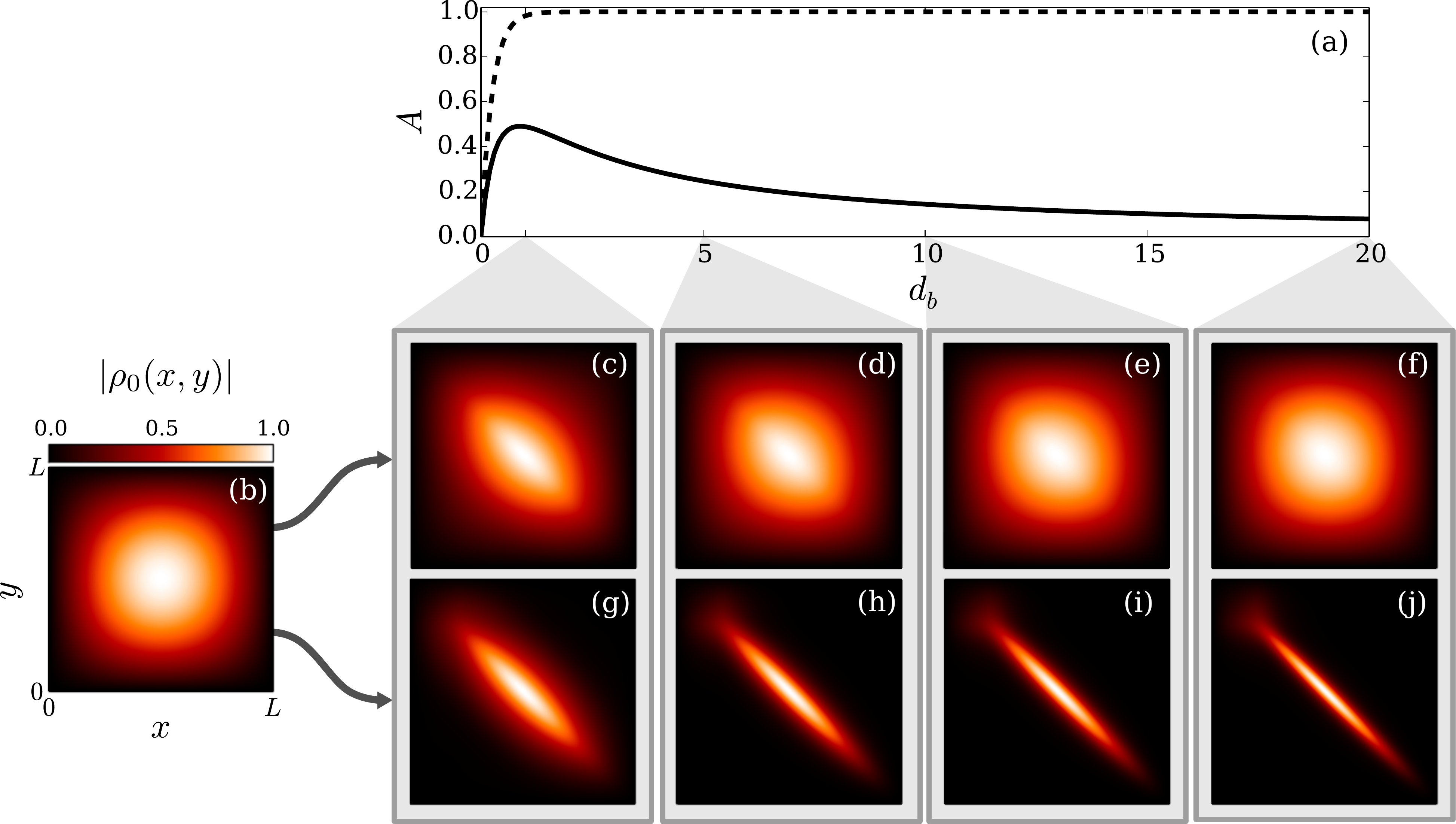}
\end{center}
\caption{\label{fig: Coherence} (a) Loss coefficient, $A$, as a function of $d_b$ (solid) compared with the conventional polariton blockade mechanism (dashed). (b) Initial density matrix, $\rho_0(x, y) = \sin(x \pi /L)\sin(y \pi /L)$, of the stored spin wave for $L=5z_b$. (c-f) Final density matrix of the stored spin wave corresponding to the present coherent polariton-switching mechanism for various indicated values of $d_b$. Panels (g-j) show the final density matrix for the dissipative polariton blockade, which is shown to cause much stronger decoherence.}
\end{figure*}

Another distinguishing feature of the described polariton switching mechanism is that it maintains EIT conditions and therefore operates at inherently suppressed photon losses. In Fig.\ref{fig: Coherence}(a) we show the loss $A = 1 - |T_1(0)|^2 - |R_1(0)|^2$ as a function of $d_b$. Remarkably, absorption decreases with increasing $d_b$, even though the target photon is resonantly coupled to the medium. This in turn permits to work under conditions of strong light-matter coupling, and stands in marked contrast to conventional Rydberg-EIT schemes where resonant photon coupling implies large interaction-induced losses $A \approx 1 - \exp[-4d_b]$ \cite{Gorshkov2011, Murray2016a, Baur2014, Tiarks2014, Gorniaczyk2014, Gorniaczyk2016} [see Fig.\ref{fig: Coherence}(a)] and one requires large single photon-detunings for coherent nonlinear operations \cite{Gorshkov2011, Firstenberg2013, Murray2016, Hofferberth2016}.

Absorption in the current situation originates from the fact that the target photon does not switch fully adiabatically between the slow and stationary light polariton solutions. Specifically, this means that the target photon partially populates the bright state polariton branches depicted in Fig.\ref{fig: dispersion relations}. However, the associated energy cost of $\sim G^2/\gamma=c l_{\rm abs}^{-1}$ ensures that this population is suppressed with increasing $d_b$ and leads to the observed decrease of the loss coefficient $A$. 

Thus far we have focussed on the dynamics of the target photon, where it was sufficient to consider a localized gate excitation at a given position in the medium. Storage of the gate photon, however, generates a spatially delocalised collective spin wave excitation, and the preservation of its coherent nature is essential for subsequent photon retrieval. Respectively, the retrieval efficiency is typically diminished by dissipative interactions with the incident target photon \cite{Murray2016a, Li2015a, Gorniaczyk2016}, as we shall discuss below.

To this end we consider the dynamics of the spin wave density matrix $\hat{\rho}(t) \coloneqq \int_0^{L}dx\int_0^{L}dy \rho(x, y, t) \hat{S}_g^{\dagger}(x, t)|0\rangle \langle 0| \hat{S}_g(y, t)$, where the complex elements $\rho(x, y, t)$ indicate the spatial coherence between spin wave components at positions $x$ and $y$. Using the theoretical framework developed in \cite{Murray2016a}, the final spin wave state after the interaction with the target photon can be obtained exactly in the CW limit and is given by
\begin{equation}
\label{eq: spin wave density matrix}
\begin{split}
\rho(x, y) = & \left\{ 1 + i d_b \left[ \frac{1}{1 + \nu(L,x)} \right] \left[ \frac{1}{1 + \nu^*(L,y)} \right] \right. \\
& \left. \times \int_0^L dz \frac{(z-x)^6 - (z-y)^6}{\left[ (z-x)^6 + 2i \right] \left[ (z-y)^6 -2i \right]} \right\} \rho_0(x, y),
\end{split}
\end{equation}
where $\rho_0(x, y)$ denotes the initial (pure) state of the stored gate excitation. A detailed derivation of eq.(\ref{eq: spin wave density matrix}) is presented in Appendix \ref{appendix: Spin wave decoherence dynamics}.

While the spin wave density, $\rho(x, x)$,  remains unaffected \cite{Murray2016a}, target photon scattering results in partial decoherence, i.e. a reduction of the off-diagonal elements of $\rho(x, y)$. Assuming that the length of the medium is significantly longer than the extent of the spin wave state, one can show that $\rho(x, y) \approx (1 - A) \rho_0(x, y)$ at large distances $|x-y|\gg z_b$. Indeed this shows explicitly that spin wave decoherence is directly related to the nonlinear photon losses and, therefore, can be greatly suppressed by increasing $d_b$ in the present approach.

Figure \ref{fig: Coherence} illustrates this difference between the present and conventional Rydberg-EIT schemes for an explicit initial spin wave with $\rho_0(x, y) = \sin(x \pi /L)\sin(y \pi /L)$ [see Fig.\ref{fig: Coherence}(b)]. In Fig.\ref{fig: Coherence}(c-f) we show the final spin wave state according to eq.(\ref{eq: spin wave density matrix}) for various values of $d_b$. As expected, the suppressed photon losses, shown by solid line in Fig.\ref{fig: Coherence}(a), only cause little decoherence reflected in a marginal deformation of $\rho(x, y)$ which reduces with increasing $d_b$. In Fig.\ref{fig: Coherence}(g-j), we plot the final spin wave state corresponding to conventional Rydberg-EIT conditions {\cite{Murray2016a, Li2015a, Gorniaczyk2016}. In this case, the virtually complete scattering of the target field, as indicated by the dashed line in Fig.\ref{fig: Coherence}(a), causes strong decoherence that turns the initial spin wave into a near-classical distribution of the stored excitation.

\section{Applications}

The demonstrated conditional reflection of the target field realises a quantum nonlinear photon router \cite{Xia2013, Shomroni2014} in which a single gate photon can be used to control or redirect the flow of target photons between the two optical modes $\hat{\mathcal{E}}_{\rightarrow}$ and $\hat{\mathcal{E}}_{\rightarrow}$. This capability facilitates a broad range of functionalities.

Firstly, the ability to modify the transmissive properties of the medium via gate storage has immediate practical applications in the context of optical switching \cite{Hwang2009, Tiecke2014, Chang2007, Bajcsy2009, Nozaki2010, Volz2012, Bose2012, Chen2013, OShea2013}. Classical switching only requires a gate-photon induced blocking of the target field transmission (either via dissipative scattering or coherent reflection), for which we can defined a fidelity $\mathcal{F}_{\rm switch}^{\rm (classical)} = 1 - |T_1(0)|^2$. In previous Rydberg-EIT schemes \cite{Gorshkov2011,Baur2014, Tiarks2014, Gorniaczyk2014, Gorniaczyk2016, Murray2016a} relying on a dissipative polariton blockade the switching fidelity is given by the nonlinear loss coefficient $\mathcal{F}_{\rm switch}^{\rm (classical)} \approx \exp [-4 d_b]$ \cite{Gorshkov2011}. 

However, we emphasise that this purely dissipative nonlinearity fundamentally prevents the realization of a quantum switch, since the underlying photon scattering completely decoheres any quantum superposition state of the gate excitation. On the contrary, this does not affect the proposed routing approach, which does not rely on photon scattering. The quantum switching fidelity in this case is by the coherent reflection coefficient according to $\mathcal{F}_{\rm switch}^{\rm (quantum)} = |R_1(0)|^2$.

\begin{figure*}[t]
\begin{center}
\includegraphics[width=0.9\textwidth]{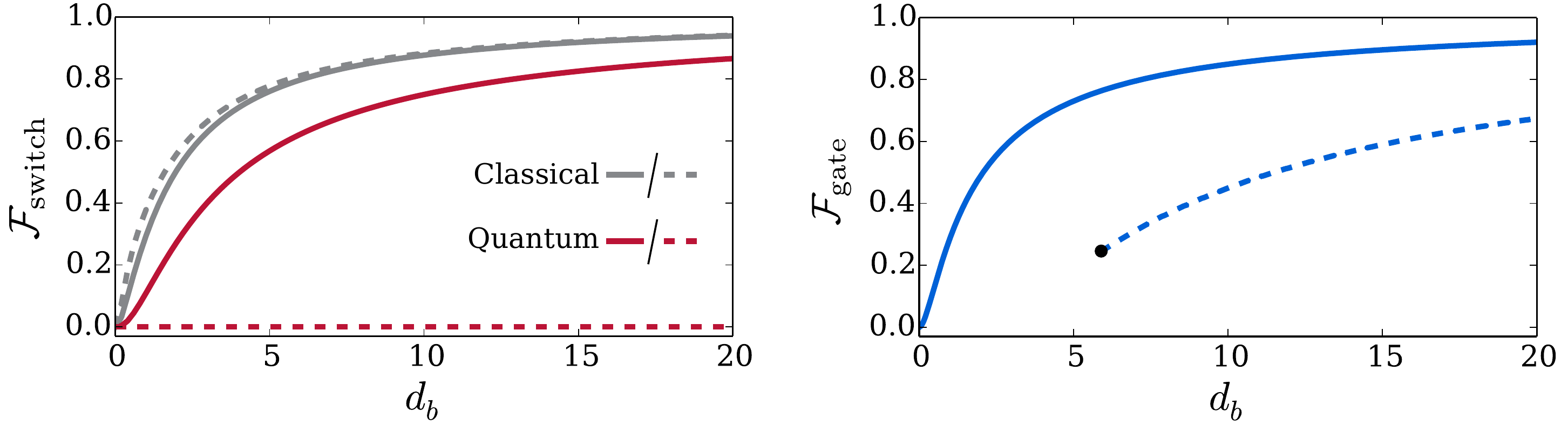}
\end{center}
\caption{\label{fig: Fidelities} (a) Overall switch fidelity, taking into account both the target field switching as well as the gate photon storage and retrieval efficiencies. The grey lines show the fidelity of a classical switch realised for a dissipative polariton blockade nonlinearity (dashed) and the coherent polariton switching nonlinearity (solid). The red lines show the corresponding fidelities for a quantum transistor. (b) Fidelity of a two-photon phase gate realised with the proposed photon router (solid) compared with the corresponding fidelity of a $\pi$-phase gate based on the dispersive phase shift obtained from the polariton blockade mechanism (dashed). The black dot indicates the critical value of $d_b \sim 6$ below which a dispersive phase shift of $\pi$ is fundamentally impossible with the polariton blockade mechanism under EIT conditions.}
\end{figure*}

The ability to retrieve the stored gate photon presents another performance aspect relevant for optical transistor operation or applications for non-destructive photon detection \cite{Baur2014, Tiarks2014, Gorniaczyk2014, Gorniaczyk2016}. This factor is critically determined by dissipative  scattering of target photons, as discussed above. To account for the finite retrieval efficiency, we adopt the strategy of \cite{Murray2016a} and optimise the gate storage and retrieval efficiency $\eta$ in the presence of such decoherence processes by shaping the spatial profile of the stored spin wave mode. We consider the target field in the CW limit, and describe the induced spin wave decoherence by eq.(\ref{eq: spin wave density matrix}). Spin wave decoherence for conventional Rydberg-EIT is calculated according to \cite{Murray2016a}. The overall transistor fidelity is then given by the product of $\eta$ with the corresponding switching fidelity discussed above. 

In Fig.\ref{fig: Fidelities}(a), we compare the different switch fidelities for the proposed photon router to the conventional approach based on nonlinear  photon scattering. For classical operations, the fidelities are nearly identical since the apparent gain in the storage and retrieval efficiency due to reduced spin wave decoherence is compensated by the higher switch fidelity, $\mathcal{F}_{\rm switch}^{\rm (classical)} \approx \exp [-4 d_b]$, in the dissipative case. The proposed routing mechanism, however, still yields a remarkably high fidelity for quantum-transistor operation, which, as described above, strictly vanishes for conventional Rydberg-EIT. 

As already pointed out in eq.(\ref{eq: R1}), the present approach permits to imprint any relative phase between the applied control fields onto the reflected target photon. In this way, the router can perform robust two-photon phase gate operations. The imprinted phase is largely independent of $d_b$, in contrast to previous schemes \cite{Gorshkov2011, Tiarks2016} based on conventional Rydberg-EIT. There one requires a large single photon detuning $\Delta$ to suppress the absorptive contribution to the nonlinear response, along with a large value of $d_b$ in order to account for the reduced single atom-coupling and achieve a significant phase shift. In this off-resonant limit, $\phi \approx - 2 d_b  (\gamma/\Delta)$, and, choosing $(\gamma/\Delta)$ to achieve a $\pi$-phase shift, the associated fidelity scales as $\mathcal{F}_{\rm gate}\sim\exp[-5\pi/4d_b]$ \cite{Gorshkov2011}. In Fig.\ref{fig: Fidelities}(b), we compare the fidelity of the present scheme, $\mathcal{F}_{\rm gate}=|R(0)|^2$, with that of the dispersive phase gate discussed above. Again, one finds a significant gain by the proposed routing mechanism for experimentally relevant parameters of moderate $d_b$.

\section{\label{sec: Experimental considerations} Experimental considerations}
Finally, we will discuss a concrete physical implementation of the described physics, using $^{87}$Rb atoms as the most relevant example for  current experiments \cite{Pritchard2010, Peyronel2012, Gorniaczyk2014, Tiarks2014}. Slow-light polaritons involving Rydberg states are being employed in a growing number of experiments \cite{Hofferberth2016,Murray2016} and stationary-light polaritons as emerging from the double-$\Lambda$ coupling scheme of Fig.\ref{fig: Schematic}(a) have also been demonstrated experimentally \cite{Lin2009,Chen2012,Peters2012}. Following this approach, the atomic cloud may be initialised in the $|g\rangle\equiv|5S_{1/2},F=1,m_F=0\rangle$ ground state. One can then choose the low-lying excited states $|p_{\rightarrow}\rangle$ and $|p_{\leftarrow}\rangle$ from the $|5P_{3/2},F=2\rangle$ hyperfine manifold. In this case, choosing the $\hat{\mathcal{E}}_{\rightarrow}$ and $\hat{\mathcal{E}}_{\leftarrow}$ fields to have $\sigma^-$ and $\sigma^+$ polarisations respectively, $\hat{\mathcal{E}}_{\rightarrow}$ will drive the $|g\rangle \to |p_{\rightarrow}\rangle\equiv|5P_{3/2},F=2,m_F=-1\rangle$ transition, while $\hat{\mathcal{E}}_{\leftarrow}$ will drive the $|g\rangle \to |p_{\leftarrow}\rangle\equiv|5P_{3/2},F=2,m_F=+1\rangle$ transition. Stationary light conditions are established by coupling these excited states back to the $|d\rangle \equiv |5S_{3/2},F=2, m_F=0\rangle$ hyperfine ground state, by the counter-propagating classical fields with Rabi frequencies $\Omega_{\rightarrow}=\Omega$ and $\Omega_{\leftarrow}=\Omega e^{i\phi}$, as indicated in Fig.\ref{fig: Schematic}(a). We note that using $m_F=0$ ground states for $|g\rangle$ and $|d\rangle$ ensures that the classical Rabi frequencies have equal magnitude, along with the coupling strengths of $\hat{\mathcal{E}}_{\rightarrow}$ and $\hat{\mathcal{E}}_{\leftarrow}$. Finally, taking an $nS_{1/2}$ Rydberg state, $\sigma^+$-polarised light can be used to drive the $|p_{\leftarrow}\rangle\to|s\rangle=|nS_{1/2}, J=1/2, m_J=1/2\rangle$ transition.

Due to the degeneracy of the involved hyperfine and fine structure manifolds, the optical fields would resonantly drive a number of additional transitions. These additional couplings can, however, be suppressed by applying a magnetic field along the light propagation axis. The resulting  Zeeman shifts can be made sufficiently strong to isolate the desired transitions upon adjusting the field frequencies accordingly to maintain a resonant coupling. The resulting slight frequency shifts of $\hat{\mathcal{E}}_{\leftarrow}$ with respect to $\hat{\mathcal{E}}_{\rightarrow}$ can be straightforwardly compensated with standard techniques or by sending the reflected light through an identical medium without the Rydberg state coupling. This permits to reconvert $\hat{\mathcal{E}}_{\leftarrow}$ into $\hat{\mathcal{E}}_{\rightarrow}$ with an efficiency, $\sim d^2/(1+d)^2$, that is much larger than that of the photon router. 

In Fig.\ref{fig: Fidelity overlap} we show the photon router fidelity 
\begin{equation}\label{eq: fidelity pulse}
\mathcal{F} = \left| \int_{-\infty}^{\infty} d \omega R_1(\omega) |\mathcal{E}_0(\omega)|^2 \right|,
\end{equation}
for realistic experimental parameters of the described coupling scheme and different pulse lengths of the incoming target photon. Its spectrum, $\mathcal{E}_0(\omega)$, is normalised according to $\int_{-\infty}^{\infty} d \omega  |\mathcal{E}_0(\omega)|^2 = 1$. Despite the frequency asymmetry of $R_1(\omega)$ [see Fig.\ref{fig: Spinwave_present}], the fidelity can approach the ideal limit $|R_1(\omega=0)|$ for realistic pulse lengths used in current experiments \cite{Gorniaczyk2014, Tiarks2014, Tiarks2016}.

\begin{figure}
\begin{center}
\includegraphics[width=0.9\columnwidth]{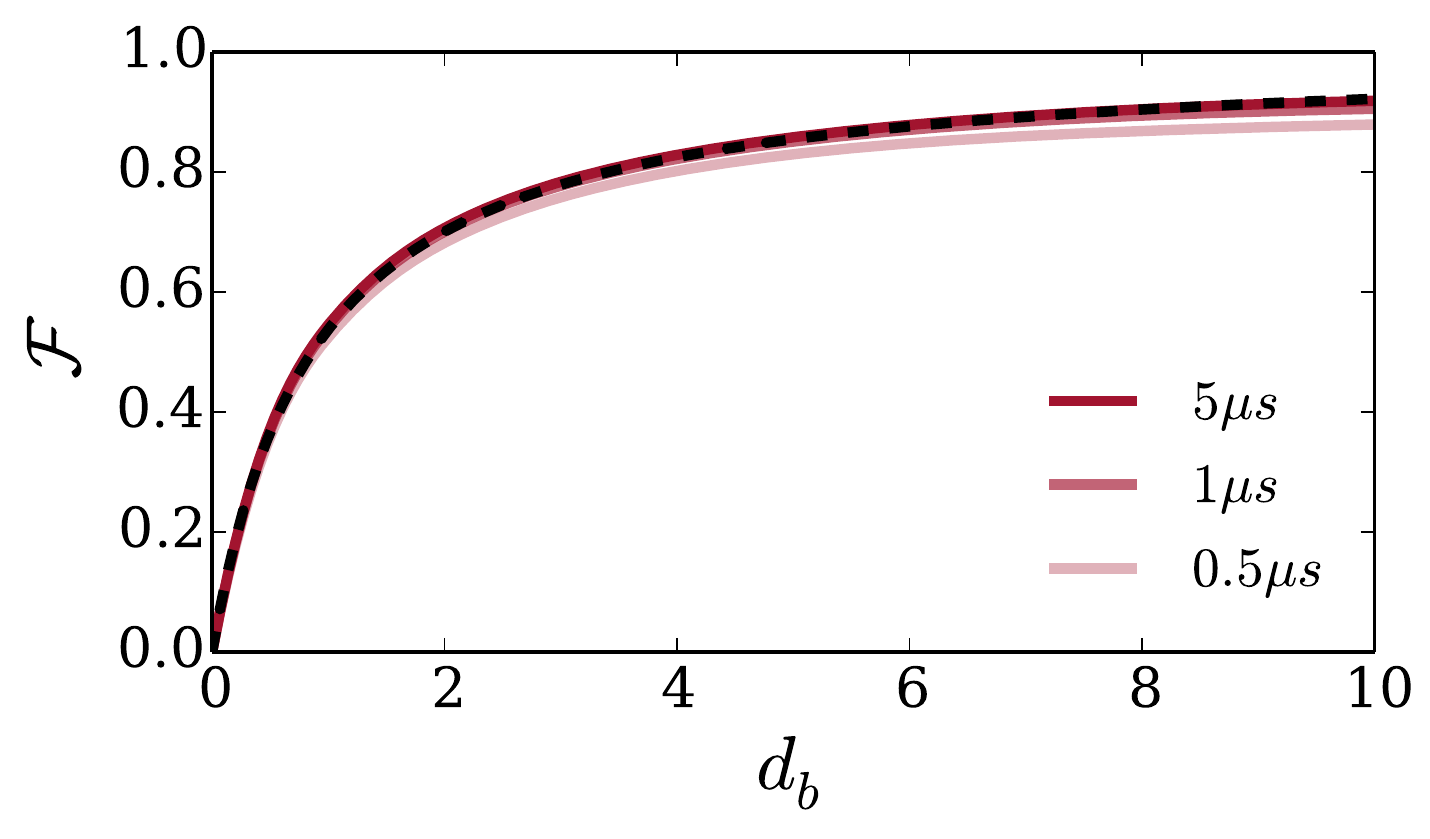}
\end{center}
\caption{\label{fig: Fidelity overlap} Fidelity, eq.(\ref{eq: fidelity pulse}) for different indicated durations of a Gaussian incoming target pulse, $\mathcal{E}_0$. The total optical depth is fixed at $2d=50$, $\gamma/2\pi = 3.05 \text{MHz}$ and the classical Rabi frequencies are $\Omega/2\pi = 5 \text{MHz}$ and $\Omega_S/2\pi = 20 \text{MHz}$. The gate excitation is located at the centre of the medium and generates an interaction region with $z_b = \sqrt[6]{C_6 \gamma / \Omega_S^2} \sim 8.7 \mu m$, as obtained for the $100S_{1/2}$ state of $^{87}\text{Rb}$ atoms \cite{Singer2005}.}
\end{figure}

Considering the finite spatial extent of both the target and the gate photon, one needs to account for another effect, namely entanglement between the target field and the gate excitation emerging from the Rydberg-Rydberg atom interaction. More specifically, the reflection time of the  target photon becomes correlated with the spatial position of the gate excitation in the atom cloud. While this presents another decoherence mechanism for the collective gate excitation, its actual effect is greatly suppressed as long as the target pulse length exceeds the characteristic difference of possible reflection times. The maximum time difference can be estimated as $\tau \approx d (\gamma/\Omega^2 + \gamma/\Omega_S^2)$ \cite{Fleischhauer2002}, which is $\sim0.5 \mu s$ for the example in Fig.\ref{fig: Fidelity overlap}. This is an order of magnitude shorter than the largest pulse length in Fig.\ref{fig: Fidelity overlap} and typical pulse lengths used in current experiments \cite{Tiarks2014,Tiarks2016}.

\section{Summary and conclusions}

In summary, we have worked out a new approach to engineering effective photon interactions via interactions between particles of an EIT medium. The basic principle is based on a modification rather than a breaking of EIT conditions to achieve a nonlinear alteration of light propagation under low-loss conditions. We have presented a specific implementation using laser-driven Rydberg-atom ensembles which realizes an effective photon-photon interaction that is reflective in character and highly coherent. 

We demonstrated that in this way a single photon acts like a mirror with a robust and continuously tuneable reflection phase, and discussed a number of applications entailed by such a quantum nonlinear photon router. Here, the enhanced coherence properties of the developed polariton-switching mechanism offer a significant performance gain compared to existing approaches based on the Rydberg-blockade of dark-state polaritons \cite{Friedler2005,Gorshkov2011,Hofferberth2016, Murray2016}. This in turn permits to achieve a strong optical response and high operational fidelities already at moderate values of $d_b$ that are well within the domain of accessible atomic densities, where additional decoherence processes \cite{Gaj2014} can be kept at a minimum.

While we have focussed in this work on the interaction of a single propagating photon with a single stored gate excitation, the interaction of multiple freely propagating photons also holds interesting perspectives. The described type of nonlinearity could be used to coherently filter out highly non-classical states of light from a classical light source, such as single-photon states \cite{Peyronel2012,Gorshkov2013} or strongly correlated trains of photons \cite{Zeuthen2016}. As compared to the dissipative nonlinearity based on the polariton-blockade considered in previous work \cite{Peyronel2012,Gorshkov2013,Zeuthen2016}, the present coherent nonlinear reflection mechanism might, for example, require significantly lower optical densities for generating spatially ordered photons \cite{Zeuthen2016} and, thereby, make such exotic states of light accessible with present experimental capabilities. 

Moreover, the broken left-right symmetry of the underlying coupling scheme in Fig.\ref{fig: Schematic}(a) suggests the emergence of chiral behaviour, in which the optical response of the medium is strongly dependent on the propagation direction of light \cite{Petersen2014, Mitsch2014, Pichler2015} in a highly nonlinear fashion. For example, a left moving photon can be made more reflective than a right moving photon, such that a head-on collision of two photons would generate a pair of co-propagating photons with a strong bias in one direction. The availability of such unsual types of interactions combined with the freely tuneable reflection phase suggests intriguing perspectives for the collective engineering of non-classical multi-photon states or the exploration of exotic many-body physics with photons in future theoretical and experimental work.

\section*{Acknowledgments}

We thank M. Baghery, H. Gorniaczyk, M. Gullans, D. Paredes-Barato and E. Zeuthen for valuable discussions.
This work was funded by the EU through the U-FET grant number 512862 (HAIRS) and the H2020-FETPROACT-2014 grant number 640378 (RYSQ), by the DFG through the SPP 1929, and by the DNRF through a Niels Bohr Professorship.

\appendix

\section{\label{appendix: Photon propagation equations} Photon propagation equations}

Here, we outline the derivation of the propagation equations (\ref{eq: MatEq}) and (\ref{eq: Propagation matrix}) and provide explicit expressions for the used susceptibilities. Starting from the Heisenberg equations deriving from the Hamiltonian eq.(\ref{eq: Hamiltonian}), one obtains equations of motion for all relevant two-body amplitudes describing a stored gate excitation at position $x$, and a target excitation at position $z$. Using the shorthand notation $\mathcal{E}_{\rightarrow} \equiv \mathcal{E}_{\rightarrow}(z, x, t)$ etc. for convenience, these coupled equations of motion can be written as
\begin{align}
\label{eq: Heisenberg Er}
\partial_t \mathcal{E}_{\rightarrow} & = -c\partial_z\mathcal{E}_{\rightarrow} - i G P_{\rightarrow}, \\
\label{eq: Heisenberg El}
\partial_t \mathcal{E}_{\leftarrow} & = c\partial_z\mathcal{E}_{\leftarrow} - i G P_{\leftarrow}, \\
\partial_t P_{\rightarrow} & = - i G \mathcal{E}_{\rightarrow} - i\Omega D - \gamma P_{\rightarrow}, \\
\partial_t P_{\leftarrow} & = - i G \mathcal{E}_{\leftarrow} - i\Omega e^{-i\phi}  D -i\Omega_S S - \gamma P_{\leftarrow}, \\
\partial_t D & = - i \Omega \left(P_{\rightarrow} + e^{i\phi} P_{\leftarrow} \right), \\
\label{eq: Heisenberg S}
\partial_t S & = - i \Omega_S P_{\leftarrow} - i V(z - x) S.
\end{align}
Transforming to frequency space, we then obtain a series of equations for $\tilde{\mathcal{E}}_{\rightarrow}(z, x, \omega) = (2 \pi)^{-1/2} \int_{-\infty}^{\infty} dt e^{-i \omega t} \mathcal{E}_{\rightarrow}(z, x, t)$ etc. Upon solving for $\tilde{P}_{\rightarrow}(z, x, \omega)$ and $\tilde{P}_{\leftarrow}(z, x, \omega)$ in terms of the photonic amplitudes, and inserting these into eqs.(\ref{eq: Heisenberg Er}-\ref{eq: Heisenberg El}), one immediately arrives at a closed system of equations for $\tilde{\mathcal{E}}_{\rightarrow}(z, x, \omega)$ and $\tilde{\mathcal{E}}_{\leftarrow}(z, x, \omega)$ as given by eqs.(\ref{eq: MatEq}) and (\ref{eq: Propagation matrix}). The susceptibility functions occuring in eq.(\ref{eq: Propagation matrix}) are explicitly given by 
\begin{align}
\chi_{\rightarrow}(z, \omega) & = -\frac{\omega}{c} + \frac{G^2}{c} \frac{\left( \xi - \frac{\Omega_S^2}{\omega - V(z)} \right)  }{\xi \left(\xi - \frac{\Omega_S^2}{\omega - V(z)} \right)  - \frac{\Omega^4}{\omega^2} } , \\
\chi_{\leftarrow}(z, \omega) & = \frac{\omega}{c} - \frac{G^2}{c} \frac{\xi }{\xi \left(\xi - \frac{\Omega_S^2}{\omega - V(z)} \right)  - \frac{\Omega^4}{\omega^2} }, \\
\chi(z, \omega) & = \frac{G^2}{c} \frac{\Omega^2}{\omega} \frac{1}{\xi \left( \xi - \frac{\Omega_S^2}{\omega - V(z)} \right)  - \frac{\Omega^4}{\omega^2} },
\end{align}
where we have introduced $\xi = \omega + i\gamma - \Omega^2/\omega$.

\section{\label{appendix: Spin wave decoherence dynamics} Spin wave decoherence dynamics}
In this appendix we outline the solution to the scattering induced decoherence dynamics of the stored gate spin wave, i.e. the derivation of  eq.(\ref{eq: spin wave density matrix}). To this end, we begin with the equation of motion for the density matrix elements, $\rho(x, y, t)$, of the stored spin wave, which can be expressed as
\begin{equation}
\label{eq: rho equation of motion}
\begin{split}
\partial_t \rho(x, y, t) = i \int_0^L & dz \left[ V(z - x) - V(z - y) \right] \\
& \phantom{xx}  \times S^*(z, x, t)S(z, y, t).
\end{split}
\end{equation}
$S(z, x, t)$ is again the two-body probability amplitude to have an $|s\rangle$ Rydberg excitation at position $z$, and a stored gate excitation at position $x$. Imposing the initial condition $\rho(x, y, 0) = \rho_0(x, y)$, the general solution to eq.(\ref{eq: rho equation of motion}) is given by
\begin{equation}
\label{eq: rho solution}
\begin{split}
\rho(x, y, t) = & \left\{1 + i  \int_0^L dz \left[ V(z - x) - V(z - y) \right] \right. \\
 & \left. \times \int_{-\infty}^t d\tau S^*(z, x, \tau)S(z, y, \tau) \right\} \rho_0(x, y).
\end{split}
\end{equation}
Solving the target field dynamics governed by eqs.(\ref{eq: Heisenberg Er}-\ref{eq: Heisenberg S}) in the CW limit, one can obtain a solution for the Rydberg spin wave amplitude $S(z, x)$ in terms of the photonic amplitudes
\begin{equation}
S(z, x) = \frac{G \Omega_S}{\gamma} \frac{\mathcal{E}_{\leftarrow}(z, x) - \mathcal{E}_{\rightarrow}(z, x)}{\Omega_S^2/\gamma + 2 i V(z - x)} .
\end{equation}
Using the solutions, eqs.(\ref{eq: Er(z)}-\ref{eq: El(z)}), for $\mathcal{E}_{\rightarrow}(z, x)$ and $\mathcal{E}_{\leftarrow}(z, x)$, the target spinwave amplitude can be written as
\begin{equation}
S(z, x) = - \frac{G \Omega_S}{\gamma} \frac{1}{\Omega_S^2/\gamma + 2 i V(z - x)} \frac{1}{1 + \nu(L, x)} \mathcal{E}_0
\end{equation} 
where the amplitude $\mathcal{E}_0$ describes the incoming target photon, with the normalisation $\int_{-\infty}^{\infty}dt|\mathcal{E}_0|^2 = 1/c$. Inserting this result into eq.(\ref{eq: rho solution}) and carrying out the time integration, one finally arrives at the result given in eq.(\ref{eq: spin wave density matrix}), where distances have been rescaled by $z_b$, defined according to $V(z_b)=\Omega_S^2/\gamma$. 

Let us finally consider eq.(\ref{eq: spin wave density matrix}) in the limit $|x - y| \gg z_b$. In this case, the spatial integral has two separate contributions around $z \sim x$ and $z \sim y$, such that the solution for $\rho(x, y)$ can be approximated by
\begin{equation}
\rho(x, y) \approx \left\{ 1 -  \frac{\nu(L,x) + \nu^*(L,y)}{\left[ 1 + \nu(L,x)\right] \left[ 1 + \nu^*(L,y) \right]} \right\} \rho_0(x, y).
\end{equation}
Assuming further that the medium length $L$ greatly exceeds the spatial extent of the gate spin wave, we can use $\nu(L,x)=\nu(L,y)=d_b\nu$ and obtain the expression 
\begin{align}
\rho(x, y) & \approx \left\{ 1 -  \frac{2d_b\text{Re}[\nu]}{|1 + d_b\nu|^2}  \right\} \rho_0(x, y) \\
& = (1 - A) \rho_0(x, y),
\end{align}
discussed in the main text, where $A$ is the loss coefficient introduced in section \ref{sec:coherence}.


\begin{thebibliography}{98}%
\makeatletter
\providecommand \@ifxundefined [1]{%
 \@ifx{#1\undefined}
}%
\providecommand \@ifnum [1]{%
 \ifnum #1\expandafter \@firstoftwo
 \else \expandafter \@secondoftwo
 \fi
}%
\providecommand \@ifx [1]{%
 \ifx #1\expandafter \@firstoftwo
 \else \expandafter \@secondoftwo
 \fi
}%
\providecommand \natexlab [1]{#1}%
\providecommand \enquote  [1]{``#1''}%
\providecommand \bibnamefont  [1]{#1}%
\providecommand \bibfnamefont [1]{#1}%
\providecommand \citenamefont [1]{#1}%
\providecommand \href@noop [0]{\@secondoftwo}%
\providecommand \href [0]{\begingroup \@sanitize@url \@href}%
\providecommand \@href[1]{\@@startlink{#1}\@@href}%
\providecommand \@@href[1]{\endgroup#1\@@endlink}%
\providecommand \@sanitize@url [0]{\catcode `\\12\catcode `\$12\catcode
  `\&12\catcode `\#12\catcode `\^12\catcode `\_12\catcode `\%12\relax}%
\providecommand \@@startlink[1]{}%
\providecommand \@@endlink[0]{}%
\providecommand \url  [0]{\begingroup\@sanitize@url \@url }%
\providecommand \@url [1]{\endgroup\@href {#1}{\urlprefix }}%
\providecommand \urlprefix  [0]{URL }%
\providecommand \Eprint [0]{\href }%
\providecommand \doibase [0]{http://dx.doi.org/}%
\providecommand \selectlanguage [0]{\@gobble}%
\providecommand \bibinfo  [0]{\@secondoftwo}%
\providecommand \bibfield  [0]{\@secondoftwo}%
\providecommand \translation [1]{[#1]}%
\providecommand \BibitemOpen [0]{}%
\providecommand \bibitemStop [0]{}%
\providecommand \bibitemNoStop [0]{.\EOS\space}%
\providecommand \EOS [0]{\spacefactor3000\relax}%
\providecommand \BibitemShut  [1]{\csname bibitem#1\endcsname}%
\let\auto@bib@innerbib\@empty
\bibitem [{\citenamefont {O'Brien}(2007)}]{OBrien2007}%
  \BibitemOpen
  \bibfield  {author} {\bibinfo {author} {\bibfnamefont {J.~L.}\ \bibnamefont
  {O'Brien}},\ }\href
  {http://science.sciencemag.org/content/318/5856/1567.abstract} {\bibfield
  {journal} {\bibinfo  {journal} {Science}\ }\textbf {\bibinfo {volume}
  {318}},\ \bibinfo {pages} {1567 LP } (\bibinfo {year} {2007})}\BibitemShut
  {NoStop}%
\bibitem [{\citenamefont {O'Brien}\ \emph {et~al.}(2009)\citenamefont
  {O'Brien}, \citenamefont {Furusawa},\ and\ \citenamefont
  {Vu{\v{c}}kovi{\'{c}}}}]{OBrien2009}%
  \BibitemOpen
  \bibfield  {author} {\bibinfo {author} {\bibfnamefont {J.~L.}\ \bibnamefont
  {O'Brien}}, \bibinfo {author} {\bibfnamefont {A.}~\bibnamefont {Furusawa}}, \
  and\ \bibinfo {author} {\bibfnamefont {J.}~\bibnamefont
  {Vu{\v{c}}kovi{\'{c}}}},\ }\href {\doibase 10.1038/nphoton.2009.229}
  {\bibfield  {journal} {\bibinfo  {journal} {Nature Photonics}\ }\textbf
  {\bibinfo {volume} {3}},\ \bibinfo {pages} {687} (\bibinfo {year}
  {2009})}\BibitemShut {NoStop}%
\bibitem [{\citenamefont {Kok}\ \emph {et~al.}(2007)\citenamefont {Kok},
  \citenamefont {Munro}, \citenamefont {Nemoto}, \citenamefont {Ralph},
  \citenamefont {Dowling},\ and\ \citenamefont {Milburn}}]{Kok2007}%
  \BibitemOpen
  \bibfield  {author} {\bibinfo {author} {\bibfnamefont {P.}~\bibnamefont
  {Kok}}, \bibinfo {author} {\bibfnamefont {W.~J.}\ \bibnamefont {Munro}},
  \bibinfo {author} {\bibfnamefont {K.}~\bibnamefont {Nemoto}}, \bibinfo
  {author} {\bibfnamefont {T.~C.}\ \bibnamefont {Ralph}}, \bibinfo {author}
  {\bibfnamefont {J.~P.}\ \bibnamefont {Dowling}}, \ and\ \bibinfo {author}
  {\bibfnamefont {G.~J.}\ \bibnamefont {Milburn}},\ }\href {\doibase
  10.1103/RevModPhys.79.135} {\bibfield  {journal} {\bibinfo  {journal}
  {Reviews of Modern Physics}\ }\textbf {\bibinfo {volume} {79}},\ \bibinfo
  {pages} {135} (\bibinfo {year} {2007})}\BibitemShut {NoStop}%
\bibitem [{\citenamefont {Kimble}(2008)}]{Kimble2008}%
  \BibitemOpen
  \bibfield  {author} {\bibinfo {author} {\bibfnamefont {H.~J.}\ \bibnamefont
  {Kimble}},\ }\href {\doibase 10.1038/nature07127} {\bibfield  {journal}
  {\bibinfo  {journal} {Nature}\ }\textbf {\bibinfo {volume} {453}},\ \bibinfo
  {pages} {1023} (\bibinfo {year} {2008})}\BibitemShut {NoStop}%
\bibitem [{\citenamefont {Chang}\ \emph {et~al.}(2014)\citenamefont {Chang},
  \citenamefont {Vuleti{\'{c}}},\ and\ \citenamefont {Lukin}}]{Chang2014}%
  \BibitemOpen
  \bibfield  {author} {\bibinfo {author} {\bibfnamefont {D.~E.}\ \bibnamefont
  {Chang}}, \bibinfo {author} {\bibfnamefont {V.}~\bibnamefont
  {Vuleti{\'{c}}}}, \ and\ \bibinfo {author} {\bibfnamefont {M.~D.}\
  \bibnamefont {Lukin}},\ }\href {\doibase 10.1038/nphoton.2014.192} {\bibfield
   {journal} {\bibinfo  {journal} {Nature Photonics}\ }\textbf {\bibinfo
  {volume} {8}},\ \bibinfo {pages} {685} (\bibinfo {year} {2014})}\BibitemShut
  {NoStop}%
\bibitem [{\citenamefont {Reiserer}\ and\ \citenamefont
  {Rempe}(2015)}]{Reiserer2015}%
  \BibitemOpen
  \bibfield  {author} {\bibinfo {author} {\bibfnamefont {A.}~\bibnamefont
  {Reiserer}}\ and\ \bibinfo {author} {\bibfnamefont {G.}~\bibnamefont
  {Rempe}},\ }\href {\doibase 10.1103/RevModPhys.87.1379} {\bibfield  {journal}
  {\bibinfo  {journal} {Reviews of Modern Physics}\ }\textbf {\bibinfo {volume}
  {87}},\ \bibinfo {pages} {1379} (\bibinfo {year} {2015})}\BibitemShut
  {NoStop}%
\bibitem [{\citenamefont {Reiserer}\ \emph {et~al.}(2013)\citenamefont
  {Reiserer}, \citenamefont {Ritter},\ and\ \citenamefont
  {Rempe}}]{Reiserer2013}%
  \BibitemOpen
  \bibfield  {author} {\bibinfo {author} {\bibfnamefont {A.}~\bibnamefont
  {Reiserer}}, \bibinfo {author} {\bibfnamefont {S.}~\bibnamefont {Ritter}}, \
  and\ \bibinfo {author} {\bibfnamefont {G.}~\bibnamefont {Rempe}},\ }\href
  {\doibase 10.1126/science.1246164} {\bibfield  {journal} {\bibinfo  {journal}
  {Science (New York, N.Y.)}\ }\textbf {\bibinfo {volume} {342}},\ \bibinfo
  {pages} {1349} (\bibinfo {year} {2013})}\BibitemShut {NoStop}%
\bibitem [{\citenamefont {Reiserer}\ \emph {et~al.}(2014)\citenamefont
  {Reiserer}, \citenamefont {Kalb}, \citenamefont {Rempe},\ and\ \citenamefont
  {Ritter}}]{Reiserer2014}%
  \BibitemOpen
  \bibfield  {author} {\bibinfo {author} {\bibfnamefont {A.}~\bibnamefont
  {Reiserer}}, \bibinfo {author} {\bibfnamefont {N.}~\bibnamefont {Kalb}},
  \bibinfo {author} {\bibfnamefont {G.}~\bibnamefont {Rempe}}, \ and\ \bibinfo
  {author} {\bibfnamefont {S.}~\bibnamefont {Ritter}},\ }\href {\doibase
  10.1038/nature13177} {\bibfield  {journal} {\bibinfo  {journal} {Nature}\
  }\textbf {\bibinfo {volume} {508}},\ \bibinfo {pages} {237} (\bibinfo {year}
  {2014})}\BibitemShut {NoStop}%
\bibitem [{\citenamefont {Shahmoon}\ and\ \citenamefont
  {Kurizki}(2014)}]{Shahmoon2014}%
  \BibitemOpen
  \bibfield  {author} {\bibinfo {author} {\bibfnamefont {E.}~\bibnamefont
  {Shahmoon}}\ and\ \bibinfo {author} {\bibfnamefont {G.}~\bibnamefont
  {Kurizki}},\ }\href {\doibase 10.1103/PhysRevA.89.043419} {\bibfield
  {journal} {\bibinfo  {journal} {Phys. Rev. A}\ }\textbf {\bibinfo {volume}
  {89}},\ \bibinfo {pages} {043419} (\bibinfo {year} {2014})}\BibitemShut
  {NoStop}%
\bibitem [{\citenamefont {Lodahl}\ \emph {et~al.}(2015)\citenamefont {Lodahl},
  \citenamefont {Mahmoodian},\ and\ \citenamefont {Stobbe}}]{Lodahl2015}%
  \BibitemOpen
  \bibfield  {author} {\bibinfo {author} {\bibfnamefont {P.}~\bibnamefont
  {Lodahl}}, \bibinfo {author} {\bibfnamefont {S.}~\bibnamefont {Mahmoodian}},
  \ and\ \bibinfo {author} {\bibfnamefont {S.}~\bibnamefont {Stobbe}},\ }\href
  {\doibase 10.1103/RevModPhys.87.347} {\bibfield  {journal} {\bibinfo
  {journal} {Reviews of Modern Physics}\ }\textbf {\bibinfo {volume} {87}},\
  \bibinfo {pages} {347} (\bibinfo {year} {2015})}\BibitemShut {NoStop}%
\bibitem [{\citenamefont {Javadi}\ \emph {et~al.}(2015)\citenamefont {Javadi},
  \citenamefont {S{\"{o}}llner}, \citenamefont {Arcari}, \citenamefont
  {Hansen}, \citenamefont {Midolo}, \citenamefont {Mahmoodian}, \citenamefont
  {Kir{\v{s}}anske}, \citenamefont {Pregnolato}, \citenamefont {Lee},
  \citenamefont {Song}, \citenamefont {Stobbe},\ and\ \citenamefont
  {Lodahl}}]{Javadi2015}%
  \BibitemOpen
  \bibfield  {author} {\bibinfo {author} {\bibfnamefont {A.}~\bibnamefont
  {Javadi}}, \bibinfo {author} {\bibfnamefont {I.}~\bibnamefont
  {S{\"{o}}llner}}, \bibinfo {author} {\bibfnamefont {M.}~\bibnamefont
  {Arcari}}, \bibinfo {author} {\bibfnamefont {S.~L.}\ \bibnamefont {Hansen}},
  \bibinfo {author} {\bibfnamefont {L.}~\bibnamefont {Midolo}}, \bibinfo
  {author} {\bibfnamefont {S.}~\bibnamefont {Mahmoodian}}, \bibinfo {author}
  {\bibfnamefont {G.}~\bibnamefont {Kir{\v{s}}anske}}, \bibinfo {author}
  {\bibfnamefont {T.}~\bibnamefont {Pregnolato}}, \bibinfo {author}
  {\bibfnamefont {E.~H.}\ \bibnamefont {Lee}}, \bibinfo {author} {\bibfnamefont
  {J.~D.}\ \bibnamefont {Song}}, \bibinfo {author} {\bibfnamefont
  {S.}~\bibnamefont {Stobbe}}, \ and\ \bibinfo {author} {\bibfnamefont
  {P.}~\bibnamefont {Lodahl}},\ }\href {\doibase 10.1038/ncomms9655} {\bibfield
   {journal} {\bibinfo  {journal} {Nature Communications}\ }\textbf {\bibinfo
  {volume} {6}},\ \bibinfo {pages} {8655} (\bibinfo {year} {2015})}\BibitemShut
  {NoStop}%
\bibitem [{\citenamefont {S{\"{o}}llner}\ \emph {et~al.}(2015)\citenamefont
  {S{\"{o}}llner}, \citenamefont {Mahmoodian}, \citenamefont {Hansen},
  \citenamefont {Midolo}, \citenamefont {Javadi}, \citenamefont
  {Kir{\v{s}}anske}, \citenamefont {Pregnolato}, \citenamefont {El-Ella},
  \citenamefont {Lee}, \citenamefont {Song}, \citenamefont {Stobbe},\ and\
  \citenamefont {Lodahl}}]{Sollner2015}%
  \BibitemOpen
  \bibfield  {author} {\bibinfo {author} {\bibfnamefont {I.}~\bibnamefont
  {S{\"{o}}llner}}, \bibinfo {author} {\bibfnamefont {S.}~\bibnamefont
  {Mahmoodian}}, \bibinfo {author} {\bibfnamefont {S.~L.}\ \bibnamefont
  {Hansen}}, \bibinfo {author} {\bibfnamefont {L.}~\bibnamefont {Midolo}},
  \bibinfo {author} {\bibfnamefont {A.}~\bibnamefont {Javadi}}, \bibinfo
  {author} {\bibfnamefont {G.}~\bibnamefont {Kir{\v{s}}anske}}, \bibinfo
  {author} {\bibfnamefont {T.}~\bibnamefont {Pregnolato}}, \bibinfo {author}
  {\bibfnamefont {H.}~\bibnamefont {El-Ella}}, \bibinfo {author} {\bibfnamefont
  {E.~H.}\ \bibnamefont {Lee}}, \bibinfo {author} {\bibfnamefont {J.~D.}\
  \bibnamefont {Song}}, \bibinfo {author} {\bibfnamefont {S.}~\bibnamefont
  {Stobbe}}, \ and\ \bibinfo {author} {\bibfnamefont {P.}~\bibnamefont
  {Lodahl}},\ }\href {\doibase 10.1038/nnano.2015.159} {\bibfield  {journal}
  {\bibinfo  {journal} {Nature Nanotechnology}\ }\textbf {\bibinfo {volume}
  {10}},\ \bibinfo {pages} {775} (\bibinfo {year} {2015})}\BibitemShut
  {NoStop}%
\bibitem [{\citenamefont {Hwang}\ \emph {et~al.}(2009)\citenamefont {Hwang},
  \citenamefont {Pototschnig}, \citenamefont {Lettow}, \citenamefont {Zumofen},
  \citenamefont {Renn}, \citenamefont {G{\"{o}}tzinger},\ and\ \citenamefont
  {Sandoghdar}}]{Hwang2009}%
  \BibitemOpen
  \bibfield  {author} {\bibinfo {author} {\bibfnamefont {J.}~\bibnamefont
  {Hwang}}, \bibinfo {author} {\bibfnamefont {M.}~\bibnamefont {Pototschnig}},
  \bibinfo {author} {\bibfnamefont {R.}~\bibnamefont {Lettow}}, \bibinfo
  {author} {\bibfnamefont {G.}~\bibnamefont {Zumofen}}, \bibinfo {author}
  {\bibfnamefont {A.}~\bibnamefont {Renn}}, \bibinfo {author} {\bibfnamefont
  {S.}~\bibnamefont {G{\"{o}}tzinger}}, \ and\ \bibinfo {author} {\bibfnamefont
  {V.}~\bibnamefont {Sandoghdar}},\ }\href {\doibase 10.1038/nature08134}
  {\bibfield  {journal} {\bibinfo  {journal} {Nature}\ }\textbf {\bibinfo
  {volume} {460}},\ \bibinfo {pages} {76} (\bibinfo {year} {2009})}\BibitemShut
  {NoStop}%
\bibitem [{\citenamefont {Maser}\ \emph {et~al.}(2016)\citenamefont {Maser},
  \citenamefont {Gmeiner}, \citenamefont {Utikal}, \citenamefont
  {G{\"{o}}tzinger},\ and\ \citenamefont {Sandoghdar}}]{Maser2016}%
  \BibitemOpen
  \bibfield  {author} {\bibinfo {author} {\bibfnamefont {A.}~\bibnamefont
  {Maser}}, \bibinfo {author} {\bibfnamefont {B.}~\bibnamefont {Gmeiner}},
  \bibinfo {author} {\bibfnamefont {T.}~\bibnamefont {Utikal}}, \bibinfo
  {author} {\bibfnamefont {S.}~\bibnamefont {G{\"{o}}tzinger}}, \ and\ \bibinfo
  {author} {\bibfnamefont {V.}~\bibnamefont {Sandoghdar}},\ }\href {\doibase
  10.1038/nphoton.2016.63} {\bibfield  {journal} {\bibinfo  {journal} {Nature
  Photonics}\ }\textbf {\bibinfo {volume} {10}},\ \bibinfo {pages} {450}
  (\bibinfo {year} {2016})}\BibitemShut {NoStop}%
\bibitem [{\citenamefont {Shomroni}\ \emph {et~al.}(2014)\citenamefont
  {Shomroni}, \citenamefont {Rosenblum}, \citenamefont {Lovsky}, \citenamefont
  {Bechler}, \citenamefont {Guendelman},\ and\ \citenamefont
  {Dayan}}]{Shomroni2014}%
  \BibitemOpen
  \bibfield  {author} {\bibinfo {author} {\bibfnamefont {I.}~\bibnamefont
  {Shomroni}}, \bibinfo {author} {\bibfnamefont {S.}~\bibnamefont {Rosenblum}},
  \bibinfo {author} {\bibfnamefont {Y.}~\bibnamefont {Lovsky}}, \bibinfo
  {author} {\bibfnamefont {O.}~\bibnamefont {Bechler}}, \bibinfo {author}
  {\bibfnamefont {G.}~\bibnamefont {Guendelman}}, \ and\ \bibinfo {author}
  {\bibfnamefont {B.}~\bibnamefont {Dayan}},\ }\href {\doibase
  10.1126/science.1254699} {\bibfield  {journal} {\bibinfo  {journal} {Science
  (New York, N.Y.)}\ }\textbf {\bibinfo {volume} {345}},\ \bibinfo {pages}
  {903} (\bibinfo {year} {2014})}\BibitemShut {NoStop}%
\bibitem [{\citenamefont {Thompson}\ \emph {et~al.}(2013)\citenamefont
  {Thompson}, \citenamefont {Tiecke}, \citenamefont {de~Leon}, \citenamefont
  {Feist}, \citenamefont {Akimov}, \citenamefont {Gullans}, \citenamefont
  {Zibrov}, \citenamefont {Vuleti{\'{c}}},\ and\ \citenamefont
  {Lukin}}]{Thompson2013}%
  \BibitemOpen
  \bibfield  {author} {\bibinfo {author} {\bibfnamefont {J.~D.}\ \bibnamefont
  {Thompson}}, \bibinfo {author} {\bibfnamefont {T.~G.}\ \bibnamefont
  {Tiecke}}, \bibinfo {author} {\bibfnamefont {N.~P.}\ \bibnamefont {de~Leon}},
  \bibinfo {author} {\bibfnamefont {J.}~\bibnamefont {Feist}}, \bibinfo
  {author} {\bibfnamefont {A.~V.}\ \bibnamefont {Akimov}}, \bibinfo {author}
  {\bibfnamefont {M.}~\bibnamefont {Gullans}}, \bibinfo {author} {\bibfnamefont
  {A.~S.}\ \bibnamefont {Zibrov}}, \bibinfo {author} {\bibfnamefont
  {V.}~\bibnamefont {Vuleti{\'{c}}}}, \ and\ \bibinfo {author} {\bibfnamefont
  {M.~D.}\ \bibnamefont {Lukin}},\ }\href {\doibase 10.1126/science.1237125}
  {\bibfield  {journal} {\bibinfo  {journal} {Science (New York, N.Y.)}\
  }\textbf {\bibinfo {volume} {340}},\ \bibinfo {pages} {1202} (\bibinfo {year}
  {2013})}\BibitemShut {NoStop}%
\bibitem [{\citenamefont {Tiecke}\ \emph {et~al.}(2014)\citenamefont {Tiecke},
  \citenamefont {Thompson}, \citenamefont {de~Leon}, \citenamefont {Liu},
  \citenamefont {Vuleti{\'{c}}},\ and\ \citenamefont {Lukin}}]{Tiecke2014}%
  \BibitemOpen
  \bibfield  {author} {\bibinfo {author} {\bibfnamefont {T.~G.}\ \bibnamefont
  {Tiecke}}, \bibinfo {author} {\bibfnamefont {J.~D.}\ \bibnamefont
  {Thompson}}, \bibinfo {author} {\bibfnamefont {N.~P.}\ \bibnamefont
  {de~Leon}}, \bibinfo {author} {\bibfnamefont {L.~R.}\ \bibnamefont {Liu}},
  \bibinfo {author} {\bibfnamefont {V.}~\bibnamefont {Vuleti{\'{c}}}}, \ and\
  \bibinfo {author} {\bibfnamefont {M.~D.}\ \bibnamefont {Lukin}},\ }\href
  {\doibase 10.1038/nature13188} {\bibfield  {journal} {\bibinfo  {journal}
  {Nature}\ }\textbf {\bibinfo {volume} {508}},\ \bibinfo {pages} {241}
  (\bibinfo {year} {2014})}\BibitemShut {NoStop}%
\bibitem [{\citenamefont {Goban}\ \emph {et~al.}(2014)\citenamefont {Goban},
  \citenamefont {Hung}, \citenamefont {Yu}, \citenamefont {Hood}, \citenamefont
  {Muniz}, \citenamefont {Lee}, \citenamefont {Martin}, \citenamefont
  {McClung}, \citenamefont {Choi}, \citenamefont {Chang}, \citenamefont
  {Painter},\ and\ \citenamefont {Kimble}}]{Goban2014}%
  \BibitemOpen
  \bibfield  {author} {\bibinfo {author} {\bibfnamefont {A.}~\bibnamefont
  {Goban}}, \bibinfo {author} {\bibfnamefont {C.-L.}\ \bibnamefont {Hung}},
  \bibinfo {author} {\bibfnamefont {S.-P.}\ \bibnamefont {Yu}}, \bibinfo
  {author} {\bibfnamefont {J.~D.}\ \bibnamefont {Hood}}, \bibinfo {author}
  {\bibfnamefont {J.~A.}\ \bibnamefont {Muniz}}, \bibinfo {author}
  {\bibfnamefont {J.~H.}\ \bibnamefont {Lee}}, \bibinfo {author} {\bibfnamefont
  {M.~J.}\ \bibnamefont {Martin}}, \bibinfo {author} {\bibfnamefont {A.~C.}\
  \bibnamefont {McClung}}, \bibinfo {author} {\bibfnamefont {K.~S.}\
  \bibnamefont {Choi}}, \bibinfo {author} {\bibfnamefont {D.~E.}\ \bibnamefont
  {Chang}}, \bibinfo {author} {\bibfnamefont {O.}~\bibnamefont {Painter}}, \
  and\ \bibinfo {author} {\bibfnamefont {H.~J.}\ \bibnamefont {Kimble}},\
  }\href {\doibase 10.1038/ncomms4808} {\bibfield  {journal} {\bibinfo
  {journal} {Nature communications}\ }\textbf {\bibinfo {volume} {5}},\
  \bibinfo {pages} {3808} (\bibinfo {year} {2014})}\BibitemShut {NoStop}%
\bibitem [{\citenamefont {Hofferberth}\ \emph {et~al.}(2016)\citenamefont
  {Hofferberth}, \citenamefont {Adams},\ and\ \citenamefont
  {S}}]{Hofferberth2016}%
  \BibitemOpen
  \bibfield  {author} {\bibinfo {author} {\bibfnamefont {O.~F.}\ \bibnamefont
  {Hofferberth}}, \bibinfo {author} {\bibfnamefont {C.~S.}\ \bibnamefont
  {Adams}}, \ and\ \bibinfo {author} {\bibnamefont {S}},\ }\href
  {http://stacks.iop.org/0953-4075/49/i=15/a=152003} {\bibfield  {journal}
  {\bibinfo  {journal} {Journal of Physics B: Atomic, Molecular and Optical
  Physics}\ }\textbf {\bibinfo {volume} {49}},\ \bibinfo {pages} {152003}
  (\bibinfo {year} {2016})}\BibitemShut {NoStop}%
\bibitem [{\citenamefont {Murray}\ and\ \citenamefont
  {Pohl}(2016)}]{Murray2016}%
  \BibitemOpen
  \bibfield  {author} {\bibinfo {author} {\bibfnamefont {C.}~\bibnamefont
  {Murray}}\ and\ \bibinfo {author} {\bibfnamefont {T.}~\bibnamefont {Pohl}},\
  }in\ \href {\doibase 10.1016/bs.aamop.2016.04.005} {\emph {\bibinfo
  {booktitle} {Advances In Atomic, Molecular, and Optical Physics}}},\
  Vol.~\bibinfo {volume} {65}\ (\bibinfo  {publisher} {Academic Press},\
  \bibinfo {year} {2016})\ pp.\ \bibinfo {pages} {321--372}\BibitemShut
  {NoStop}%
\bibitem [{\citenamefont {Douglas}\ \emph {et~al.}(2015)\citenamefont
  {Douglas}, \citenamefont {Habibian}, \citenamefont {Hung}, \citenamefont
  {Gorshkov}, \citenamefont {Kimble},\ and\ \citenamefont
  {Chang}}]{Douglas2015}%
  \BibitemOpen
  \bibfield  {author} {\bibinfo {author} {\bibfnamefont {J.~S.}\ \bibnamefont
  {Douglas}}, \bibinfo {author} {\bibfnamefont {H.}~\bibnamefont {Habibian}},
  \bibinfo {author} {\bibfnamefont {C.-L.}\ \bibnamefont {Hung}}, \bibinfo
  {author} {\bibfnamefont {A.~V.}\ \bibnamefont {Gorshkov}}, \bibinfo {author}
  {\bibfnamefont {H.~J.}\ \bibnamefont {Kimble}}, \ and\ \bibinfo {author}
  {\bibfnamefont {D.~E.}\ \bibnamefont {Chang}},\ }\href {\doibase
  10.1038/nphoton.2015.57} {\bibfield  {journal} {\bibinfo  {journal} {Nature
  Photonics}\ }\textbf {\bibinfo {volume} {9}},\ \bibinfo {pages} {326}
  (\bibinfo {year} {2015})}\BibitemShut {NoStop}%
\bibitem [{\citenamefont {Shi}\ \emph {et~al.}(2015)\citenamefont {Shi},
  \citenamefont {Chang},\ and\ \citenamefont {Cirac}}]{Shi2015}%
  \BibitemOpen
  \bibfield  {author} {\bibinfo {author} {\bibfnamefont {T.}~\bibnamefont
  {Shi}}, \bibinfo {author} {\bibfnamefont {D.~E.}\ \bibnamefont {Chang}}, \
  and\ \bibinfo {author} {\bibfnamefont {J.~I.}\ \bibnamefont {Cirac}},\ }\href
  {\doibase 10.1103/PhysRevA.92.053834} {\bibfield  {journal} {\bibinfo
  {journal} {Phys. Rev. A}\ }\textbf {\bibinfo {volume} {92}},\ \bibinfo
  {pages} {053834} (\bibinfo {year} {2015})}\BibitemShut {NoStop}%
\bibitem [{\citenamefont {Shahmoon}\ \emph {et~al.}(2016)\citenamefont
  {Shahmoon}, \citenamefont {Gri{\v{s}}ins}, \citenamefont {Stimming},
  \citenamefont {Mazets},\ and\ \citenamefont {Kurizki}}]{Shahmoon2016}%
  \BibitemOpen
  \bibfield  {author} {\bibinfo {author} {\bibfnamefont {E.}~\bibnamefont
  {Shahmoon}}, \bibinfo {author} {\bibfnamefont {P.}~\bibnamefont
  {Gri{\v{s}}ins}}, \bibinfo {author} {\bibfnamefont {H.~P.}\ \bibnamefont
  {Stimming}}, \bibinfo {author} {\bibfnamefont {I.}~\bibnamefont {Mazets}}, \
  and\ \bibinfo {author} {\bibfnamefont {G.}~\bibnamefont {Kurizki}},\ }\href
  {\doibase 10.1364/OPTICA.3.000725} {\bibfield  {journal} {\bibinfo  {journal}
  {Optica}\ }\textbf {\bibinfo {volume} {3}},\ \bibinfo {pages} {725} (\bibinfo
  {year} {2016})}\BibitemShut {NoStop}%
\bibitem [{\citenamefont {Douglas}\ \emph {et~al.}(2016)\citenamefont
  {Douglas}, \citenamefont {Caneva},\ and\ \citenamefont
  {Chang}}]{Douglas2016}%
  \BibitemOpen
  \bibfield  {author} {\bibinfo {author} {\bibfnamefont {J.~S.}\ \bibnamefont
  {Douglas}}, \bibinfo {author} {\bibfnamefont {T.}~\bibnamefont {Caneva}}, \
  and\ \bibinfo {author} {\bibfnamefont {D.~E.}\ \bibnamefont {Chang}},\ }\href
  {\doibase 10.1103/PhysRevX.6.031017} {\bibfield  {journal} {\bibinfo
  {journal} {Physical Review X}\ }\textbf {\bibinfo {volume} {6}},\ \bibinfo
  {pages} {031017} (\bibinfo {year} {2016})}\BibitemShut {NoStop}%
\bibitem [{\citenamefont {Singer}\ \emph {et~al.}(2004)\citenamefont {Singer},
  \citenamefont {Reetz-Lamour}, \citenamefont {Amthor}, \citenamefont
  {Marcassa},\ and\ \citenamefont {Weidem\"uller}}]{Singer2004}%
  \BibitemOpen
  \bibfield  {author} {\bibinfo {author} {\bibfnamefont {K.}~\bibnamefont
  {Singer}}, \bibinfo {author} {\bibfnamefont {M.}~\bibnamefont
  {Reetz-Lamour}}, \bibinfo {author} {\bibfnamefont {T.}~\bibnamefont
  {Amthor}}, \bibinfo {author} {\bibfnamefont {L.~G.}\ \bibnamefont
  {Marcassa}}, \ and\ \bibinfo {author} {\bibfnamefont {M.}~\bibnamefont
  {Weidem\"uller}},\ }\href {\doibase 10.1103/PhysRevLett.93.163001} {\bibfield
   {journal} {\bibinfo  {journal} {Phys. Rev. Lett.}\ }\textbf {\bibinfo
  {volume} {93}},\ \bibinfo {pages} {163001} (\bibinfo {year}
  {2004})}\BibitemShut {NoStop}%
\bibitem [{\citenamefont {Tong}\ \emph {et~al.}(2004)\citenamefont {Tong},
  \citenamefont {Farooqi}, \citenamefont {Stanojevic}, \citenamefont
  {Krishnan}, \citenamefont {Zhang}, \citenamefont {C\^ot\'e}, \citenamefont
  {Eyler},\ and\ \citenamefont {Gould}}]{Tong2004}%
  \BibitemOpen
  \bibfield  {author} {\bibinfo {author} {\bibfnamefont {D.}~\bibnamefont
  {Tong}}, \bibinfo {author} {\bibfnamefont {S.~M.}\ \bibnamefont {Farooqi}},
  \bibinfo {author} {\bibfnamefont {J.}~\bibnamefont {Stanojevic}}, \bibinfo
  {author} {\bibfnamefont {S.}~\bibnamefont {Krishnan}}, \bibinfo {author}
  {\bibfnamefont {Y.~P.}\ \bibnamefont {Zhang}}, \bibinfo {author}
  {\bibfnamefont {R.}~\bibnamefont {C\^ot\'e}}, \bibinfo {author}
  {\bibfnamefont {E.~E.}\ \bibnamefont {Eyler}}, \ and\ \bibinfo {author}
  {\bibfnamefont {P.~L.}\ \bibnamefont {Gould}},\ }\href {\doibase
  10.1103/PhysRevLett.93.063001} {\bibfield  {journal} {\bibinfo  {journal}
  {Phys. Rev. Lett.}\ }\textbf {\bibinfo {volume} {93}},\ \bibinfo {pages}
  {063001} (\bibinfo {year} {2004})}\BibitemShut {NoStop}%
\bibitem [{\citenamefont {Liebisch}\ \emph {et~al.}(2005)\citenamefont
  {Liebisch}, \citenamefont {Reinhard}, \citenamefont {Berman},\ and\
  \citenamefont {Raithel}}]{Liebisch2005}%
  \BibitemOpen
  \bibfield  {author} {\bibinfo {author} {\bibfnamefont {T.~C.}\ \bibnamefont
  {Liebisch}}, \bibinfo {author} {\bibfnamefont {A.}~\bibnamefont {Reinhard}},
  \bibinfo {author} {\bibfnamefont {P.~R.}\ \bibnamefont {Berman}}, \ and\
  \bibinfo {author} {\bibfnamefont {G.}~\bibnamefont {Raithel}},\ }\href
  {\doibase 10.1103/PhysRevLett.95.253002} {\bibfield  {journal} {\bibinfo
  {journal} {Phys. Rev. Lett.}\ }\textbf {\bibinfo {volume} {95}},\ \bibinfo
  {pages} {253002} (\bibinfo {year} {2005})}\BibitemShut {NoStop}%
\bibitem [{\citenamefont {Heidemann}\ \emph {et~al.}(2007)\citenamefont
  {Heidemann}, \citenamefont {Raitzsch}, \citenamefont {Bendkowsky},
  \citenamefont {Butscher}, \citenamefont {L\"ow}, \citenamefont {Santos},\
  and\ \citenamefont {Pfau}}]{Heidemann2007}%
  \BibitemOpen
  \bibfield  {author} {\bibinfo {author} {\bibfnamefont {R.}~\bibnamefont
  {Heidemann}}, \bibinfo {author} {\bibfnamefont {U.}~\bibnamefont {Raitzsch}},
  \bibinfo {author} {\bibfnamefont {V.}~\bibnamefont {Bendkowsky}}, \bibinfo
  {author} {\bibfnamefont {B.}~\bibnamefont {Butscher}}, \bibinfo {author}
  {\bibfnamefont {R.}~\bibnamefont {L\"ow}}, \bibinfo {author} {\bibfnamefont
  {L.}~\bibnamefont {Santos}}, \ and\ \bibinfo {author} {\bibfnamefont
  {T.}~\bibnamefont {Pfau}},\ }\href {\doibase 10.1103/PhysRevLett.99.163601}
  {\bibfield  {journal} {\bibinfo  {journal} {Phys. Rev. Lett.}\ }\textbf
  {\bibinfo {volume} {99}},\ \bibinfo {pages} {163601} (\bibinfo {year}
  {2007})}\BibitemShut {NoStop}%
\bibitem [{\citenamefont {Isenhower}\ \emph {et~al.}(2010)\citenamefont
  {Isenhower}, \citenamefont {Urban}, \citenamefont {Zhang}, \citenamefont
  {Gill}, \citenamefont {Henage}, \citenamefont {Johnson}, \citenamefont
  {Walker},\ and\ \citenamefont {Saffman}}]{Isenhower2010}%
  \BibitemOpen
  \bibfield  {author} {\bibinfo {author} {\bibfnamefont {L.}~\bibnamefont
  {Isenhower}}, \bibinfo {author} {\bibfnamefont {E.}~\bibnamefont {Urban}},
  \bibinfo {author} {\bibfnamefont {X.~L.}\ \bibnamefont {Zhang}}, \bibinfo
  {author} {\bibfnamefont {A.~T.}\ \bibnamefont {Gill}}, \bibinfo {author}
  {\bibfnamefont {T.}~\bibnamefont {Henage}}, \bibinfo {author} {\bibfnamefont
  {T.~A.}\ \bibnamefont {Johnson}}, \bibinfo {author} {\bibfnamefont {T.~G.}\
  \bibnamefont {Walker}}, \ and\ \bibinfo {author} {\bibfnamefont
  {M.}~\bibnamefont {Saffman}},\ }\href {\doibase
  10.1103/PhysRevLett.104.010503} {\bibfield  {journal} {\bibinfo  {journal}
  {Phys. Rev. Lett.}\ }\textbf {\bibinfo {volume} {104}},\ \bibinfo {pages}
  {010503} (\bibinfo {year} {2010})}\BibitemShut {NoStop}%
\bibitem [{\citenamefont {Wilk}\ \emph {et~al.}(2010)\citenamefont {Wilk},
  \citenamefont {Ga\"etan}, \citenamefont {Evellin}, \citenamefont {Wolters},
  \citenamefont {Miroshnychenko}, \citenamefont {Grangier},\ and\ \citenamefont
  {Browaeys}}]{Wilk2010}%
  \BibitemOpen
  \bibfield  {author} {\bibinfo {author} {\bibfnamefont {T.}~\bibnamefont
  {Wilk}}, \bibinfo {author} {\bibfnamefont {A.}~\bibnamefont {Ga\"etan}},
  \bibinfo {author} {\bibfnamefont {C.}~\bibnamefont {Evellin}}, \bibinfo
  {author} {\bibfnamefont {J.}~\bibnamefont {Wolters}}, \bibinfo {author}
  {\bibfnamefont {Y.}~\bibnamefont {Miroshnychenko}}, \bibinfo {author}
  {\bibfnamefont {P.}~\bibnamefont {Grangier}}, \ and\ \bibinfo {author}
  {\bibfnamefont {A.}~\bibnamefont {Browaeys}},\ }\href {\doibase
  10.1103/PhysRevLett.104.010502} {\bibfield  {journal} {\bibinfo  {journal}
  {Phys. Rev. Lett.}\ }\textbf {\bibinfo {volume} {104}},\ \bibinfo {pages}
  {010502} (\bibinfo {year} {2010})}\BibitemShut {NoStop}%
\bibitem [{\citenamefont {Viteau}\ \emph {et~al.}(2011)\citenamefont {Viteau},
  \citenamefont {Bason}, \citenamefont {Radogostowicz}, \citenamefont
  {Malossi}, \citenamefont {Ciampini}, \citenamefont {Morsch},\ and\
  \citenamefont {Arimondo}}]{Viteau2011}%
  \BibitemOpen
  \bibfield  {author} {\bibinfo {author} {\bibfnamefont {M.}~\bibnamefont
  {Viteau}}, \bibinfo {author} {\bibfnamefont {M.~G.}\ \bibnamefont {Bason}},
  \bibinfo {author} {\bibfnamefont {J.}~\bibnamefont {Radogostowicz}}, \bibinfo
  {author} {\bibfnamefont {N.}~\bibnamefont {Malossi}}, \bibinfo {author}
  {\bibfnamefont {D.}~\bibnamefont {Ciampini}}, \bibinfo {author}
  {\bibfnamefont {O.}~\bibnamefont {Morsch}}, \ and\ \bibinfo {author}
  {\bibfnamefont {E.}~\bibnamefont {Arimondo}},\ }\href {\doibase
  10.1103/PhysRevLett.107.060402} {\bibfield  {journal} {\bibinfo  {journal}
  {Phys. Rev. Lett.}\ }\textbf {\bibinfo {volume} {107}},\ \bibinfo {pages}
  {060402} (\bibinfo {year} {2011})}\BibitemShut {NoStop}%
\bibitem [{\citenamefont {Ebert}\ \emph {et~al.}(2014)\citenamefont {Ebert},
  \citenamefont {Gill}, \citenamefont {Gibbons}, \citenamefont {Zhang},
  \citenamefont {Saffman},\ and\ \citenamefont {Walker}}]{Ebert2014}%
  \BibitemOpen
  \bibfield  {author} {\bibinfo {author} {\bibfnamefont {M.}~\bibnamefont
  {Ebert}}, \bibinfo {author} {\bibfnamefont {A.}~\bibnamefont {Gill}},
  \bibinfo {author} {\bibfnamefont {M.}~\bibnamefont {Gibbons}}, \bibinfo
  {author} {\bibfnamefont {X.}~\bibnamefont {Zhang}}, \bibinfo {author}
  {\bibfnamefont {M.}~\bibnamefont {Saffman}}, \ and\ \bibinfo {author}
  {\bibfnamefont {T.~G.}\ \bibnamefont {Walker}},\ }\href {\doibase
  10.1103/PhysRevLett.112.043602} {\bibfield  {journal} {\bibinfo  {journal}
  {Phys. Rev. Lett.}\ }\textbf {\bibinfo {volume} {112}},\ \bibinfo {pages}
  {043602} (\bibinfo {year} {2014})}\BibitemShut {NoStop}%
\bibitem [{\citenamefont {Schempp}\ \emph {et~al.}(2014)\citenamefont
  {Schempp}, \citenamefont {G\"unter}, \citenamefont {Robert-de Saint-Vincent},
  \citenamefont {Hofmann}, \citenamefont {Breyel}, \citenamefont {Komnik},
  \citenamefont {Sch\"onleber}, \citenamefont {G\"arttner}, \citenamefont
  {Evers}, \citenamefont {Whitlock},\ and\ \citenamefont
  {Weidem\"uller}}]{Schempp2014}%
  \BibitemOpen
  \bibfield  {author} {\bibinfo {author} {\bibfnamefont {H.}~\bibnamefont
  {Schempp}}, \bibinfo {author} {\bibfnamefont {G.}~\bibnamefont {G\"unter}},
  \bibinfo {author} {\bibfnamefont {M.}~\bibnamefont {Robert-de
  Saint-Vincent}}, \bibinfo {author} {\bibfnamefont {C.~S.}\ \bibnamefont
  {Hofmann}}, \bibinfo {author} {\bibfnamefont {D.}~\bibnamefont {Breyel}},
  \bibinfo {author} {\bibfnamefont {A.}~\bibnamefont {Komnik}}, \bibinfo
  {author} {\bibfnamefont {D.~W.}\ \bibnamefont {Sch\"onleber}}, \bibinfo
  {author} {\bibfnamefont {M.}~\bibnamefont {G\"arttner}}, \bibinfo {author}
  {\bibfnamefont {J.}~\bibnamefont {Evers}}, \bibinfo {author} {\bibfnamefont
  {S.}~\bibnamefont {Whitlock}}, \ and\ \bibinfo {author} {\bibfnamefont
  {M.}~\bibnamefont {Weidem\"uller}},\ }\href {\doibase
  10.1103/PhysRevLett.112.013002} {\bibfield  {journal} {\bibinfo  {journal}
  {Phys. Rev. Lett.}\ }\textbf {\bibinfo {volume} {112}},\ \bibinfo {pages}
  {013002} (\bibinfo {year} {2014})}\BibitemShut {NoStop}%
\bibitem [{\citenamefont {Malossi}\ \emph {et~al.}(2014)\citenamefont
  {Malossi}, \citenamefont {Valado}, \citenamefont {Scotto}, \citenamefont
  {Huillery}, \citenamefont {Pillet}, \citenamefont {Ciampini}, \citenamefont
  {Arimondo},\ and\ \citenamefont {Morsch}}]{Malossi2014}%
  \BibitemOpen
  \bibfield  {author} {\bibinfo {author} {\bibfnamefont {N.}~\bibnamefont
  {Malossi}}, \bibinfo {author} {\bibfnamefont {M.~M.}\ \bibnamefont {Valado}},
  \bibinfo {author} {\bibfnamefont {S.}~\bibnamefont {Scotto}}, \bibinfo
  {author} {\bibfnamefont {P.}~\bibnamefont {Huillery}}, \bibinfo {author}
  {\bibfnamefont {P.}~\bibnamefont {Pillet}}, \bibinfo {author} {\bibfnamefont
  {D.}~\bibnamefont {Ciampini}}, \bibinfo {author} {\bibfnamefont
  {E.}~\bibnamefont {Arimondo}}, \ and\ \bibinfo {author} {\bibfnamefont
  {O.}~\bibnamefont {Morsch}},\ }\href {\doibase
  10.1103/PhysRevLett.113.023006} {\bibfield  {journal} {\bibinfo  {journal}
  {Phys. Rev. Lett.}\ }\textbf {\bibinfo {volume} {113}},\ \bibinfo {pages}
  {023006} (\bibinfo {year} {2014})}\BibitemShut {NoStop}%
\bibitem [{\citenamefont {Urvoy}\ \emph {et~al.}(2015)\citenamefont {Urvoy},
  \citenamefont {Ripka}, \citenamefont {Lesanovsky}, \citenamefont {Booth},
  \citenamefont {Shaffer}, \citenamefont {Pfau},\ and\ \citenamefont
  {L\"ow}}]{Urvoy2015}%
  \BibitemOpen
  \bibfield  {author} {\bibinfo {author} {\bibfnamefont {A.}~\bibnamefont
  {Urvoy}}, \bibinfo {author} {\bibfnamefont {F.}~\bibnamefont {Ripka}},
  \bibinfo {author} {\bibfnamefont {I.}~\bibnamefont {Lesanovsky}}, \bibinfo
  {author} {\bibfnamefont {D.}~\bibnamefont {Booth}}, \bibinfo {author}
  {\bibfnamefont {J.~P.}\ \bibnamefont {Shaffer}}, \bibinfo {author}
  {\bibfnamefont {T.}~\bibnamefont {Pfau}}, \ and\ \bibinfo {author}
  {\bibfnamefont {R.}~\bibnamefont {L\"ow}},\ }\href {\doibase
  10.1103/PhysRevLett.114.203002} {\bibfield  {journal} {\bibinfo  {journal}
  {Phys. Rev. Lett.}\ }\textbf {\bibinfo {volume} {114}},\ \bibinfo {pages}
  {203002} (\bibinfo {year} {2015})}\BibitemShut {NoStop}%
\bibitem [{\citenamefont {Ebert}\ \emph {et~al.}(2015)\citenamefont {Ebert},
  \citenamefont {Kwon}, \citenamefont {Walker},\ and\ \citenamefont
  {Saffman}}]{Ebert2015}%
  \BibitemOpen
  \bibfield  {author} {\bibinfo {author} {\bibfnamefont {M.}~\bibnamefont
  {Ebert}}, \bibinfo {author} {\bibfnamefont {M.}~\bibnamefont {Kwon}},
  \bibinfo {author} {\bibfnamefont {T.~G.}\ \bibnamefont {Walker}}, \ and\
  \bibinfo {author} {\bibfnamefont {M.}~\bibnamefont {Saffman}},\ }\href
  {\doibase 10.1103/PhysRevLett.115.093601} {\bibfield  {journal} {\bibinfo
  {journal} {Phys. Rev. Lett.}\ }\textbf {\bibinfo {volume} {115}},\ \bibinfo
  {pages} {093601} (\bibinfo {year} {2015})}\BibitemShut {NoStop}%
\bibitem [{\citenamefont {Faoro}\ \emph {et~al.}(2016)\citenamefont {Faoro},
  \citenamefont {Simonelli}, \citenamefont {Archimi}, \citenamefont {Masella},
  \citenamefont {Valado}, \citenamefont {Arimondo}, \citenamefont {Mannella},
  \citenamefont {Ciampini},\ and\ \citenamefont {Morsch}}]{Faoro2016}%
  \BibitemOpen
  \bibfield  {author} {\bibinfo {author} {\bibfnamefont {R.}~\bibnamefont
  {Faoro}}, \bibinfo {author} {\bibfnamefont {C.}~\bibnamefont {Simonelli}},
  \bibinfo {author} {\bibfnamefont {M.}~\bibnamefont {Archimi}}, \bibinfo
  {author} {\bibfnamefont {G.}~\bibnamefont {Masella}}, \bibinfo {author}
  {\bibfnamefont {M.~M.}\ \bibnamefont {Valado}}, \bibinfo {author}
  {\bibfnamefont {E.}~\bibnamefont {Arimondo}}, \bibinfo {author}
  {\bibfnamefont {R.}~\bibnamefont {Mannella}}, \bibinfo {author}
  {\bibfnamefont {D.}~\bibnamefont {Ciampini}}, \ and\ \bibinfo {author}
  {\bibfnamefont {O.}~\bibnamefont {Morsch}},\ }\href {\doibase
  10.1103/PhysRevA.93.030701} {\bibfield  {journal} {\bibinfo  {journal} {Phys.
  Rev. A}\ }\textbf {\bibinfo {volume} {93}},\ \bibinfo {pages} {030701}
  (\bibinfo {year} {2016})}\BibitemShut {NoStop}%
\bibitem [{\citenamefont {Saffman}\ \emph {et~al.}(2010)\citenamefont
  {Saffman}, \citenamefont {Walker},\ and\ \citenamefont
  {M{\o}lmer}}]{Saffman2010}%
  \BibitemOpen
  \bibfield  {author} {\bibinfo {author} {\bibfnamefont {M.}~\bibnamefont
  {Saffman}}, \bibinfo {author} {\bibfnamefont {T.~G.}\ \bibnamefont {Walker}},
  \ and\ \bibinfo {author} {\bibfnamefont {K.}~\bibnamefont {M{\o}lmer}},\
  }\href {\doibase 10.1103/RevModPhys.82.2313} {\bibfield  {journal} {\bibinfo
  {journal} {Reviews of Modern Physics}\ }\textbf {\bibinfo {volume} {82}},\
  \bibinfo {pages} {2313} (\bibinfo {year} {2010})}\BibitemShut {NoStop}%
\bibitem [{\citenamefont {Fleischhauer}\ \emph {et~al.}(2005)\citenamefont
  {Fleischhauer}, \citenamefont {Imamoglu},\ and\ \citenamefont
  {Marangos}}]{Fleischhauer2005}%
  \BibitemOpen
  \bibfield  {author} {\bibinfo {author} {\bibfnamefont {M.}~\bibnamefont
  {Fleischhauer}}, \bibinfo {author} {\bibfnamefont {A.}~\bibnamefont
  {Imamoglu}}, \ and\ \bibinfo {author} {\bibfnamefont {J.~P.}\ \bibnamefont
  {Marangos}},\ }\href {\doibase 10.1103/RevModPhys.77.633} {\bibfield
  {journal} {\bibinfo  {journal} {Reviews of Modern Physics}\ }\textbf
  {\bibinfo {volume} {77}},\ \bibinfo {pages} {633} (\bibinfo {year}
  {2005})}\BibitemShut {NoStop}%
\bibitem [{\citenamefont {Friedler}\ \emph {et~al.}(2005)\citenamefont
  {Friedler}, \citenamefont {Petrosyan}, \citenamefont {Fleischhauer},\ and\
  \citenamefont {Kurizki}}]{Friedler2005}%
  \BibitemOpen
  \bibfield  {author} {\bibinfo {author} {\bibfnamefont {I.}~\bibnamefont
  {Friedler}}, \bibinfo {author} {\bibfnamefont {D.}~\bibnamefont {Petrosyan}},
  \bibinfo {author} {\bibfnamefont {M.}~\bibnamefont {Fleischhauer}}, \ and\
  \bibinfo {author} {\bibfnamefont {G.}~\bibnamefont {Kurizki}},\ }\href
  {\doibase 10.1103/PhysRevA.72.043803} {\bibfield  {journal} {\bibinfo
  {journal} {Physical Review A}\ }\textbf {\bibinfo {volume} {72}},\ \bibinfo
  {pages} {043803} (\bibinfo {year} {2005})}\BibitemShut {NoStop}%
\bibitem [{\citenamefont {Pritchard}\ \emph {et~al.}(2010)\citenamefont
  {Pritchard}, \citenamefont {Maxwell}, \citenamefont {Gauguet}, \citenamefont
  {Weatherill}, \citenamefont {Jones},\ and\ \citenamefont
  {Adams}}]{Pritchard2010}%
  \BibitemOpen
  \bibfield  {author} {\bibinfo {author} {\bibfnamefont {J.~D.}\ \bibnamefont
  {Pritchard}}, \bibinfo {author} {\bibfnamefont {D.}~\bibnamefont {Maxwell}},
  \bibinfo {author} {\bibfnamefont {A.}~\bibnamefont {Gauguet}}, \bibinfo
  {author} {\bibfnamefont {K.~J.}\ \bibnamefont {Weatherill}}, \bibinfo
  {author} {\bibfnamefont {M.~P.~A.}\ \bibnamefont {Jones}}, \ and\ \bibinfo
  {author} {\bibfnamefont {C.~S.}\ \bibnamefont {Adams}},\ }\href {\doibase
  10.1103/PhysRevLett.105.193603} {\bibfield  {journal} {\bibinfo  {journal}
  {Physical review letters}\ }\textbf {\bibinfo {volume} {105}},\ \bibinfo
  {pages} {193603} (\bibinfo {year} {2010})}\BibitemShut {NoStop}%
\bibitem [{\citenamefont {Schempp}\ \emph {et~al.}(2010)\citenamefont
  {Schempp}, \citenamefont {G\"unter}, \citenamefont {Hofmann}, \citenamefont
  {Giese}, \citenamefont {Saliba}, \citenamefont {DePaola}, \citenamefont
  {Amthor}, \citenamefont {Weidem\"uller}, \citenamefont
  {Sevin\ifmmode~\mbox{\c{c}}\else \c{c}\fi{}li},\ and\ \citenamefont
  {Pohl}}]{Schempp2010}%
  \BibitemOpen
  \bibfield  {author} {\bibinfo {author} {\bibfnamefont {H.}~\bibnamefont
  {Schempp}}, \bibinfo {author} {\bibfnamefont {G.}~\bibnamefont {G\"unter}},
  \bibinfo {author} {\bibfnamefont {C.~S.}\ \bibnamefont {Hofmann}}, \bibinfo
  {author} {\bibfnamefont {C.}~\bibnamefont {Giese}}, \bibinfo {author}
  {\bibfnamefont {S.~D.}\ \bibnamefont {Saliba}}, \bibinfo {author}
  {\bibfnamefont {B.~D.}\ \bibnamefont {DePaola}}, \bibinfo {author}
  {\bibfnamefont {T.}~\bibnamefont {Amthor}}, \bibinfo {author} {\bibfnamefont
  {M.}~\bibnamefont {Weidem\"uller}}, \bibinfo {author} {\bibfnamefont
  {S.}~\bibnamefont {Sevin\ifmmode~\mbox{\c{c}}\else \c{c}\fi{}li}}, \ and\
  \bibinfo {author} {\bibfnamefont {T.}~\bibnamefont {Pohl}},\ }\href {\doibase
  10.1103/PhysRevLett.104.173602} {\bibfield  {journal} {\bibinfo  {journal}
  {Phys. Rev. Lett.}\ }\textbf {\bibinfo {volume} {104}},\ \bibinfo {pages}
  {173602} (\bibinfo {year} {2010})}\BibitemShut {NoStop}%
\bibitem [{\citenamefont {Ates}\ \emph {et~al.}(2011)\citenamefont {Ates},
  \citenamefont {Sevin{\c{c}}li},\ and\ \citenamefont {Pohl}}]{Ates2011}%
  \BibitemOpen
  \bibfield  {author} {\bibinfo {author} {\bibfnamefont {C.}~\bibnamefont
  {Ates}}, \bibinfo {author} {\bibfnamefont {S.}~\bibnamefont
  {Sevin{\c{c}}li}}, \ and\ \bibinfo {author} {\bibfnamefont {T.}~\bibnamefont
  {Pohl}},\ }\href {\doibase 10.1103/PhysRevA.83.041802} {\bibfield  {journal}
  {\bibinfo  {journal} {Physical Review A}\ }\textbf {\bibinfo {volume} {83}},\
  \bibinfo {pages} {041802} (\bibinfo {year} {2011})}\BibitemShut {NoStop}%
\bibitem [{\citenamefont {Sevin{\c{c}}li}\ \emph {et~al.}(2011)\citenamefont
  {Sevin{\c{c}}li}, \citenamefont {Henkel}, \citenamefont {Ates},\ and\
  \citenamefont {Pohl}}]{Sevincli2011}%
  \BibitemOpen
  \bibfield  {author} {\bibinfo {author} {\bibfnamefont {S.}~\bibnamefont
  {Sevin{\c{c}}li}}, \bibinfo {author} {\bibfnamefont {N.}~\bibnamefont
  {Henkel}}, \bibinfo {author} {\bibfnamefont {C.}~\bibnamefont {Ates}}, \ and\
  \bibinfo {author} {\bibfnamefont {T.}~\bibnamefont {Pohl}},\ }\href {\doibase
  10.1103/PhysRevLett.107.153001} {\bibfield  {journal} {\bibinfo  {journal}
  {Physical review letters}\ }\textbf {\bibinfo {volume} {107}},\ \bibinfo
  {pages} {153001} (\bibinfo {year} {2011})}\BibitemShut {NoStop}%
\bibitem [{\citenamefont {Petrosyan}\ \emph {et~al.}(2011)\citenamefont
  {Petrosyan}, \citenamefont {Otterbach},\ and\ \citenamefont
  {Fleischhauer}}]{Petrosyan2011}%
  \BibitemOpen
  \bibfield  {author} {\bibinfo {author} {\bibfnamefont {D.}~\bibnamefont
  {Petrosyan}}, \bibinfo {author} {\bibfnamefont {J.}~\bibnamefont
  {Otterbach}}, \ and\ \bibinfo {author} {\bibfnamefont {M.}~\bibnamefont
  {Fleischhauer}},\ }\href {\doibase 10.1103/PhysRevLett.107.213601} {\bibfield
   {journal} {\bibinfo  {journal} {Physical review letters}\ }\textbf {\bibinfo
  {volume} {107}},\ \bibinfo {pages} {213601} (\bibinfo {year}
  {2011})}\BibitemShut {NoStop}%
\bibitem [{\citenamefont {Gorshkov}\ \emph {et~al.}(2011)\citenamefont
  {Gorshkov}, \citenamefont {Otterbach}, \citenamefont {Fleischhauer},
  \citenamefont {Pohl},\ and\ \citenamefont {Lukin}}]{Gorshkov2011}%
  \BibitemOpen
  \bibfield  {author} {\bibinfo {author} {\bibfnamefont {A.~V.}\ \bibnamefont
  {Gorshkov}}, \bibinfo {author} {\bibfnamefont {J.}~\bibnamefont {Otterbach}},
  \bibinfo {author} {\bibfnamefont {M.}~\bibnamefont {Fleischhauer}}, \bibinfo
  {author} {\bibfnamefont {T.}~\bibnamefont {Pohl}}, \ and\ \bibinfo {author}
  {\bibfnamefont {M.~D.}\ \bibnamefont {Lukin}},\ }\href {\doibase
  10.1103/PhysRevLett.107.133602} {\bibfield  {journal} {\bibinfo  {journal}
  {Physical review letters}\ }\textbf {\bibinfo {volume} {107}},\ \bibinfo
  {pages} {133602} (\bibinfo {year} {2011})}\BibitemShut {NoStop}%
\bibitem [{\citenamefont {Gorshkov}\ \emph {et~al.}(2013)\citenamefont
  {Gorshkov}, \citenamefont {Nath},\ and\ \citenamefont {Pohl}}]{Gorshkov2013}%
  \BibitemOpen
  \bibfield  {author} {\bibinfo {author} {\bibfnamefont {A.~V.}\ \bibnamefont
  {Gorshkov}}, \bibinfo {author} {\bibfnamefont {R.}~\bibnamefont {Nath}}, \
  and\ \bibinfo {author} {\bibfnamefont {T.}~\bibnamefont {Pohl}},\ }\href
  {\doibase 10.1103/PhysRevLett.110.153601} {\bibfield  {journal} {\bibinfo
  {journal} {Physical review letters}\ }\textbf {\bibinfo {volume} {110}},\
  \bibinfo {pages} {153601} (\bibinfo {year} {2013})}\BibitemShut {NoStop}%
\bibitem [{\citenamefont {Petrosyan}(2016)}]{Petrosyan2016}%
  \BibitemOpen
  \bibfield  {author} {\bibinfo {author} {\bibfnamefont {D.}~\bibnamefont
  {Petrosyan}},\ }\href {http://arxiv.org/abs/1607.00621} {\  (\bibinfo {year}
  {2016})},\ \Eprint {http://arxiv.org/abs/1607.00621} {arXiv:1607.00621}
  \BibitemShut {NoStop}%
\bibitem [{\citenamefont {Das}\ \emph {et~al.}(2016)\citenamefont {Das},
  \citenamefont {Grankin}, \citenamefont {Iakoupov}, \citenamefont {Brion},
  \citenamefont {Borregaard}, \citenamefont {Boddeda}, \citenamefont {Usmani},
  \citenamefont {Ourjoumtsev}, \citenamefont {Grangier},\ and\ \citenamefont
  {S{\o}rensen}}]{Das2016}%
  \BibitemOpen
  \bibfield  {author} {\bibinfo {author} {\bibfnamefont {S.}~\bibnamefont
  {Das}}, \bibinfo {author} {\bibfnamefont {A.}~\bibnamefont {Grankin}},
  \bibinfo {author} {\bibfnamefont {I.}~\bibnamefont {Iakoupov}}, \bibinfo
  {author} {\bibfnamefont {E.}~\bibnamefont {Brion}}, \bibinfo {author}
  {\bibfnamefont {J.}~\bibnamefont {Borregaard}}, \bibinfo {author}
  {\bibfnamefont {R.}~\bibnamefont {Boddeda}}, \bibinfo {author} {\bibfnamefont
  {I.}~\bibnamefont {Usmani}}, \bibinfo {author} {\bibfnamefont
  {A.}~\bibnamefont {Ourjoumtsev}}, \bibinfo {author} {\bibfnamefont
  {P.}~\bibnamefont {Grangier}}, \ and\ \bibinfo {author} {\bibfnamefont
  {A.~S.}\ \bibnamefont {S{\o}rensen}},\ }\href {\doibase
  10.1103/PhysRevA.93.040303} {\bibfield  {journal} {\bibinfo  {journal}
  {Physical Review A}\ }\textbf {\bibinfo {volume} {93}},\ \bibinfo {pages}
  {040303} (\bibinfo {year} {2016})}\BibitemShut {NoStop}%
\bibitem [{\citenamefont {Fleischhauer}\ and\ \citenamefont
  {Lukin}(2000)}]{Fleischhauer2000}%
  \BibitemOpen
  \bibfield  {author} {\bibinfo {author} {\bibfnamefont {M.}~\bibnamefont
  {Fleischhauer}}\ and\ \bibinfo {author} {\bibfnamefont {M.}~\bibnamefont
  {Lukin}},\ }\href {\doibase 10.1103/PhysRevLett.84.5094} {\bibfield
  {journal} {\bibinfo  {journal} {Physical review letters}\ }\textbf {\bibinfo
  {volume} {84}},\ \bibinfo {pages} {5094} (\bibinfo {year}
  {2000})}\BibitemShut {NoStop}%
\bibitem [{\citenamefont {Fleischhauer}\ and\ \citenamefont
  {Lukin}(2002)}]{Fleischhauer2002}%
  \BibitemOpen
  \bibfield  {author} {\bibinfo {author} {\bibfnamefont {M.}~\bibnamefont
  {Fleischhauer}}\ and\ \bibinfo {author} {\bibfnamefont {M.~D.}\ \bibnamefont
  {Lukin}},\ }\href {\doibase 10.1103/PhysRevA.65.022314} {\bibfield  {journal}
  {\bibinfo  {journal} {Physical Review A}\ }\textbf {\bibinfo {volume} {65}},\
  \bibinfo {pages} {022314} (\bibinfo {year} {2002})}\BibitemShut {NoStop}%
\bibitem [{\citenamefont {Dudin}\ and\ \citenamefont
  {Kuzmich}(2012)}]{Dudin2012}%
  \BibitemOpen
  \bibfield  {author} {\bibinfo {author} {\bibfnamefont {Y.~O.}\ \bibnamefont
  {Dudin}}\ and\ \bibinfo {author} {\bibfnamefont {A.}~\bibnamefont
  {Kuzmich}},\ }\href {\doibase 10.1126/science.1217901} {\bibfield  {journal}
  {\bibinfo  {journal} {Science (New York, N.Y.)}\ }\textbf {\bibinfo {volume}
  {336}},\ \bibinfo {pages} {887} (\bibinfo {year} {2012})}\BibitemShut
  {NoStop}%
\bibitem [{\citenamefont {Peyronel}\ \emph {et~al.}(2012)\citenamefont
  {Peyronel}, \citenamefont {Firstenberg}, \citenamefont {Liang}, \citenamefont
  {Hofferberth}, \citenamefont {Gorshkov}, \citenamefont {Pohl}, \citenamefont
  {Lukin},\ and\ \citenamefont {Vuleti{\'{c}}}}]{Peyronel2012}%
  \BibitemOpen
  \bibfield  {author} {\bibinfo {author} {\bibfnamefont {T.}~\bibnamefont
  {Peyronel}}, \bibinfo {author} {\bibfnamefont {O.}~\bibnamefont
  {Firstenberg}}, \bibinfo {author} {\bibfnamefont {Q.-Y.}\ \bibnamefont
  {Liang}}, \bibinfo {author} {\bibfnamefont {S.}~\bibnamefont {Hofferberth}},
  \bibinfo {author} {\bibfnamefont {A.~V.}\ \bibnamefont {Gorshkov}}, \bibinfo
  {author} {\bibfnamefont {T.}~\bibnamefont {Pohl}}, \bibinfo {author}
  {\bibfnamefont {M.~D.}\ \bibnamefont {Lukin}}, \ and\ \bibinfo {author}
  {\bibfnamefont {V.}~\bibnamefont {Vuleti{\'{c}}}},\ }\href {\doibase
  10.1038/nature11361} {\bibfield  {journal} {\bibinfo  {journal} {Nature}\
  }\textbf {\bibinfo {volume} {488}},\ \bibinfo {pages} {57} (\bibinfo {year}
  {2012})}\BibitemShut {NoStop}%
\bibitem [{\citenamefont {Firstenberg}\ \emph {et~al.}(2013)\citenamefont
  {Firstenberg}, \citenamefont {Peyronel}, \citenamefont {Liang}, \citenamefont
  {Gorshkov}, \citenamefont {Lukin},\ and\ \citenamefont
  {Vuleti{\'{c}}}}]{Firstenberg2013}%
  \BibitemOpen
  \bibfield  {author} {\bibinfo {author} {\bibfnamefont {O.}~\bibnamefont
  {Firstenberg}}, \bibinfo {author} {\bibfnamefont {T.}~\bibnamefont
  {Peyronel}}, \bibinfo {author} {\bibfnamefont {Q.-Y.}\ \bibnamefont {Liang}},
  \bibinfo {author} {\bibfnamefont {A.~V.}\ \bibnamefont {Gorshkov}}, \bibinfo
  {author} {\bibfnamefont {M.~D.}\ \bibnamefont {Lukin}}, \ and\ \bibinfo
  {author} {\bibfnamefont {V.}~\bibnamefont {Vuleti{\'{c}}}},\ }\href {\doibase
  10.1038/nature12512} {\bibfield  {journal} {\bibinfo  {journal} {Nature}\
  }\textbf {\bibinfo {volume} {502}},\ \bibinfo {pages} {71} (\bibinfo {year}
  {2013})}\BibitemShut {NoStop}%
\bibitem [{\citenamefont {Parigi}\ \emph {et~al.}(2012)\citenamefont {Parigi},
  \citenamefont {Bimbard}, \citenamefont {Stanojevic}, \citenamefont
  {Hilliard}, \citenamefont {Nogrette}, \citenamefont {Tualle-Brouri},
  \citenamefont {Ourjoumtsev},\ and\ \citenamefont {Grangier}}]{Parigi2012}%
  \BibitemOpen
  \bibfield  {author} {\bibinfo {author} {\bibfnamefont {V.}~\bibnamefont
  {Parigi}}, \bibinfo {author} {\bibfnamefont {E.}~\bibnamefont {Bimbard}},
  \bibinfo {author} {\bibfnamefont {J.}~\bibnamefont {Stanojevic}}, \bibinfo
  {author} {\bibfnamefont {A.~J.}\ \bibnamefont {Hilliard}}, \bibinfo {author}
  {\bibfnamefont {F.}~\bibnamefont {Nogrette}}, \bibinfo {author}
  {\bibfnamefont {R.}~\bibnamefont {Tualle-Brouri}}, \bibinfo {author}
  {\bibfnamefont {A.}~\bibnamefont {Ourjoumtsev}}, \ and\ \bibinfo {author}
  {\bibfnamefont {P.}~\bibnamefont {Grangier}},\ }\href {\doibase
  10.1103/PhysRevLett.109.233602} {\bibfield  {journal} {\bibinfo  {journal}
  {Physical review letters}\ }\textbf {\bibinfo {volume} {109}},\ \bibinfo
  {pages} {233602} (\bibinfo {year} {2012})}\BibitemShut {NoStop}%
\bibitem [{\citenamefont {Maxwell}\ \emph {et~al.}(2013)\citenamefont
  {Maxwell}, \citenamefont {Szwer}, \citenamefont {Paredes-Barato},
  \citenamefont {Busche}, \citenamefont {Pritchard}, \citenamefont {Gauguet},
  \citenamefont {Weatherill}, \citenamefont {Jones},\ and\ \citenamefont
  {Adams}}]{Maxwell2013}%
  \BibitemOpen
  \bibfield  {author} {\bibinfo {author} {\bibfnamefont {D.}~\bibnamefont
  {Maxwell}}, \bibinfo {author} {\bibfnamefont {D.~J.}\ \bibnamefont {Szwer}},
  \bibinfo {author} {\bibfnamefont {D.}~\bibnamefont {Paredes-Barato}},
  \bibinfo {author} {\bibfnamefont {H.}~\bibnamefont {Busche}}, \bibinfo
  {author} {\bibfnamefont {J.~D.}\ \bibnamefont {Pritchard}}, \bibinfo {author}
  {\bibfnamefont {A.}~\bibnamefont {Gauguet}}, \bibinfo {author} {\bibfnamefont
  {K.~J.}\ \bibnamefont {Weatherill}}, \bibinfo {author} {\bibfnamefont
  {M.~P.~A.}\ \bibnamefont {Jones}}, \ and\ \bibinfo {author} {\bibfnamefont
  {C.~S.}\ \bibnamefont {Adams}},\ }\href {\doibase
  10.1103/PhysRevLett.110.103001} {\bibfield  {journal} {\bibinfo  {journal}
  {Physical review letters}\ }\textbf {\bibinfo {volume} {110}},\ \bibinfo
  {pages} {103001} (\bibinfo {year} {2013})}\BibitemShut {NoStop}%
\bibitem [{\citenamefont {Maxwell}\ \emph {et~al.}(2014)\citenamefont
  {Maxwell}, \citenamefont {Szwer}, \citenamefont {Paredes-Barato},
  \citenamefont {Busche}, \citenamefont {Pritchard}, \citenamefont {Gauguet},
  \citenamefont {Jones},\ and\ \citenamefont {Adams}}]{Maxwell2014}%
  \BibitemOpen
  \bibfield  {author} {\bibinfo {author} {\bibfnamefont {D.}~\bibnamefont
  {Maxwell}}, \bibinfo {author} {\bibfnamefont {D.~J.}\ \bibnamefont {Szwer}},
  \bibinfo {author} {\bibfnamefont {D.}~\bibnamefont {Paredes-Barato}},
  \bibinfo {author} {\bibfnamefont {H.}~\bibnamefont {Busche}}, \bibinfo
  {author} {\bibfnamefont {J.~D.}\ \bibnamefont {Pritchard}}, \bibinfo {author}
  {\bibfnamefont {A.}~\bibnamefont {Gauguet}}, \bibinfo {author} {\bibfnamefont
  {M.~P.~A.}\ \bibnamefont {Jones}}, \ and\ \bibinfo {author} {\bibfnamefont
  {C.~S.}\ \bibnamefont {Adams}},\ }\href {\doibase 10.1103/PhysRevA.89.043827}
  {\bibfield  {journal} {\bibinfo  {journal} {Physical Review A}\ }\textbf
  {\bibinfo {volume} {89}},\ \bibinfo {pages} {043827} (\bibinfo {year}
  {2014})}\BibitemShut {NoStop}%
\bibitem [{\citenamefont {Baur}\ \emph {et~al.}(2014)\citenamefont {Baur},
  \citenamefont {Tiarks}, \citenamefont {Rempe},\ and\ \citenamefont
  {D{\"{u}}rr}}]{Baur2014}%
  \BibitemOpen
  \bibfield  {author} {\bibinfo {author} {\bibfnamefont {S.}~\bibnamefont
  {Baur}}, \bibinfo {author} {\bibfnamefont {D.}~\bibnamefont {Tiarks}},
  \bibinfo {author} {\bibfnamefont {G.}~\bibnamefont {Rempe}}, \ and\ \bibinfo
  {author} {\bibfnamefont {S.}~\bibnamefont {D{\"{u}}rr}},\ }\href {\doibase
  10.1103/PhysRevLett.112.073901} {\bibfield  {journal} {\bibinfo  {journal}
  {Physical review letters}\ }\textbf {\bibinfo {volume} {112}},\ \bibinfo
  {pages} {073901} (\bibinfo {year} {2014})}\BibitemShut {NoStop}%
\bibitem [{\citenamefont {Tiarks}\ \emph {et~al.}(2014)\citenamefont {Tiarks},
  \citenamefont {Baur}, \citenamefont {Schneider}, \citenamefont {D{\"{u}}rr},\
  and\ \citenamefont {Rempe}}]{Tiarks2014}%
  \BibitemOpen
  \bibfield  {author} {\bibinfo {author} {\bibfnamefont {D.}~\bibnamefont
  {Tiarks}}, \bibinfo {author} {\bibfnamefont {S.}~\bibnamefont {Baur}},
  \bibinfo {author} {\bibfnamefont {K.}~\bibnamefont {Schneider}}, \bibinfo
  {author} {\bibfnamefont {S.}~\bibnamefont {D{\"{u}}rr}}, \ and\ \bibinfo
  {author} {\bibfnamefont {G.}~\bibnamefont {Rempe}},\ }\href {\doibase
  10.1103/PhysRevLett.113.053602} {\bibfield  {journal} {\bibinfo  {journal}
  {Physical review letters}\ }\textbf {\bibinfo {volume} {113}},\ \bibinfo
  {pages} {053602} (\bibinfo {year} {2014})}\BibitemShut {NoStop}%
\bibitem [{\citenamefont {Gorniaczyk}\ \emph {et~al.}(2014)\citenamefont
  {Gorniaczyk}, \citenamefont {Tresp}, \citenamefont {Schmidt}, \citenamefont
  {Fedder},\ and\ \citenamefont {Hofferberth}}]{Gorniaczyk2014}%
  \BibitemOpen
  \bibfield  {author} {\bibinfo {author} {\bibfnamefont {H.}~\bibnamefont
  {Gorniaczyk}}, \bibinfo {author} {\bibfnamefont {C.}~\bibnamefont {Tresp}},
  \bibinfo {author} {\bibfnamefont {J.}~\bibnamefont {Schmidt}}, \bibinfo
  {author} {\bibfnamefont {H.}~\bibnamefont {Fedder}}, \ and\ \bibinfo {author}
  {\bibfnamefont {S.}~\bibnamefont {Hofferberth}},\ }\href {\doibase
  10.1103/PhysRevLett.113.053601} {\bibfield  {journal} {\bibinfo  {journal}
  {Physical review letters}\ }\textbf {\bibinfo {volume} {113}},\ \bibinfo
  {pages} {053601} (\bibinfo {year} {2014})}\BibitemShut {NoStop}%
\bibitem [{\citenamefont {Gorniaczyk}\ \emph {et~al.}(2016)\citenamefont
  {Gorniaczyk}, \citenamefont {Tresp}, \citenamefont {Bienias}, \citenamefont
  {Paris-Mandoki}, \citenamefont {Li}, \citenamefont {Mirgorodskiy},
  \citenamefont {B{\"{u}}chler}, \citenamefont {Lesanovsky},\ and\
  \citenamefont {Hofferberth}}]{Gorniaczyk2016}%
  \BibitemOpen
  \bibfield  {author} {\bibinfo {author} {\bibfnamefont {H.}~\bibnamefont
  {Gorniaczyk}}, \bibinfo {author} {\bibfnamefont {C.}~\bibnamefont {Tresp}},
  \bibinfo {author} {\bibfnamefont {P.}~\bibnamefont {Bienias}}, \bibinfo
  {author} {\bibfnamefont {A.}~\bibnamefont {Paris-Mandoki}}, \bibinfo {author}
  {\bibfnamefont {W.}~\bibnamefont {Li}}, \bibinfo {author} {\bibfnamefont
  {I.}~\bibnamefont {Mirgorodskiy}}, \bibinfo {author} {\bibfnamefont {H.~P.}\
  \bibnamefont {B{\"{u}}chler}}, \bibinfo {author} {\bibfnamefont
  {I.}~\bibnamefont {Lesanovsky}}, \ and\ \bibinfo {author} {\bibfnamefont
  {S.}~\bibnamefont {Hofferberth}},\ }\href {\doibase 10.1038/ncomms12480}
  {\bibfield  {journal} {\bibinfo  {journal} {Nature Communications}\ }\textbf
  {\bibinfo {volume} {7}},\ \bibinfo {pages} {12480} (\bibinfo {year}
  {2016})}\BibitemShut {NoStop}%
\bibitem [{\citenamefont {Tiarks}\ \emph {et~al.}(2016)\citenamefont {Tiarks},
  \citenamefont {Schmidt}, \citenamefont {Rempe},\ and\ \citenamefont
  {Du rr}}]{Tiarks2016}%
  \BibitemOpen
  \bibfield  {author} {\bibinfo {author} {\bibfnamefont {D.}~\bibnamefont
  {Tiarks}}, \bibinfo {author} {\bibfnamefont {S.}~\bibnamefont {Schmidt}},
  \bibinfo {author} {\bibfnamefont {G.}~\bibnamefont {Rempe}}, \ and\ \bibinfo
  {author} {\bibfnamefont {S.}~\bibnamefont {Du rr}},\ }\href {\doibase
  10.1126/sciadv.1600036} {\bibfield  {journal} {\bibinfo  {journal} {Science
  Advances}\ }\textbf {\bibinfo {volume} {2}},\ \bibinfo {pages} {e1600036}
  (\bibinfo {year} {2016})}\BibitemShut {NoStop}%
\bibitem [{\citenamefont {Tresp}\ \emph {et~al.}(2016)\citenamefont {Tresp},
  \citenamefont {Zimmer}, \citenamefont {Mirgorodskiy}, \citenamefont
  {Gorniaczyk}, \citenamefont {Paris-Mandoki},\ and\ \citenamefont
  {Hofferberth}}]{Tresp2016a}%
  \BibitemOpen
  \bibfield  {author} {\bibinfo {author} {\bibfnamefont {C.}~\bibnamefont
  {Tresp}}, \bibinfo {author} {\bibfnamefont {C.}~\bibnamefont {Zimmer}},
  \bibinfo {author} {\bibfnamefont {I.}~\bibnamefont {Mirgorodskiy}}, \bibinfo
  {author} {\bibfnamefont {H.}~\bibnamefont {Gorniaczyk}}, \bibinfo {author}
  {\bibfnamefont {A.}~\bibnamefont {Paris-Mandoki}}, \ and\ \bibinfo {author}
  {\bibfnamefont {S.}~\bibnamefont {Hofferberth}},\ }\href {\doibase
  10.1103/PhysRevLett.117.223001} {\bibfield  {journal} {\bibinfo  {journal}
  {Physical Review Letters}\ }\textbf {\bibinfo {volume} {117}},\ \bibinfo
  {pages} {223001} (\bibinfo {year} {2016})}\BibitemShut {NoStop}%
\bibitem [{\citenamefont {G\"unter}\ \emph {et~al.}(2012)\citenamefont
  {G\"unter}, \citenamefont {Robert-de Saint-Vincent}, \citenamefont {Schempp},
  \citenamefont {Hofmann}, \citenamefont {Whitlock},\ and\ \citenamefont
  {Weidem\"uller}}]{Gunter2012}%
  \BibitemOpen
  \bibfield  {author} {\bibinfo {author} {\bibfnamefont {G.}~\bibnamefont
  {G\"unter}}, \bibinfo {author} {\bibfnamefont {M.}~\bibnamefont {Robert-de
  Saint-Vincent}}, \bibinfo {author} {\bibfnamefont {H.}~\bibnamefont
  {Schempp}}, \bibinfo {author} {\bibfnamefont {C.~S.}\ \bibnamefont
  {Hofmann}}, \bibinfo {author} {\bibfnamefont {S.}~\bibnamefont {Whitlock}}, \
  and\ \bibinfo {author} {\bibfnamefont {M.}~\bibnamefont {Weidem\"uller}},\
  }\href {\doibase 10.1103/PhysRevLett.108.013002} {\bibfield  {journal}
  {\bibinfo  {journal} {Phys. Rev. Lett.}\ }\textbf {\bibinfo {volume} {108}},\
  \bibinfo {pages} {013002} (\bibinfo {year} {2012})}\BibitemShut {NoStop}%
\bibitem [{\citenamefont {G{\"u}nter}\ \emph {et~al.}(2013)\citenamefont
  {G{\"u}nter}, \citenamefont {Schempp}, \citenamefont {Robert-de
  Saint-Vincent}, \citenamefont {Gavryusev}, \citenamefont {Helmrich},
  \citenamefont {Hofmann}, \citenamefont {Whitlock},\ and\ \citenamefont
  {Weidem{\"u}ller}}]{Gunter2013}%
  \BibitemOpen
  \bibfield  {author} {\bibinfo {author} {\bibfnamefont {G.}~\bibnamefont
  {G{\"u}nter}}, \bibinfo {author} {\bibfnamefont {H.}~\bibnamefont {Schempp}},
  \bibinfo {author} {\bibfnamefont {M.}~\bibnamefont {Robert-de
  Saint-Vincent}}, \bibinfo {author} {\bibfnamefont {V.}~\bibnamefont
  {Gavryusev}}, \bibinfo {author} {\bibfnamefont {S.}~\bibnamefont {Helmrich}},
  \bibinfo {author} {\bibfnamefont {C.~S.}\ \bibnamefont {Hofmann}}, \bibinfo
  {author} {\bibfnamefont {S.}~\bibnamefont {Whitlock}}, \ and\ \bibinfo
  {author} {\bibfnamefont {M.}~\bibnamefont {Weidem{\"u}ller}},\ }\href
  {\doibase 10.1126/science.1244843} {\bibfield  {journal} {\bibinfo  {journal}
  {Science}\ }\textbf {\bibinfo {volume} {342}},\ \bibinfo {pages} {954}
  (\bibinfo {year} {2013})},\ \Eprint
  {http://arxiv.org/abs/http://science.sciencemag.org/content/342/6161/954.full.pdf}
  {http://science.sciencemag.org/content/342/6161/954.full.pdf} \BibitemShut
  {NoStop}%
\bibitem [{\citenamefont {Murray}\ \emph {et~al.}(2016)\citenamefont {Murray},
  \citenamefont {Gorshkov},\ and\ \citenamefont {Pohl}}]{Murray2016a}%
  \BibitemOpen
  \bibfield  {author} {\bibinfo {author} {\bibfnamefont {C.~R.}\ \bibnamefont
  {Murray}}, \bibinfo {author} {\bibfnamefont {A.~R.}\ \bibnamefont
  {Gorshkov}}, \ and\ \bibinfo {author} {\bibfnamefont {T.}~\bibnamefont
  {Pohl}},\ }\href {http://stacks.iop.org/1367-2630/18/i=9/a=092001} {\bibfield
   {journal} {\bibinfo  {journal} {New Journal of Physics}\ }\textbf {\bibinfo
  {volume} {18}},\ \bibinfo {pages} {92001} (\bibinfo {year}
  {2016})}\BibitemShut {NoStop}%
\bibitem [{\citenamefont {Zeuthen}\ \emph {et~al.}(2016)\citenamefont
  {Zeuthen}, \citenamefont {Gullans}, \citenamefont {Maghrebi},\ and\
  \citenamefont {Gorshkov}}]{Zeuthen2016}%
  \BibitemOpen
  \bibfield  {author} {\bibinfo {author} {\bibfnamefont {E.}~\bibnamefont
  {Zeuthen}}, \bibinfo {author} {\bibfnamefont {M.~J.}\ \bibnamefont
  {Gullans}}, \bibinfo {author} {\bibfnamefont {M.~F.}\ \bibnamefont
  {Maghrebi}}, \ and\ \bibinfo {author} {\bibfnamefont {A.~V.}\ \bibnamefont
  {Gorshkov}},\ }\href {http://arxiv.org/abs/1608.06068} {\  (\bibinfo {year}
  {2016})},\ \Eprint {http://arxiv.org/abs/1608.06068} {arXiv:1608.06068}
  \BibitemShut {NoStop}%
\bibitem [{\citenamefont {Gaj}\ \emph {et~al.}(2014)\citenamefont {Gaj},
  \citenamefont {Krupp}, \citenamefont {Balewski}, \citenamefont {L{\"{o}}w},
  \citenamefont {Hofferberth},\ and\ \citenamefont {Pfau}}]{Gaj2014}%
  \BibitemOpen
  \bibfield  {author} {\bibinfo {author} {\bibfnamefont {A.}~\bibnamefont
  {Gaj}}, \bibinfo {author} {\bibfnamefont {A.~T.}\ \bibnamefont {Krupp}},
  \bibinfo {author} {\bibfnamefont {J.~B.}\ \bibnamefont {Balewski}}, \bibinfo
  {author} {\bibfnamefont {R.}~\bibnamefont {L{\"{o}}w}}, \bibinfo {author}
  {\bibfnamefont {S.}~\bibnamefont {Hofferberth}}, \ and\ \bibinfo {author}
  {\bibfnamefont {T.}~\bibnamefont {Pfau}},\ }\href {\doibase
  10.1038/ncomms5546} {\bibfield  {journal} {\bibinfo  {journal} {Nature
  communications}\ }\textbf {\bibinfo {volume} {5}},\ \bibinfo {pages} {4546}
  (\bibinfo {year} {2014})}\BibitemShut {NoStop}%
\bibitem [{\citenamefont {Harris}\ \emph {et~al.}(1992)\citenamefont {Harris},
  \citenamefont {Field},\ and\ \citenamefont {Kasapi}}]{Harris1992}%
  \BibitemOpen
  \bibfield  {author} {\bibinfo {author} {\bibfnamefont {S.~E.}\ \bibnamefont
  {Harris}}, \bibinfo {author} {\bibfnamefont {J.~E.}\ \bibnamefont {Field}}, \
  and\ \bibinfo {author} {\bibfnamefont {A.}~\bibnamefont {Kasapi}},\ }\href
  {\doibase 10.1103/PhysRevA.46.R29} {\bibfield  {journal} {\bibinfo  {journal}
  {Physical Review A}\ }\textbf {\bibinfo {volume} {46}},\ \bibinfo {pages}
  {R29} (\bibinfo {year} {1992})}\BibitemShut {NoStop}%
\bibitem [{\citenamefont {Andr{\'{e}}}\ and\ \citenamefont
  {Lukin}(2002)}]{Andre2002}%
  \BibitemOpen
  \bibfield  {author} {\bibinfo {author} {\bibfnamefont {A.}~\bibnamefont
  {Andr{\'{e}}}}\ and\ \bibinfo {author} {\bibfnamefont {M.~D.}\ \bibnamefont
  {Lukin}},\ }\href {\doibase 10.1103/PhysRevLett.89.143602} {\bibfield
  {journal} {\bibinfo  {journal} {Physical Review Letters}\ }\textbf {\bibinfo
  {volume} {89}},\ \bibinfo {pages} {143602} (\bibinfo {year}
  {2002})}\BibitemShut {NoStop}%
\bibitem [{\citenamefont {Andr{\'{e}}}\ \emph {et~al.}(2005)\citenamefont
  {Andr{\'{e}}}, \citenamefont {Bajcsy}, \citenamefont {Zibrov},\ and\
  \citenamefont {Lukin}}]{Andre2005}%
  \BibitemOpen
  \bibfield  {author} {\bibinfo {author} {\bibfnamefont {A.}~\bibnamefont
  {Andr{\'{e}}}}, \bibinfo {author} {\bibfnamefont {M.}~\bibnamefont {Bajcsy}},
  \bibinfo {author} {\bibfnamefont {A.~S.}\ \bibnamefont {Zibrov}}, \ and\
  \bibinfo {author} {\bibfnamefont {M.~D.}\ \bibnamefont {Lukin}},\ }\href
  {\doibase 10.1103/PhysRevLett.94.063902} {\bibfield  {journal} {\bibinfo
  {journal} {Physical Review Letters}\ }\textbf {\bibinfo {volume} {94}},\
  \bibinfo {pages} {063902} (\bibinfo {year} {2005})}\BibitemShut {NoStop}%
\bibitem [{\citenamefont {Fleischhauer}\ \emph {et~al.}(2008)\citenamefont
  {Fleischhauer}, \citenamefont {Otterbach},\ and\ \citenamefont
  {Unanyan}}]{Fleischhauer2008}%
  \BibitemOpen
  \bibfield  {author} {\bibinfo {author} {\bibfnamefont {M.}~\bibnamefont
  {Fleischhauer}}, \bibinfo {author} {\bibfnamefont {J.}~\bibnamefont
  {Otterbach}}, \ and\ \bibinfo {author} {\bibfnamefont {R.~G.}\ \bibnamefont
  {Unanyan}},\ }\href {\doibase 10.1103/PhysRevLett.101.163601} {\bibfield
  {journal} {\bibinfo  {journal} {Physical Review Letters}\ }\textbf {\bibinfo
  {volume} {101}},\ \bibinfo {pages} {163601} (\bibinfo {year}
  {2008})}\BibitemShut {NoStop}%
\bibitem [{\citenamefont {Zimmer}\ \emph {et~al.}(2008)\citenamefont {Zimmer},
  \citenamefont {Otterbach}, \citenamefont {Unanyan}, \citenamefont {Shore},\
  and\ \citenamefont {Fleischhauer}}]{Zimmer2008}%
  \BibitemOpen
  \bibfield  {author} {\bibinfo {author} {\bibfnamefont {F.~E.}\ \bibnamefont
  {Zimmer}}, \bibinfo {author} {\bibfnamefont {J.}~\bibnamefont {Otterbach}},
  \bibinfo {author} {\bibfnamefont {R.~G.}\ \bibnamefont {Unanyan}}, \bibinfo
  {author} {\bibfnamefont {B.~W.}\ \bibnamefont {Shore}}, \ and\ \bibinfo
  {author} {\bibfnamefont {M.}~\bibnamefont {Fleischhauer}},\ }\href {\doibase
  10.1103/PhysRevA.77.063823} {\bibfield  {journal} {\bibinfo  {journal}
  {Physical Review A}\ }\textbf {\bibinfo {volume} {77}},\ \bibinfo {pages}
  {063823} (\bibinfo {year} {2008})}\BibitemShut {NoStop}%
\bibitem [{\citenamefont {Hafezi}\ \emph {et~al.}(2012)\citenamefont {Hafezi},
  \citenamefont {Chang}, \citenamefont {Gritsev}, \citenamefont {Demler},\ and\
  \citenamefont {Lukin}}]{Hafezi2012}%
  \BibitemOpen
  \bibfield  {author} {\bibinfo {author} {\bibfnamefont {M.}~\bibnamefont
  {Hafezi}}, \bibinfo {author} {\bibfnamefont {D.~E.}\ \bibnamefont {Chang}},
  \bibinfo {author} {\bibfnamefont {V.}~\bibnamefont {Gritsev}}, \bibinfo
  {author} {\bibfnamefont {E.}~\bibnamefont {Demler}}, \ and\ \bibinfo {author}
  {\bibfnamefont {M.~D.}\ \bibnamefont {Lukin}},\ }\href {\doibase
  10.1103/PhysRevA.85.013822} {\bibfield  {journal} {\bibinfo  {journal}
  {Physical Review A}\ }\textbf {\bibinfo {volume} {85}},\ \bibinfo {pages}
  {013822} (\bibinfo {year} {2012})}\BibitemShut {NoStop}%
\bibitem [{\citenamefont {Iakoupov}\ \emph {et~al.}(2016)\citenamefont
  {Iakoupov}, \citenamefont {Ott}, \citenamefont {Chang},\ and\ \citenamefont
  {S{\o}rensen}}]{Iakoupov2016}%
  \BibitemOpen
  \bibfield  {author} {\bibinfo {author} {\bibfnamefont {I.}~\bibnamefont
  {Iakoupov}}, \bibinfo {author} {\bibfnamefont {J.~R.}\ \bibnamefont {Ott}},
  \bibinfo {author} {\bibfnamefont {D.~E.}\ \bibnamefont {Chang}}, \ and\
  \bibinfo {author} {\bibfnamefont {A.~S.}\ \bibnamefont {S{\o}rensen}},\
  }\href {\doibase 10.1103/PhysRevA.94.053824} {\bibfield  {journal} {\bibinfo
  {journal} {Physical Review A}\ }\textbf {\bibinfo {volume} {94}},\ \bibinfo
  {pages} {053824} (\bibinfo {year} {2016})}\BibitemShut {NoStop}%
\bibitem [{\citenamefont {Hau}\ \emph {et~al.}(1999)\citenamefont {Hau},
  \citenamefont {Harris}, \citenamefont {Dutton},\ and\ \citenamefont
  {Behroozi}}]{Hau1999}%
  \BibitemOpen
  \bibfield  {author} {\bibinfo {author} {\bibfnamefont {L.~V.}\ \bibnamefont
  {Hau}}, \bibinfo {author} {\bibfnamefont {S.~E.}\ \bibnamefont {Harris}},
  \bibinfo {author} {\bibfnamefont {Z.}~\bibnamefont {Dutton}}, \ and\ \bibinfo
  {author} {\bibfnamefont {C.~H.}\ \bibnamefont {Behroozi}},\ }\href {\doibase
  10.1038/17561} {\bibfield  {journal} {\bibinfo  {journal} {Nature}\ }\textbf
  {\bibinfo {volume} {397}},\ \bibinfo {pages} {594} (\bibinfo {year}
  {1999})}\BibitemShut {NoStop}%
\bibitem [{\citenamefont {Kash}\ \emph {et~al.}(1999)\citenamefont {Kash},
  \citenamefont {Sautenkov}, \citenamefont {Zibrov}, \citenamefont {Hollberg},
  \citenamefont {Welch}, \citenamefont {Lukin}, \citenamefont {Rostovtsev},
  \citenamefont {Fry},\ and\ \citenamefont {Scully}}]{Kash1999}%
  \BibitemOpen
  \bibfield  {author} {\bibinfo {author} {\bibfnamefont {M.~M.}\ \bibnamefont
  {Kash}}, \bibinfo {author} {\bibfnamefont {V.~A.}\ \bibnamefont {Sautenkov}},
  \bibinfo {author} {\bibfnamefont {A.~S.}\ \bibnamefont {Zibrov}}, \bibinfo
  {author} {\bibfnamefont {L.}~\bibnamefont {Hollberg}}, \bibinfo {author}
  {\bibfnamefont {G.~R.}\ \bibnamefont {Welch}}, \bibinfo {author}
  {\bibfnamefont {M.~D.}\ \bibnamefont {Lukin}}, \bibinfo {author}
  {\bibfnamefont {Y.}~\bibnamefont {Rostovtsev}}, \bibinfo {author}
  {\bibfnamefont {E.~S.}\ \bibnamefont {Fry}}, \ and\ \bibinfo {author}
  {\bibfnamefont {M.~O.}\ \bibnamefont {Scully}},\ }\href {\doibase
  10.1103/PhysRevLett.82.5229} {\bibfield  {journal} {\bibinfo  {journal}
  {Physical Review Letters}\ }\textbf {\bibinfo {volume} {82}},\ \bibinfo
  {pages} {5229} (\bibinfo {year} {1999})}\BibitemShut {NoStop}%
\bibitem [{\citenamefont {Bajcsy}\ \emph {et~al.}(2003)\citenamefont {Bajcsy},
  \citenamefont {Zibrov},\ and\ \citenamefont {Lukin}}]{Bajcsy2003}%
  \BibitemOpen
  \bibfield  {author} {\bibinfo {author} {\bibfnamefont {M.}~\bibnamefont
  {Bajcsy}}, \bibinfo {author} {\bibfnamefont {A.~S.}\ \bibnamefont {Zibrov}},
  \ and\ \bibinfo {author} {\bibfnamefont {M.~D.}\ \bibnamefont {Lukin}},\
  }\href {\doibase 10.1038/nature02176} {\bibfield  {journal} {\bibinfo
  {journal} {Nature}\ }\textbf {\bibinfo {volume} {426}},\ \bibinfo {pages}
  {638} (\bibinfo {year} {2003})}\BibitemShut {NoStop}%
\bibitem [{\citenamefont {Lin}\ \emph {et~al.}(2009)\citenamefont {Lin},
  \citenamefont {Liao}, \citenamefont {Peters}, \citenamefont {Chou},
  \citenamefont {Wang}, \citenamefont {Cho}, \citenamefont {Kuan},\ and\
  \citenamefont {Yu}}]{Lin2009}%
  \BibitemOpen
  \bibfield  {author} {\bibinfo {author} {\bibfnamefont {Y.-W.}\ \bibnamefont
  {Lin}}, \bibinfo {author} {\bibfnamefont {W.-T.}\ \bibnamefont {Liao}},
  \bibinfo {author} {\bibfnamefont {T.}~\bibnamefont {Peters}}, \bibinfo
  {author} {\bibfnamefont {H.-C.}\ \bibnamefont {Chou}}, \bibinfo {author}
  {\bibfnamefont {J.-S.}\ \bibnamefont {Wang}}, \bibinfo {author}
  {\bibfnamefont {H.-W.}\ \bibnamefont {Cho}}, \bibinfo {author} {\bibfnamefont
  {P.-C.}\ \bibnamefont {Kuan}}, \ and\ \bibinfo {author} {\bibfnamefont
  {I.~A.}\ \bibnamefont {Yu}},\ }\href {\doibase
  10.1103/PhysRevLett.102.213601} {\bibfield  {journal} {\bibinfo  {journal}
  {Physical Review Letters}\ }\textbf {\bibinfo {volume} {102}},\ \bibinfo
  {pages} {213601} (\bibinfo {year} {2009})}\BibitemShut {NoStop}%
\bibitem [{\citenamefont {Everett}\ \emph {et~al.}(2016)\citenamefont
  {Everett}, \citenamefont {Campbell}, \citenamefont {Cho}, \citenamefont
  {Vernaz-Gris}, \citenamefont {Higginbottom}, \citenamefont {Pinel},
  \citenamefont {Robins}, \citenamefont {Lam},\ and\ \citenamefont
  {Buchler}}]{Everett2016}%
  \BibitemOpen
  \bibfield  {author} {\bibinfo {author} {\bibfnamefont {J.}~\bibnamefont
  {Everett}}, \bibinfo {author} {\bibfnamefont {G.~T.}\ \bibnamefont
  {Campbell}}, \bibinfo {author} {\bibfnamefont {Y.-W.}\ \bibnamefont {Cho}},
  \bibinfo {author} {\bibfnamefont {P.}~\bibnamefont {Vernaz-Gris}}, \bibinfo
  {author} {\bibfnamefont {D.}~\bibnamefont {Higginbottom}}, \bibinfo {author}
  {\bibfnamefont {O.}~\bibnamefont {Pinel}}, \bibinfo {author} {\bibfnamefont
  {N.~P.}\ \bibnamefont {Robins}}, \bibinfo {author} {\bibfnamefont {P.~K.}\
  \bibnamefont {Lam}}, \ and\ \bibinfo {author} {\bibfnamefont {B.~C.}\
  \bibnamefont {Buchler}},\ }\href {\doibase 10.1038/nphys3901} {\bibfield
  {journal} {\bibinfo  {journal} {Nature Physics}\ } (\bibinfo {year} {2016}),\
  10.1038/nphys3901}\BibitemShut {NoStop}%
\bibitem [{\citenamefont {Novikova}\ \emph {et~al.}(2007)\citenamefont
  {Novikova}, \citenamefont {Gorshkov}, \citenamefont {Phillips}, \citenamefont
  {S{\o}rensen}, \citenamefont {Lukin},\ and\ \citenamefont
  {Walsworth}}]{Novikova2007}%
  \BibitemOpen
  \bibfield  {author} {\bibinfo {author} {\bibfnamefont {I.}~\bibnamefont
  {Novikova}}, \bibinfo {author} {\bibfnamefont {A.~V.}\ \bibnamefont
  {Gorshkov}}, \bibinfo {author} {\bibfnamefont {D.~F.}\ \bibnamefont
  {Phillips}}, \bibinfo {author} {\bibfnamefont {A.~S.}\ \bibnamefont
  {S{\o}rensen}}, \bibinfo {author} {\bibfnamefont {M.~D.}\ \bibnamefont
  {Lukin}}, \ and\ \bibinfo {author} {\bibfnamefont {R.~L.}\ \bibnamefont
  {Walsworth}},\ }\href {\doibase 10.1103/PhysRevLett.98.243602} {\bibfield
  {journal} {\bibinfo  {journal} {Physical review letters}\ }\textbf {\bibinfo
  {volume} {98}},\ \bibinfo {pages} {243602} (\bibinfo {year}
  {2007})}\BibitemShut {NoStop}%
\bibitem [{\citenamefont {Gorshkov}\ \emph
  {et~al.}(2007{\natexlab{a}})\citenamefont {Gorshkov}, \citenamefont
  {Andr{\'{e}}}, \citenamefont {Fleischhauer}, \citenamefont {S{\o}rensen},\
  and\ \citenamefont {Lukin}}]{Gorshkov2007}%
  \BibitemOpen
  \bibfield  {author} {\bibinfo {author} {\bibfnamefont {A.~V.}\ \bibnamefont
  {Gorshkov}}, \bibinfo {author} {\bibfnamefont {A.}~\bibnamefont
  {Andr{\'{e}}}}, \bibinfo {author} {\bibfnamefont {M.}~\bibnamefont
  {Fleischhauer}}, \bibinfo {author} {\bibfnamefont {A.~S.}\ \bibnamefont
  {S{\o}rensen}}, \ and\ \bibinfo {author} {\bibfnamefont {M.~D.}\ \bibnamefont
  {Lukin}},\ }\href {\doibase 10.1103/PhysRevLett.98.123601} {\bibfield
  {journal} {\bibinfo  {journal} {Physical review letters}\ }\textbf {\bibinfo
  {volume} {98}},\ \bibinfo {pages} {123601} (\bibinfo {year}
  {2007}{\natexlab{a}})}\BibitemShut {NoStop}%
\bibitem [{\citenamefont {Gorshkov}\ \emph
  {et~al.}(2007{\natexlab{b}})\citenamefont {Gorshkov}, \citenamefont
  {Andr{\'{e}}}, \citenamefont {Lukin},\ and\ \citenamefont
  {S{\o}rensen}}]{Gorshkov2007a}%
  \BibitemOpen
  \bibfield  {author} {\bibinfo {author} {\bibfnamefont {A.~V.}\ \bibnamefont
  {Gorshkov}}, \bibinfo {author} {\bibfnamefont {A.}~\bibnamefont
  {Andr{\'{e}}}}, \bibinfo {author} {\bibfnamefont {M.~D.}\ \bibnamefont
  {Lukin}}, \ and\ \bibinfo {author} {\bibfnamefont {A.~S.}\ \bibnamefont
  {S{\o}rensen}},\ }\href {\doibase 10.1103/PhysRevA.76.033805} {\bibfield
  {journal} {\bibinfo  {journal} {Physical Review A}\ }\textbf {\bibinfo
  {volume} {76}},\ \bibinfo {pages} {033805} (\bibinfo {year}
  {2007}{\natexlab{b}})}\BibitemShut {NoStop}%
\bibitem [{\citenamefont {Li}\ and\ \citenamefont
  {Lesanovsky}(2015)}]{Li2015a}%
  \BibitemOpen
  \bibfield  {author} {\bibinfo {author} {\bibfnamefont {W.}~\bibnamefont
  {Li}}\ and\ \bibinfo {author} {\bibfnamefont {I.}~\bibnamefont
  {Lesanovsky}},\ }\href {\doibase 10.1103/PhysRevA.92.043828} {\bibfield
  {journal} {\bibinfo  {journal} {Physical Review A}\ }\textbf {\bibinfo
  {volume} {92}},\ \bibinfo {pages} {043828} (\bibinfo {year}
  {2015})}\BibitemShut {NoStop}%
\bibitem [{\citenamefont {Xia}\ and\ \citenamefont {Twamley}(2013)}]{Xia2013}%
  \BibitemOpen
  \bibfield  {author} {\bibinfo {author} {\bibfnamefont {K.}~\bibnamefont
  {Xia}}\ and\ \bibinfo {author} {\bibfnamefont {J.}~\bibnamefont {Twamley}},\
  }\href {\doibase 10.1103/PhysRevX.3.031013} {\bibfield  {journal} {\bibinfo
  {journal} {Physical Review X}\ }\textbf {\bibinfo {volume} {3}},\ \bibinfo
  {pages} {031013} (\bibinfo {year} {2013})}\BibitemShut {NoStop}%
\bibitem [{\citenamefont {Chang}\ \emph {et~al.}(2007)\citenamefont {Chang},
  \citenamefont {S{\o}rensen}, \citenamefont {Demler},\ and\ \citenamefont
  {Lukin}}]{Chang2007}%
  \BibitemOpen
  \bibfield  {author} {\bibinfo {author} {\bibfnamefont {D.~E.}\ \bibnamefont
  {Chang}}, \bibinfo {author} {\bibfnamefont {A.~S.}\ \bibnamefont
  {S{\o}rensen}}, \bibinfo {author} {\bibfnamefont {E.~A.}\ \bibnamefont
  {Demler}}, \ and\ \bibinfo {author} {\bibfnamefont {M.~D.}\ \bibnamefont
  {Lukin}},\ }\href {\doibase 10.1038/nphys708} {\bibfield  {journal} {\bibinfo
   {journal} {Nature Physics}\ }\textbf {\bibinfo {volume} {3}},\ \bibinfo
  {pages} {807} (\bibinfo {year} {2007})}\BibitemShut {NoStop}%
\bibitem [{\citenamefont {Bajcsy}\ \emph {et~al.}(2009)\citenamefont {Bajcsy},
  \citenamefont {Hofferberth}, \citenamefont {Balic}, \citenamefont {Peyronel},
  \citenamefont {Hafezi}, \citenamefont {Zibrov}, \citenamefont {Vuletic},\
  and\ \citenamefont {Lukin}}]{Bajcsy2009}%
  \BibitemOpen
  \bibfield  {author} {\bibinfo {author} {\bibfnamefont {M.}~\bibnamefont
  {Bajcsy}}, \bibinfo {author} {\bibfnamefont {S.}~\bibnamefont {Hofferberth}},
  \bibinfo {author} {\bibfnamefont {V.}~\bibnamefont {Balic}}, \bibinfo
  {author} {\bibfnamefont {T.}~\bibnamefont {Peyronel}}, \bibinfo {author}
  {\bibfnamefont {M.}~\bibnamefont {Hafezi}}, \bibinfo {author} {\bibfnamefont
  {A.~S.}\ \bibnamefont {Zibrov}}, \bibinfo {author} {\bibfnamefont
  {V.}~\bibnamefont {Vuletic}}, \ and\ \bibinfo {author} {\bibfnamefont
  {M.~D.}\ \bibnamefont {Lukin}},\ }\href {\doibase
  10.1103/PhysRevLett.102.203902} {\bibfield  {journal} {\bibinfo  {journal}
  {Physical review letters}\ }\textbf {\bibinfo {volume} {102}},\ \bibinfo
  {pages} {203902} (\bibinfo {year} {2009})}\BibitemShut {NoStop}%
\bibitem [{\citenamefont {Nozaki}\ \emph {et~al.}(2010)\citenamefont {Nozaki},
  \citenamefont {Tanabe}, \citenamefont {Shinya}, \citenamefont {Matsuo},
  \citenamefont {Sato}, \citenamefont {Taniyama},\ and\ \citenamefont
  {Notomi}}]{Nozaki2010}%
  \BibitemOpen
  \bibfield  {author} {\bibinfo {author} {\bibfnamefont {K.}~\bibnamefont
  {Nozaki}}, \bibinfo {author} {\bibfnamefont {T.}~\bibnamefont {Tanabe}},
  \bibinfo {author} {\bibfnamefont {A.}~\bibnamefont {Shinya}}, \bibinfo
  {author} {\bibfnamefont {S.}~\bibnamefont {Matsuo}}, \bibinfo {author}
  {\bibfnamefont {T.}~\bibnamefont {Sato}}, \bibinfo {author} {\bibfnamefont
  {H.}~\bibnamefont {Taniyama}}, \ and\ \bibinfo {author} {\bibfnamefont
  {M.}~\bibnamefont {Notomi}},\ }\href {\doibase 10.1038/nphoton.2010.89}
  {\bibfield  {journal} {\bibinfo  {journal} {Nature Photonics}\ }\textbf
  {\bibinfo {volume} {4}},\ \bibinfo {pages} {477} (\bibinfo {year}
  {2010})}\BibitemShut {NoStop}%
\bibitem [{\citenamefont {Volz}\ \emph {et~al.}(2012)\citenamefont {Volz},
  \citenamefont {Reinhard}, \citenamefont {Winger}, \citenamefont {Badolato},
  \citenamefont {Hennessy}, \citenamefont {Hu},\ and\ \citenamefont
  {Imamoglu}}]{Volz2012}%
  \BibitemOpen
  \bibfield  {author} {\bibinfo {author} {\bibfnamefont {T.}~\bibnamefont
  {Volz}}, \bibinfo {author} {\bibfnamefont {A.}~\bibnamefont {Reinhard}},
  \bibinfo {author} {\bibfnamefont {M.}~\bibnamefont {Winger}}, \bibinfo
  {author} {\bibfnamefont {A.}~\bibnamefont {Badolato}}, \bibinfo {author}
  {\bibfnamefont {K.~J.}\ \bibnamefont {Hennessy}}, \bibinfo {author}
  {\bibfnamefont {E.~L.}\ \bibnamefont {Hu}}, \ and\ \bibinfo {author}
  {\bibfnamefont {A.}~\bibnamefont {Imamoglu}},\ }\href {\doibase
  10.1038/nphoton.2012.181} {\bibfield  {journal} {\bibinfo  {journal} {Nature
  Photonics}\ }\textbf {\bibinfo {volume} {6}},\ \bibinfo {pages} {607}
  (\bibinfo {year} {2012})}\BibitemShut {NoStop}%
\bibitem [{\citenamefont {Bose}\ \emph {et~al.}(2012)\citenamefont {Bose},
  \citenamefont {Sridharan}, \citenamefont {Kim}, \citenamefont {Solomon},\
  and\ \citenamefont {Waks}}]{Bose2012}%
  \BibitemOpen
  \bibfield  {author} {\bibinfo {author} {\bibfnamefont {R.}~\bibnamefont
  {Bose}}, \bibinfo {author} {\bibfnamefont {D.}~\bibnamefont {Sridharan}},
  \bibinfo {author} {\bibfnamefont {H.}~\bibnamefont {Kim}}, \bibinfo {author}
  {\bibfnamefont {G.~S.}\ \bibnamefont {Solomon}}, \ and\ \bibinfo {author}
  {\bibfnamefont {E.}~\bibnamefont {Waks}},\ }\href {\doibase
  10.1103/PhysRevLett.108.227402} {\bibfield  {journal} {\bibinfo  {journal}
  {Physical review letters}\ }\textbf {\bibinfo {volume} {108}},\ \bibinfo
  {pages} {227402} (\bibinfo {year} {2012})}\BibitemShut {NoStop}%
\bibitem [{\citenamefont {Chen}\ \emph {et~al.}(2013)\citenamefont {Chen},
  \citenamefont {Beck}, \citenamefont {B{\"{u}}cker}, \citenamefont {Gullans},
  \citenamefont {Lukin}, \citenamefont {Tanji-Suzuki},\ and\ \citenamefont
  {Vuleti{\'{c}}}}]{Chen2013}%
  \BibitemOpen
  \bibfield  {author} {\bibinfo {author} {\bibfnamefont {W.}~\bibnamefont
  {Chen}}, \bibinfo {author} {\bibfnamefont {K.~M.}\ \bibnamefont {Beck}},
  \bibinfo {author} {\bibfnamefont {R.}~\bibnamefont {B{\"{u}}cker}}, \bibinfo
  {author} {\bibfnamefont {M.}~\bibnamefont {Gullans}}, \bibinfo {author}
  {\bibfnamefont {M.~D.}\ \bibnamefont {Lukin}}, \bibinfo {author}
  {\bibfnamefont {H.}~\bibnamefont {Tanji-Suzuki}}, \ and\ \bibinfo {author}
  {\bibfnamefont {V.}~\bibnamefont {Vuleti{\'{c}}}},\ }\href {\doibase
  10.1126/science.1238169} {\bibfield  {journal} {\bibinfo  {journal} {Science
  (New York, N.Y.)}\ }\textbf {\bibinfo {volume} {341}},\ \bibinfo {pages}
  {768} (\bibinfo {year} {2013})}\BibitemShut {NoStop}%
\bibitem [{\citenamefont {O'Shea}\ \emph {et~al.}(2013)\citenamefont {O'Shea},
  \citenamefont {Junge}, \citenamefont {Volz},\ and\ \citenamefont
  {Rauschenbeutel}}]{OShea2013}%
  \BibitemOpen
  \bibfield  {author} {\bibinfo {author} {\bibfnamefont {D.}~\bibnamefont
  {O'Shea}}, \bibinfo {author} {\bibfnamefont {C.}~\bibnamefont {Junge}},
  \bibinfo {author} {\bibfnamefont {J.}~\bibnamefont {Volz}}, \ and\ \bibinfo
  {author} {\bibfnamefont {A.}~\bibnamefont {Rauschenbeutel}},\ }\href
  {\doibase 10.1103/PhysRevLett.111.193601} {\bibfield  {journal} {\bibinfo
  {journal} {Physical review letters}\ }\textbf {\bibinfo {volume} {111}},\
  \bibinfo {pages} {193601} (\bibinfo {year} {2013})}\BibitemShut {NoStop}%
\bibitem [{\citenamefont {Chen}\ \emph {et~al.}(2012)\citenamefont {Chen},
  \citenamefont {Lee}, \citenamefont {Hung}, \citenamefont {Chen},
  \citenamefont {Chen},\ and\ \citenamefont {Yu}}]{Chen2012}%
  \BibitemOpen
  \bibfield  {author} {\bibinfo {author} {\bibfnamefont {Y.-H.}\ \bibnamefont
  {Chen}}, \bibinfo {author} {\bibfnamefont {M.-J.}\ \bibnamefont {Lee}},
  \bibinfo {author} {\bibfnamefont {W.}~\bibnamefont {Hung}}, \bibinfo {author}
  {\bibfnamefont {Y.-C.}\ \bibnamefont {Chen}}, \bibinfo {author}
  {\bibfnamefont {Y.-F.}\ \bibnamefont {Chen}}, \ and\ \bibinfo {author}
  {\bibfnamefont {I.~A.}\ \bibnamefont {Yu}},\ }\href {\doibase
  10.1103/PhysRevLett.108.173603} {\bibfield  {journal} {\bibinfo  {journal}
  {Phys. Rev. Lett.}\ }\textbf {\bibinfo {volume} {108}},\ \bibinfo {pages}
  {173603} (\bibinfo {year} {2012})}\BibitemShut {NoStop}%
\bibitem [{\citenamefont {Peters}\ \emph {et~al.}(2012)\citenamefont {Peters},
  \citenamefont {Su}, \citenamefont {Chen}, \citenamefont {Wang}, \citenamefont
  {Gou},\ and\ \citenamefont {Yu}}]{Peters2012}%
  \BibitemOpen
  \bibfield  {author} {\bibinfo {author} {\bibfnamefont {T.}~\bibnamefont
  {Peters}}, \bibinfo {author} {\bibfnamefont {S.-W.}\ \bibnamefont {Su}},
  \bibinfo {author} {\bibfnamefont {Y.-H.}\ \bibnamefont {Chen}}, \bibinfo
  {author} {\bibfnamefont {J.-S.}\ \bibnamefont {Wang}}, \bibinfo {author}
  {\bibfnamefont {S.-C.}\ \bibnamefont {Gou}}, \ and\ \bibinfo {author}
  {\bibfnamefont {I.~A.}\ \bibnamefont {Yu}},\ }\href {\doibase
  10.1103/PhysRevA.85.023838} {\bibfield  {journal} {\bibinfo  {journal} {Phys.
  Rev. A}\ }\textbf {\bibinfo {volume} {85}},\ \bibinfo {pages} {023838}
  (\bibinfo {year} {2012})}\BibitemShut {NoStop}%
\bibitem [{\citenamefont {Singer}\ \emph {et~al.}(2005)\citenamefont {Singer},
  \citenamefont {Stanojevic}, \citenamefont {Weidem{\"{u}}ller},\ and\
  \citenamefont {C{\^{o}}t{\'{e}}}}]{Singer2005}%
  \BibitemOpen
  \bibfield  {author} {\bibinfo {author} {\bibfnamefont {K.}~\bibnamefont
  {Singer}}, \bibinfo {author} {\bibfnamefont {J.}~\bibnamefont {Stanojevic}},
  \bibinfo {author} {\bibfnamefont {M.}~\bibnamefont {Weidem{\"{u}}ller}}, \
  and\ \bibinfo {author} {\bibfnamefont {R.}~\bibnamefont {C{\^{o}}t{\'{e}}}},\
  }\href {\doibase 10.1088/0953-4075/38/2/021} {\bibfield  {journal} {\bibinfo
  {journal} {Journal of Physics B: Atomic, Molecular and Optical Physics}\
  }\textbf {\bibinfo {volume} {38}},\ \bibinfo {pages} {S295} (\bibinfo {year}
  {2005})}\BibitemShut {NoStop}%
\bibitem [{\citenamefont {Petersen}\ \emph {et~al.}(2014)\citenamefont
  {Petersen}, \citenamefont {Volz},\ and\ \citenamefont
  {Rauschenbeutel}}]{Petersen2014}%
  \BibitemOpen
  \bibfield  {author} {\bibinfo {author} {\bibfnamefont {J.}~\bibnamefont
  {Petersen}}, \bibinfo {author} {\bibfnamefont {J.}~\bibnamefont {Volz}}, \
  and\ \bibinfo {author} {\bibfnamefont {A.}~\bibnamefont {Rauschenbeutel}},\
  }\href@noop {} {\bibfield  {journal} {\bibinfo  {journal} {Science}\ }\textbf
  {\bibinfo {volume} {346}} (\bibinfo {year} {2014})}\BibitemShut {NoStop}%
\bibitem [{\citenamefont {Mitsch}\ \emph {et~al.}(2014)\citenamefont {Mitsch},
  \citenamefont {Sayrin}, \citenamefont {Albrecht}, \citenamefont
  {Schneeweiss},\ and\ \citenamefont {Rauschenbeutel}}]{Mitsch2014}%
  \BibitemOpen
  \bibfield  {author} {\bibinfo {author} {\bibfnamefont {R.}~\bibnamefont
  {Mitsch}}, \bibinfo {author} {\bibfnamefont {C.}~\bibnamefont {Sayrin}},
  \bibinfo {author} {\bibfnamefont {B.}~\bibnamefont {Albrecht}}, \bibinfo
  {author} {\bibfnamefont {P.}~\bibnamefont {Schneeweiss}}, \ and\ \bibinfo
  {author} {\bibfnamefont {A.}~\bibnamefont {Rauschenbeutel}},\ }\href
  {\doibase 10.1038/ncomms6713} {\bibfield  {journal} {\bibinfo  {journal}
  {Nature Communications}\ }\textbf {\bibinfo {volume} {5}},\ \bibinfo {pages}
  {5713} (\bibinfo {year} {2014})}\BibitemShut {NoStop}%
\bibitem [{\citenamefont {Pichler}\ \emph {et~al.}(2015)\citenamefont
  {Pichler}, \citenamefont {Ramos}, \citenamefont {Daley},\ and\ \citenamefont
  {Zoller}}]{Pichler2015}%
  \BibitemOpen
  \bibfield  {author} {\bibinfo {author} {\bibfnamefont {H.}~\bibnamefont
  {Pichler}}, \bibinfo {author} {\bibfnamefont {T.}~\bibnamefont {Ramos}},
  \bibinfo {author} {\bibfnamefont {A.~J.}\ \bibnamefont {Daley}}, \ and\
  \bibinfo {author} {\bibfnamefont {P.}~\bibnamefont {Zoller}},\ }\href
  {\doibase 10.1103/PhysRevA.91.042116} {\bibfield  {journal} {\bibinfo
  {journal} {Physical Review A}\ }\textbf {\bibinfo {volume} {91}},\ \bibinfo
  {pages} {042116} (\bibinfo {year} {2015})}\BibitemShut {NoStop}%
\end{thebibliography}
\end{document}